%% file: ms.tex
\documentclass[preprint,12pt]{aastex62}

\shorttitle{Orion Disks}
\shortauthors{Tobin et al.}

\newcommand{\thco}{\mbox{$^{13}$CO}}
\newcommand{\twco}{\mbox{$^{12}$CO}}

\newcommand{\kms}{\mbox{km s$^{-1}$}}

\newcommand{\lsun}{\mbox{L$_{\sun}$}}
\newcommand{\msun}{\mbox{M$_{\sun}$}}
\newcommand{\mearth}{\mbox{M$_{\earth}$}}
\newcommand{\tbol}{\mbox{T$_{bol}$}}
\newcommand{\lbol}{\mbox{L$_{bol}$}}
\def\rad{r_\mathrm{disk,AD}}   
\def\rhd{r_\mathrm{disk,HD}}   
\begin{document}

\title{The VLA/ALMA Nascent Disk and Multiplicity (VANDAM) Survey of Orion Protostars. A Statistical Characterization of Class 0 and I Protostellar Disks}
\author[0000-0002-6195-0152]{John J. Tobin}
\affiliation{National Radio Astronomy Observatory, 520 Edgemont Rd., Charlottesville, VA 22903, USA}
\author{Patrick D. Sheehan}
\affiliation{National Radio Astronomy Observatory, 520 Edgemont Rd., Charlottesville, VA 22903, USA}
\author{S. Thomas Megeath}
\affiliation{Department of Physics and Astronomy, University of Toledo, Toledo, OH 43560}
\author[0000-0001-9112-6474]{Ana Karla D{\'i}az-Rodr{\'i}guez }
\affiliation{Instituto de Astrof\'{\i}sica de Andaluc\'{\i}a, CSIC, Glorieta de la Astronom\'{\i}a s/n, E-18008 Granada, Spain}
\author{Stella S. R. Offner}
\affiliation{The University of Texas at Austin, 2500 Speedway, Austin, TX USA}
\author{Nadia M. Murillo}
\affiliation{Leiden Observatory, Leiden University, P.O. Box 9513, 2300-RA Leiden, The Netherlands}
\author{Merel L. R. van 't Hoff}
\affiliation{Leiden Observatory, Leiden University, P.O. Box 9513, 2300-RA Leiden, The Netherlands}
\author{Ewine F. van Dishoeck}
\affiliation{Leiden Observatory, Leiden University, P.O. Box 9513, 2300-RA Leiden, The Netherlands}
\author[0000-0002-6737-5267]{Mayra Osorio}
\affiliation{Instituto de Astrof\'{\i}sica de Andaluc\'{\i}a, CSIC, Glorieta de la Astronom\'{\i}a s/n, E-18008 Granada, Spain}
\author[0000-0002-7506-5429]{Guillem Anglada}
\affiliation{Instituto de Astrof\'{\i}sica de Andaluc\'{\i}a, CSIC, Glorieta de la Astronom\'{\i}a s/n, E-18008 Granada, Spain}
\author[0000-0001-9800-6248]{Elise Furlan}
\affiliation{IPAC, Mail Code 314-6, Caltech, 1200 E. California Blvd., Pasadena,CA 91125, USA}
\author{Amelia M. Stutz}
\affiliation{Departmento de Astronom\'{i}a, Universidad de Concepci\'{o}n, Casilla 160-C, Concepci\'{o}n, Chile}
\author{Nickalas Reynolds}
\affiliation{Homer L. Dodge Department of Physics and Astronomy, University of Oklahoma, 440 W. Brooks Street, Norman, OK 73019, USA}
\author[0000-0003-3682-854X]{Nicole Karnath}
\affiliation{Department of Physics and Astronomy, University of Toledo, Toledo, OH 43560}
\author[0000-0002-3747-2496]{William J. Fischer}
\affiliation{Space Telescope Science Institute, Baltimore, MD, USA}
\author{Magnus Persson}
\affiliation{Chalmers University of Technology, Department of Space, Earth and Environment, Sweden}
\author{Leslie W. Looney}
\affiliation{Department of Astronomy, University of Illinois, Urbana, IL 61801}
\author{Zhi-Yun Li}
\affiliation{Department of Astronomy, University of Virginia, Charlottesville, VA 22903}
\author{Ian Stephens}
\affiliation{Harvard-Smithsonian Center for Astrophysics, 60 Garden St, MS 78, Cambridge, MA 02138}
\author{Claire J. Chandler}
\affiliation{National Radio Astronomy Observatory, P.O. Box O, Socorro, NM 87801}
\author{Erin Cox}
\affiliation{Center for Interdisciplinary Exploration and Research in Astrophysics (CIERA) and Department of Physics and Astronomy, Northwestern University, Evanston, IL 60208, USA}
\author{Michael M. Dunham}
\affiliation{Department of Physics, State University of New York Fredonia, Fredonia, New York 14063, USA}
\affiliation{Harvard-Smithsonian Center for Astrophysics, 60 Garden St, MS 78, Cambridge, MA 02138}
\author[0000-0002-9470-2358]{{\L}ukasz Tychoniec}
\affiliation{Leiden Observatory, Leiden University, P.O. Box 9513, 2300-RA Leiden, The Netherlands}
\author{Mihkel Kama}
\affiliation{Institute of Astronomy, Madingley Road, Cambridge CB3 OHA, UK}
\author{Kaitlin Kratter}
\affiliation{University of Arizona, Steward Observatory, Tucson, AZ 85721}
\author[0000-0002-5365-1267]{Marina Kounkel}
\affiliation{Department of Physics and Astronomy, Western Washington University, 516 High St., Bellingham, WA 98225, USA}
\author{Brian Mazur}
\affiliation{Department of Physics and Astronomy, University of Toledo, Toledo, OH 43560}
\author{Luke Maud}
\affiliation{Leiden Observatory, Leiden University, P.O. Box 9513, 2300-RA Leiden, The Netherlands}
\affiliation{European Southern Observatory, Garching, Germany}
\author{Lisa Patel}
\affiliation{Homer L. Dodge Department of Physics and Astronomy, University of Oklahoma, 440 W. Brooks Street, Norman, OK 73019, USA}
\author{Laura Perez}
\affiliation{Departamento de Astronom\'ia, Universidad de Chile, Camino El Observatorio 1515, Las Condes, Santiago, Chile}
\author{Sarah I. Sadavoy}
\affiliation{Harvard-Smithsonian Center for Astrophysics, 60 Garden St, MS 78, Cambridge, MA 02138}
\author{Dominique Segura-Cox}
\affiliation{Max-Planck-Institut f{\"u}r extraterrestrische Physik, Giessenbachstrasse 1, D-85748 Garching, Germany}
\author{Rajeeb Sharma}
\affiliation{Homer L. Dodge Department of Physics and Astronomy, University of Oklahoma, 440 W. Brooks Street, Norman, OK 73019, USA}
\author{Brian Stephenson}
\affiliation{Homer L. Dodge Department of Physics and Astronomy, University of Oklahoma, 440 W. Brooks Street, Norman, OK 73019, USA}
\author{Dan M. Watson}
\affiliation{Department of Physics and Astronomy, University of Rochester, Rochester, NY 14627}
\author{Friedrich Wyrowski}
\affiliation{Max-Planck-Institut f\"ur Radioastronomie, Auf dem H\"ugel 69, 53121, Bonn, Germany}

\begin{abstract}
We have conducted a survey of 328 protostars in the Orion molecular clouds with ALMA at 0.87 mm at a resolution of $\sim$0\farcs1 (40~au), 
including observations with the VLA at 9~mm toward 148 protostars at a resolution of
$\sim$0\farcs08 (32~au). This is the largest
multi-wavelength survey of protostars at this resolution by an order of magnitude.
We use the dust continuum emission at 0.87~mm and 9~mm to measure the dust
disk radii and masses toward the Class 0, Class I, and Flat Spectrum protostars, characterizing the evolution
of these disk properties in the protostellar phase. The mean dust disk
radii for the Class 0, Class I, and Flat Spectrum protostars are 
44.9$^{+5.8}_{-3.4}$, 37.0$^{+4.9}_{-3.0}$, and 28.5$^{+3.7}_{-2.3}$~au,
respectively, and the mean protostellar dust disk masses are 
25.9$^{+7.7}_{-4.0}$, 14.9$^{+3.8}_{-2.2}$, 11.6$^{+3.5}_{-1.9}$~\mearth,
respectively. The decrease in dust disk masses is expected from disk evolution and
accretion, but the decrease in disk radii may point to the initial conditions of star
formation not leading to the systematic growth of disk radii or that radial drift 
is keeping the dust disk sizes small. At least
146 protostellar disks (35\% out of 379 detected 0.87~mm continuum sources plus 42 non-detections)
have disk radii greater than 50 au in our sample. These properties are not found to vary 
significantly between different regions within Orion.
The protostellar dust disk mass distributions are systematically
larger than that of Class II disks by a factor of $>$4, providing
evidence that the cores of giant planets may need to at least begin their formation during 
the protostellar phase.  
\end{abstract}

\section{Introduction}

The formation of stars and planets is initiated by the gravitational collapse of dense clouds
of gas and dust. In order for gravitational collapse
to proceed, other sources of support \citep[e.g., thermal pressure, 
magnetic fields, turbulence;][]{mckeeostriker2007} must
either be reduced or not significant at the onset of collapse. As the protostar is forming
within a collapsing envelope of gas and dust, a rotationally-supported 
disk is expected to form around the protostar via conservation of 
angular momentum. 
Once a disk has formed, the majority of accretion onto the star will happen through 
the disk, and the disk material is expected to provide the raw material for planet formation.

The angular momentum that drives disk formation
may originate from rotation of the core ($\sim$0.05~pc in diameter), but organized
rotation of cores is found less frequently as cores are observed with higher angular
resolution and sensitivity \citep[e.g.,][Chen et al. 2019]{tobin2011,tobin2012,tobin2018}. 
Thus, the angular momentum may not derive from organized core rotation. The origin of the 
net angular momentum is not specifically important, but within larger-scale molecular 
clouds (1 - 10~pc), the angular momentum within cores that leads to the formation of
disks likely derives from the residual core-scale turbulent motion of the
gas or gravitational torques between overdensities in the molecular cloud 
\citep{burkert2000,offner2016,kuznetsova2019}. However, in order for conservation 
of angular momentum to lead to the formation
of disks around protostars \citep[e.g.,][]{tsc1984}, magnetic fields must not be strong enough
or not coupled strongly enough to the gas to prevent the spin-up of infalling material 
as it conserves angular momentum during collapse \citep{allen2003,
mellon2008,padovani2013}. On the other hand, non-ideal magneto-hydrodynamic effects (MHD) can also dissipate the magnetic flux and
enable the formation of disks to proceed \citep[e.g.,][]{dapp2010,li2014,masson2016,hennebelle2016}, as can
turbulence and/or magnetic fields misaligned with the core rotation axis \citep{seifried2012,joos2012}.

The youngest observationally recognized protostars are those in the Class 0 phase, 
in which a dense infalling envelope of gas and dust surrounds the protostar
\citep{andre1993}. The Class I phase follows, where
the protostar is less deeply embedded, but still surrounded by an infalling envelope.
The transition between Class 0 and Class I is not exact, but 
a bolometric temperature (\tbol) of 70~K or L$_{bol}$/L$_{submm}$ $<$ 0.005 have been
adopted as the divisions between the classes. 
\tbol\ is a typical diagnostic to characterize the evolutionary state 
of a young star \citep{ladd1993,dunham2014}.
The envelope is expected to be largely dissipated by the end of the Class I phase, leaving
a disk surrounding a pre-main sequence star, also known as 
Class II YSOs, which have \tbol~$>$~650~K \citep[e.g.,][]{dunham2014}. 
Furthermore, a possible transition phase prior to becoming 
a Class II YSO, known as Flat Spectrum sources, also exists. These protostars are 
characterized by a flat spectral energy
distribution (SED) in $\lambda$F$_{\lambda}$ from $\sim$2~\micron\ to 24~\micron. 
The nature of Flat Spectrum sources with respect to Class I sources is still
unclear. Some Flat Spectrum sources are suggested to be Class II based on their lack of 
dense molecular gas \citep{vankempen2009,heiderman2015}, but SED modeling of the Flat
Spectrum sources in Orion found that they were best fit by models with an envelope
in the majority of systems \citep{furlan2016}.
The length of the protostellar phase (Class 0, I, and Flat Spectrum) combined has been
estimated to be $\sim$500~kyr and the Class 0 phase itself is estimated to last
$\sim$160~kyr \citep{dunham2014}. However, \citet{kristensen2018} used a different
set of assumptions to derive half-lives of the protostellar phase in which the Class 0,
Class I, and Flat Spectrum phases have half-lives of 74~kyr, 88~kyr, and 87~kyr, respectively, 222~kyr in total.

Disks are observed nearly ubiquitously toward the youngest stellar populations that are dominated
by Class II YSOs, and the frequency of disks within a population declines for older associations of 
YSOs \citep{hernandez2008}. This high occurrence rate of disks in later stages is an 
indication that disk formation is a universal process in star formation. These disks around pre-main-sequence stars
have been commonly referred to as protoplanetary disks or Class II disks, and to draw
distinction between disks around YSOs in the protostellar phase (Class 0, I, and Flat Spectrum), 
we will generically refer to the latter as protostellar disks.

The observed properties of disks throughout the protostellar phase will both inform us of
the conditions of their formation as well as the initial conditions for
disk evolution. The properties of Class 0 disks have been sought after with (sub)millimeter and centimeter-wave 
interferometry, and each increase in the capability of interferometers at these wavelengths has led to 
new constraints on the properties of Class 0 disks from their dust emission. 
\citet{brown2000} used a single baseline
interferometer formed by the James Clerk Maxwell Telescope (JCMT) and the Caltech Submillimeter
Observatory (CSO) to characterize the disk radii toward a number
of Class 0 protostars. \citet{looney2000} used the Berkeley Illinois Maryland Array (BIMA)
to resolve a number of Class 0, Class I, and Class II protostars, 
measuring disk radii, masses, and multiple systems.
\citet{harvey2003} used the Plateau de Bure Interferometer (PdBI) to characterize the 
unresolved disk toward B335, finding a dust disk with a radius less than 100~au and a dust mass
of $\sim$4$\times$10$^{-5}$~\msun. However, the sensitivity and resolution of these earlier instruments
was not sufficient to characterize the disks with extremely high fidelity, nor were samples large enough
to be statistically meaningful.

Larger samples of disks and higher-fidelity imaging with upgraded interferometers began with
the Submillimeter Array (SMA), using unresolved observations to infer
the masses of protostellar disks from the Class 0 to Class I phase 
\citep{jorgensen2009}. \citet{maury2010} observed
5 Class 0 protostars with the PdBI, which only had sufficient resolution to detect dust
disks with radii larger than $\sim$150~au and none were positively identified. 
\citet{chiang2012} used multi-configuration
observations with the Combined Array for Millimeter-wave Astronomy (CARMA) toward the 
Class 0 protostar L1157-mm
to identify a candidate unresolved disk with a radius smaller than $\sim$100~au. Also, \citet{enoch2011}
examined a sample of 9 candidate protostellar disks in Serpens, including a possible disk toward the 
Class 0 protostar Serpens FIRS1 \citep{enoch2009}.
Despite the improved sensitivity of these instruments, most studies were limited to characterizing
disks via dust continuum emission
with a best resolution of $\sim$0\farcs3 ($\sim$120~au).
This means that these observations only primarily probed the dust disks and not the gas disks.

Molecular line observations were possible toward some of the most nearby protostellar disks with
the previous generation of instruments. \citet{tobin2012} were able to use
CARMA to positively resolve the disk toward the Class 0 protostar L1527 IRS in the dust continuum
and identify likely Keplerian rotation from \thco\ emission, and \citet{murillo2013} detected possible
evidence of disk rotation toward VLA 1623 with the SMA, which is now recognized to be a
triple system with a circum-multiple disk \citep{harris2018, sadavoy2018}. At the same time, observations of disks
toward Class I protostars had also yielded some detections of resolved disks 
and Keplerian rotation \citep{wolf2008,takakuwa2012,launhardt2009,harsono2014,harris2018,alves2018}.

The advent of the Atacama Large Millimeter/submillimeter Array (ALMA) came on the heels of these
pioneering studies with more than an order of magnitude greater sensitivity and 
angular resolution. ALMA is leading a
revolution in the characterization of individual protostellar disks, confirming and extending 
earlier results such as the Class 0 rotationally-supported disk around L1527 IRS
\citep{ohashi2014, sakai2014, aso2017}.
Furthermore, a number of new Class 0 disks have been identified and confirmed
to be rotationally-supported \citep{murillo2013b, lindberg2014, codella2014, yen2017,alves2018},
and some very small Class 0 disks have also been identified \citep{yen2015,hsieh2019}. At the same time,
the characterization of Class I disks has been progressing \citep{yen2014, aso2015, sakai2016}.
Finally, a number of circum-binary and circum-multiple disks have been identified in both the 
Class 0 and Class I phase \citep{tobin2016b,takakuwa2014,harris2018, sadavoy2018}.

A trend that has emerged from the aforementioned studies of Class 0 disks is that, when a 
disk-like morphology is resolved in the dust continuum toward Class 0 and I protostars, this structure is a rotationally supported disk. 
Thus, if a disk-like continuum feature is well-resolved, then it is likely that this feature reflects a
rotationally-supported disk. This has enabled larger surveys that focus primarily
on continuum sensitivity to characterize larger samples of Class 0 and I disks.
\citet{seguracox2016, seguracox2018} used the data from the NSF's Karl G. Jansky Very Large Array (VLA) taken as part of the
VLA Nascent Disk and Multiplicity (VANDAM) Survey, to identify a total of 18 Class 0 disk candidates 
(out of 37 Class 0 protostars and 8 Class 0/I protostars observed), many with 
radii less than 30~au, greatly increasing the range of scales at which Class 0 disk candidates have been resolved.
Finally, \citet{maury2018} used the IRAM-PdBI to conduct a survey of 16 Class 0 protostars as part of
the Continuum And Lines in Young Protostellar Objects (CALYPSO) Survey in both lines
and continuum. The continuum observations found that 4 out of 16 protostars have evidence
for disks with radii $>$ 60 au. While these new continuum surveys are important 
for increasing the statistics, the CALYPSO survey was limited in both sensitivity 
and angular resolution (0\farcs3), while the VANDAM survey had excellent angular 
resolution ($\sim$0\farcs07), but limited surface brightness and dust
mass sensitivity due to the 9~mm wavelength of the observations.

A principal limitation of protostellar disk studies has been the sample size.
Protostars are inherently rarer than the more-evolved pre-main sequence stars with
disks, making their populations in the nearby star-forming regions small. For this
reason, Orion is an essential region to study in order to obtain a representative characterization of protostellar
disk characteristics. 
Orion is the nearest region forming massive stars and richest region of low-mass 
star formation within 500~pc. Orion is also the best
analogue for examining star and planet formation in an environment that is likely representative 
of most star formation in our Galaxy.
Studies of Orion with the \textit{Spitzer Space Telescope} and 
\textit{Herschel Space Observatory} have identified at least 428 protostar candidates in Orion (Class 0
through Flat Spectrum), in addition to 2991 more-evolved dusty young stars 
\citep[Class II and III;][]{megeath2012,furlan2016}. Therefore, while the more nearby regions like
Taurus and Perseus enable protostellar disks to be resolved in greater detail, Orion provides 
a much larger sample of protostars than the nearby star-forming regions. Orion contains nearly as many protostars
as the rest of the Gould Belt, which encompasses all the other star-forming 
regions within 500~pc \citep{dunham2015}. Orion is composed of two main molecular clouds that are
known as the Orion A and Orion B molecular clouds (see Figure \ref{overview}). 
Orion A contains the most active region of
star formation, harboring the Integral-Shaped Filament, the Trapezium, and Orion BN-KL, 
while Orion B also has massive star
formation, as well as the second (NGC 2024) and third (NGC 2068/2071) most massive 
clusters in Orion \citep{megeath2016}. The entire Orion complex spans $\sim$83~pc projected on the plane of the sky, but
the protostars are preferentially located in regions of high gas column density. Both Orion A and
Orion B contain clustered and isolated protostars, and the majority of protostars are not
in close proximity to the Orion Nebula.
Despite being a single region, there is significant distance variation across the plane of the sky.
The Orion Nebula Cluster, the southern
end of Orion A, and Orion B have typical distances of 389~pc, 443~pc, and 407~pc, 
respectively \citep{kounkel2017,kounkel2018}.

The high angular resolution and sensitivity to continuum emission
makes ALMA uniquely suited to characterize the properties of protostellar disks
for large samples such as Orion. However,
even at submillimeter wavelengths the protostellar disks can be optically thick; therefore,
VLA observations at 9~mm are crucial to examine the inner disks.
This has motivated us to conduct the VLA/ALMA Nascent Disk and Multiplicity (VANDAM) survey toward
all well-characterized protostars in the Orion A and B molecular clouds
using ALMA and with VLA observations toward all the Class 0 and the youngest Class I protostars.
We have used the ALMA and VLA
data to characterize the dust disk masses and radii toward a sample of 328 protostars
to better understand the
structure of disks throughout the entire protostellar phase. This is the largest protostellar disk survey 
to date by an order of magnitude. The ALMA and VLA observations are
described in Section 2. The results from continuum observations 
toward all sources are presented in Section 3.  
We discuss our results in Section 4 and present our 
conclusions in Section 5.

\section{Observations and Data Reduction}

\subsection{The Sample}

The sample of protostars is drawn from the \textit{Herschel} Orion 
Protostar Survey \citep[HOPS; ][]{fischer2010,stutz2013,furlan2016}.
We selected all Class 0, Class I, and Flat Spectrum protostars from the survey 
that had reliable measurements of bolometric temperature (\tbol),
bolometric luminosity (\lbol), 70~\micron\ detections,
and were not flagged as extragalactic contaminants.
From that sample of 409 HOPS protostars, we selected
320 HOPS protostars for observations with ALMA using the aforementioned criteria. We also included
a few sources that were not part of the HOPS sample but are bonafide protostars 
in Orion B \citep[HH270VLA1, HH270mms1, HH270mms2, HH212mms, HH111mms;][]{reipurth1999,choi2006,wiseman2001,lee2017},
and 3 unclassified protostellar candidates from \citet{stutz2013} (S13-021010, S13-006006, S13-038002).
This makes the total number of protostellar systems observed 328, of which 
94 are Class 0 protostars, 128 are Class I protostars, 103 
are Flat spectrum sources, and 3 were unclassified but expected to be Class 0 or I.
The luminosity range of the sample is 0.1~\lsun\ to $\sim$1400~\lsun. 
An overview image of the Orion region with the targeted protostars overlaid is 
shown in Figure \ref{overview}, and we show a plot of \lbol\ vs. \tbol\ for the
sample in Figure \ref{lbol-tbol}.

There is a distance variation on the order of $\pm$40~pc across the Orion A and B molecular clouds \citep{kounkel2017,kounkel2018}.
To mitigate its impact on our analysis, we take advantage of the availability of \textit{Gaia} data for a large sample of more evolved 
members within Orion, enabling us to estimate the distance toward each protostellar system. These distance estimates enable
more precise calculations of the physical properties of the systems and comparison of the flux densities on
a common scale. The method for estimating the distances is described in Appendix A; however, with respect to the
typical distance of 400~pc to the region, the distances are all within $\sim$10\% of this value.

\subsection{ALMA 0.87~mm Observations}

ALMA is located in northern Chile on the Chajnantor plateau at an elevation
of $\sim$5000~m. The protostars in Orion selected for observations with ALMA at 0.87~mm 
were divided into three scheduling blocks. One scheduling block contained the selected
protostars in the Orion B molecular cloud and two other scheduling blocks contained the
selected protostars in the Orion A molecular cloud. Each scheduling block was 
successfully executed three times for nine executions in total.
Six were executed in 2016 
September, and three were executed in 2017 July. The date of each observation,
number of antennas, precipitable water vapor, and maximum baseline are given in Table 1; the combined
datasets sample baseline lengths from $\sim$15~m to $\sim$3700~m. We list the 
targeted protostars in Table 2; the total time on each source was $\sim$0.9 minutes.

The correlator was configured to provide high continuum sensitivity.
We used two basebands set to low spectral resolution continuum mode,
1.875 GHz bandwidth divided into 128, 31.25~MHz channels,
centered at 333 GHz and 344 GHz.
We also observed \twco\ ($J=3\rightarrow2$) at 345.79599~GHz and \thco\ ($J=3\rightarrow2$)
at 330.58797~GHz. 
The baseband centered on \twco\ ($J=3\rightarrow2$) had a total bandwidth
of 937.5~MHz and 0.489~\kms\ channels, and the baseband centered on \thco\ ($J=3\rightarrow2$) 
had a bandwidth of 234.375~MHz with 0.128~\kms\ channels. The line-free regions of the
\twco\ and \thco\ basebands were used for additional continuum bandwidth, resulting in 
an aggregate continuum bandwidth of $\sim$4.75~GHz. 

The calibrators used for each execution are listed in Table 1. The data were manually reduced 
by the Dutch Allegro ARC Node using the Common Astronomy 
Software Application \citep[CASA][]{mcmullin2007}. The manual reduction was necessary to compensate
for variability of the quasar J0510+1800 that was used for absolute flux calibration in some
executions. The absolute flux calibration 
accuracy is expected to be $\sim$10\%, and comparisons of the observed flux densities 
for the science targets during different executions
are consistent with this level of accuracy. However, we only use
statistical uncertainties for the flux density measurements and their derived quantities
throughout the paper.

After the standard calibration, we performed up to three rounds of phase-only 
self-calibration on the continuum data to increase the S/N. The ability to self-calibrate
depends on the S/N of the data, and we only attempted self calibration when the S/N of the
emission peak was $>$10. For each successive round of self-calibration, we used solution
intervals that spanned the entire scan length for the first round, as short as 12.08~s in the second round,
and as short as 3.02~s in the third round, which was the length of a single integration. 
The solution interval was adjusted in the second and/or third round depending on the S/N of
the source and the number of flagged solutions reported. We applied the self-calibration
solutions using the CASA \textit{applycal} task using \textit{applymode=calonly} to avoid flagging data
for which a self-calibration solution did not have high enough S/N to converge on a solution in a
given round of self-calibration, but were otherwise good. Given the short total time on source, our
observations were able to reach close to the thermal noise limit and were not strongly
limited by dynamic range in most instances.

Following the continuum self-calibration, the phase solutions were then applied 
to the \twco\ and \thco\ spectral line data. The typical
root-mean-squared (RMS) noise of the continuum, \twco, and \thco\
are 0.31~mJy~beam$^{-1}$, 17.7 mJy~beam$^{-1}$ (1~\kms\ channels), 
and 33.3 mJy~beam$^{-1}$ (0.5~\kms\ channels), respectively. The spectral line observations
were averaged by two and four channels for \twco\ and \thco, respectively, to
reduce noise. The 
continuum and spectral line data cubes were imaged
using the \textit{clean} task of CASA 4.7.2 for all
self-calibration and imaging.

The aggregate continuum image was reconstructed using \textit{Briggs} weighting with a \textit{robust}
parameter of \textit{0.5}, yielding a synthesized beam of $\sim$0\farcs11 (44~au). We also made images with
\textit{robust=2, 0, and -0.5}, but we primarily use the \textit{robust=0.5} images in this paper, providing
a compromise between sensitivity and angular resolution. For protostars that are not well-detected,
we use the \textit{robust=2} images.

The continuum images are reconstructed only using
uv-points at baselines $>$25~k$\lambda$ to mitigate striping resulting from large-scale emission
that is not properly recovered. This data selection typically only removes a single baseline, and there is a gap
between the shortest baseline and where the density of uv-points increases significantly. 
We used a different approach for the spectral line data because the \twco\ and \thco\ emission
is typically much more extended than the continuum. We imaged the spectral line data
using \textit{Natural} weighting for baselines $>$50~k$\lambda$ to mitigate striping 
and with an outer taper of 500~k$\lambda$ 
applied to increase the sensitivity to extended 
structure; this yielded synthesized beams of $\sim$0\farcs25. However, we focus on the continuum
for the remainder of this paper and do not discuss the spectral line data further.

\subsection{VLA Observations}

We conducted observations with the VLA in A-configuration
between 2016 October 20 and 2017 January 07 in $\sim$100
individual observations; the observations are detailed in Table 3. We also conducted observations of
the sources in C-configuration during February and March of 2016 with $\sim$1\arcsec\ resolution, 
but these data were primarily used for A-configuration target selection and are not utilized in this paper
except for a few upper limits. The targeted fields are detailed in Table 4.

The observations
used the Ka-band receivers and the correlator was used in the wide bandwidth mode (3-bit samplers) with
one 4~GHz baseband centered at 36.9~GHz (8.1~mm) and the other baseband was centered at 29~GHz (1.05~cm).
Most observations were conducted in $\sim$2.5 hour scheduling blocks
toward a single source
with $\sim$1 hour on-source. However, a few observations
were conducted in 4 hour scheduling blocks, observing two sources, each for $\sim$ 1 hr.
In all observations, the absolute flux calibrator was 
3C48 (J0137+3309), the bandpass calibrator was 3C84 (J0319+4130),
and the complex gain calibrator was either J0552+0313 or J0541-0541 
for protostars associated with Orion B or Orion A, respectively. The observations were
conducted in fast-switching mode ($\sim$2.6 minute cycle times) to reduce phase decoherence in the high frequency observations, and
between 25 and 27 antennas were available during each observation. 
The antenna pointing corrections were updated prior to observing the flux calibrator, 
bandpass calibrator, before the first observation of the complex gain calibrator,
and after one hour had elapsed since the last pointing update.
The absolute calibration uncertainty of the VLA data is expected to be $\sim$10\%, and, similar to the
ALMA data, we only report the statistical uncertainties in this paper.

The data were reduced using the scripted version of the VLA pipeline in CASA 4.4.0. We note that some of our observations
were obtained during the period where the tropospheric delay correction was being misapplied to all VLA data;
all A-configuration data prior to 2016 November 14 were affected.
This resulted in a phase offset that was larger for lower elevations and when the angular
separation of the source to the calibrator was large. When this error was integrated over an 
entire scheduling block that included observations at elevation below 30\degr, the continuum
images would be smeared in the direction of elevation. However, 
we did not have a large separation between source and calibrator in most cases and the
data were not taken for long periods at below 30\degr\ elevation. For sources that were determined
to be strongly affected by the delay error, we utilized CASA 4.5.2 to run the VLA pipeline which incorporated a
fix for the delay error.

We performed phase-only self-calibration on HOPS-370, HOPS-384, and HOPS-361
because these fields had high enough S/N to be dynamic range limited ($>$100). 
To perform self calibration, we used two 
solution intervals of 230~s (first round) and 90~s (second round), 
which corresponded to one solution for every two scans and one solution for each scan, 
respectively. 

The continuum data for all sources were imaged using the \textit{clean} 
task in CASA 4.5.1 using 
\textit{Natural} weighting and multi-frequency synthesis with \textit{nterms=2} 
across both basebands. The final images have an RMS noise of $\sim$7-8~$\mu$Jy~beam$^{-1}$ 
and a synthesized beam of 
$\sim$0\farcs08 (32~au).

\subsection{Data Analysis}

We fit elliptical Gaussians to each detected source using the \textit{imfit} task of 
CASA 4.7.2. This enables us to measure the flux density of each source,
its size, and its orientation from the major and minor axes of the Gaussian fits. 
While Gaussian fitting has limitations, its advantage lies in its simplicity and
ability to rapidly fit a large number of sources. The principal metrics that
we aim to derive are the protostellar disk radii and masses. Other methods
used to observationally estimate disk radii include the curve of growth method used on the Lupus
survey \citep{ansdell2016} and fitting a `Nuker profile' \citep{tripathi2017}. However, these
methods are less ideal for protostellar disks.
The curve of growth method works best if
the orientation of the disk can be determined from 
its observed aspect ratio, enabling its visibility data and images to be deprojected,
and the 'Nuker profile' requires an assumption of an intensity profile.
These methods and assumptions are not always possible and/or reliable for protostellar disks,
due in large part to the surrounding envelope. Thus, these other methods 
will not necessarily lead to better results for protostellar disks.

We note that the curve of growth methodology employed by \citet{ansdell2016} defined the disk radius as the radial
point which contains 90\% of the total flux density. When compared with a Gaussian fit,
this is approximately the 2$\sigma$ point of a Gaussian.
If one considers exponentially tapered disks, with a surface density profile defined 
as $\Sigma$~$\propto$~(R/R$_C$)$^{-\gamma}$, following the discussion in \citet{bate2018} for $\gamma$~$<$~2,
R$_C$ always encompasses 63.2\% of the dust disk mass, close to the 1$\sigma$ value of a Gaussian (68\%).
R$_C$ is the critical radius, where the surface density of the disk begins to be truncated with
an exponential taper. If the disks have a power-law surface density profile (exponentially-tapered
or not), their intensity profile will not necessarily
be well-described by a Gaussian when resolved. In fact, a Gaussian can systematically underestimate the size
of an object with a power-law surface density (and intensity) profile due to a power-law decaying 
more slowly than a Gaussian. However, despite these caveats, we adopt
the 2$\sigma$ size of the deconvolved major axis
as a proxy for disk radius. Its value represents a compromise between
potentially overestimating the disk radii by using a radius defined 
by the 90\% level of the total flux density \citep[e.g.,][]{ansdell2018}
and underestimating the disk radius by using 1$\sigma$. To convert to a radius in au, we
multiply the full-width at half-maximum (FWHM) (in arcsec) by 2.0/2.355
and multiply by the estimated distance (in pc) toward the 
protostar\footnote{The FWHM of a Gaussian is equivalent to 2 (2 ln (2))$^{0.5}$$\sigma$~$\simeq$~2.355$\sigma$.}.
This radius will contain $\sim$95\% of the flux density within the fitted Gaussian.
Assuming that the submillimeter/centimeter flux density traces mass, then the 
2$\sigma$ radius may be somewhat larger than the expected $R_C$ 
for exponentially tapered disks, but the 2$\sigma$ radius can also 
systematically underestimate the full radius of the disks if they are not well-described
by Gaussians.

The integrated flux density measured with the Gaussian fit is used to analytically estimate
the mass of the protostellar disks in each detected system. We make
the assumption that the disk is isothermal and optically thin, enabling us to use the 
equation
\begin{equation}
\label{eq:dustm}
M_{dust} = \frac{D^2 F_{\nu} }{ \kappa_{\nu}B_{\nu}(T_{dust}) },
\end{equation}
where D is the estimated distance toward the protostar,
$F_{\nu}$ is the observed flux density, $B_{\nu}$ is
the Planck function, $T_{dust}$ is the dust temperature, and $\kappa_{\nu}$ is the
dust opacity at the observed wavelength. If the dust emission is not optically thin,
then the masses will be lower limits.
We adopt $\kappa_{0.87mm}$~=~1.84~cm$^2$~g$^{-1}$ 
 from \citet{ossenkopf1994}, and at 9.1~mm we adopt a dust opacity of 0.13~cm$^2$~g$^{-1}$
by extrapolating from the \citet{ossenkopf1994} dust opacity at 1.3~mm (0.899~cm$^2$~g$^{-1}$)
 assuming a dust opacity spectral index of 1.

In the literature, T$_{dust}$ is typically assumed to be 30~K for solar-luminosity protostars
 \citep{tobin2015,tobin2016a, tychoniec2018}. Given the wide range of luminosities 
for the protostars in Orion \citep[see Figure \ref{lbol-tbol};][]{fischer2017},
it is essential that we scale T$_{dust}$ using the bolometric luminosity for each system
in order to obtain more realistic dust mass measurements.
We used a grid of radiative
transfer models to calculate the appropriate average temperature to use for
protostellar disks found in systems with particular luminosities and 
radii (Appendix B). 
We note, however, that the dust emission from the disks can be optically thick, 
resulting in underestimates of the dust disk masses.

Based on these models, we adopt an average dust temperature of
\begin{equation}
T_{dust} = T_{0}\left(\frac{L_{bol}}{1~L_{\odot}}\right)^{0.25}
\end{equation}
where T$_{0}$ = 43~K, and we scale this using \lbol\ for each protostellar system.
The average dust temperature of 43~K is reasonable for a $\sim$1~\lsun\ protostar at a
radius of $\sim$50~au \citep[see Appendix B; ][]{whitney2003a,tobin2013}. 
While \citet{tazzari2017} demonstrated that the dust temperature of Class II
disks is typically independent of total luminosity, the dust temperature of disks
embedded within envelopes are not independent of luminosity due to the surrounding envelope
also illuminating the disk (see also \citealt{osorio2003} and Appendix B for further details).
Other studies have similarly employed such corrections to the average dust temperatures
to obtain more realistic mass measurements 
\citep[e.g.,][]{jorgensen2009,andrews2013,wardduong2018}. Our 3$\sigma$ detection limit at 0.87~mm 
($\sim$1~mJy~beam$^{-1}$) corresponds to $\sim$1.1~$M_{\earth}$  for a 1~\lsun\
protostar (T$_{dust}$ = 43~K), and the 3$\sigma$ limit at 9~mm ($\sim$25~$\mu$Jy~beam$^{-1}$) corresponds
to 35~$M_{\earth}$. 

\section{Results}

The ALMA and VLA continuum images reveal compact dusty structures on scales $\la$2\arcsec\
toward the sampled protostars in Orion. The observations have very limited sensitivity to 
structure larger than 2\arcsec\ due to the data being taken in high-resolution configurations
with few short baselines. The ALMA and VLA surveys detected 
the protostellar sources (i.e., dust emission from their disks and/or inner envelopes)
in their targeted fields with a small percentage of non-detections, producing a 
large sample of sources observed at high angular resolution from 
submillimeter to centimeter wavelengths. 

\subsection{Detection Statistics}

Out of 328 protostars targeted with ALMA, 94 are Class 0 protostars, 128 are 
Class I protostars, 103 are Flat Spectrum protostars, 
and 3 are unclassified but presumed protostars. 
The detection statistics are summarized in Table 5.
We detected continuum emission associated
with the protostars in 286 fields with at least S/N $>$ 3, corresponding
to a 87\% detection rate. 
The 42 non-detections correspond to 8 Class 0 protostars,
19 Class I protostars, 12 Flat spectrum protostars, and 3 unclassified but presumed protostars \citep{stutz2013}. 
However, the total number of discrete continuum sources
identified by the survey is 379 when multiple protostar systems are taken into consideration
and additional sources are detected within a field that targeted a protostar. 
Of these discrete source detections,
125 are associated with Class 0 systems, 130 are associated 
with Class I systems, 118 are associated with 
Flat spectrum systems, and 6 are unclassified. Of the unclassified sources, four are associated with
the OMC2-FIR4 core and are very likely protostellar \citep{tobin2019}, 
the other two (HOPS-72 and 2M05414483-0154357) 
are likely more-evolved YSOs due to their association with infrared sources.
HOPS-72 was classified as a potential extra-galactic contaminant 
from its \textit{Spitzer} IRS spectrum, but it is also associated with a bright near-infrared point source
and may indeed be a YSO.

The VLA A-array survey targeted 88 Class 0 protostar systems, 
10 early Class I protostars, and 4 fields in the OMC1N region that are known to 
harbor young systems \citep{teixeira2016} but do not have detections
shortward of millimeter wavelengths. The detection statistics (again S/N $>$ 3)
are also summarized in Table 5. 
The primary beam of the VLA at 9~mm ($\sim$45\arcsec)
also encompassed many additional Class I, Flat spectrum, and more-evolved YSOs. 
A total of 232 discrete continuum sources were detected
within all the VLA fields combined. Of these, 122 are associated
with Class 0 systems, 43 with Class I systems, 26 with Flat spectrum sources, and 41 are unclassified.
Within the unclassified sample, 16 are associated with OMC1N \citep{teixeira2016} and 3 are
associated with OMC2-FIR4; these are all likely to be Class 0 or I protostars. Then, 
20 are associated with near-infrared sources and are likely more-evolved YSOs. Finally, the last two 
unclassified sources have strong negative spectral indices 
with increasing frequency and are likely background quasars.
There were 46 non-detections of 9~mm continuum associated
with protostellar sources; this number includes additional continuum sources detected 
by ALMA that were not detected with the VLA. These are separated into 12 Class 0 systems (totaling 20
continuum sources), 16 Class I, 10 Flat Spectrum, and 1 unclassified source. 

The non-detections of Class 0 systems with both ALMA and the VLA are of particular interest. Neither ALMA
nor the VLA detected HOPS-38, HOPS-121, HOPS-316, HOPS-391, and HOPS-380.
HOPS-38, HOPS-121, HOPS-316, and HOPS-391 were likely misclassified
due to poor photometry (and/or blending at long wavelengths) and are likely
not protostars. However, HOPS-380
could be a low-luminosity embedded source. The
Class 0 systems HOPS-137, HOPS-285, and HOPS-396 
were also not-detected by ALMA, but these were eliminated from the VLA Orion sample because further 
inspection of their photometry lead us to doubt their status of Class 0 protostars. 
They had point-like detections in all \textit{Spitzer} IRAC and MIPS 24~\micron\ bands 
and possible contamination from extended emission to their far-infrared flux densities and/or
upper limits; they could be more-evolved YSOs with very low-mass disks.

The additional Class 0 non-detections
with the VLA were HOPS-44, HOPS-91, HOPS-256, HOPS-243, HOPS-326, HOPS-371, and HOPS-374. These
were all detected by ALMA, but did not have strong enough dust emission and/or free-free emission
to enable detection with the VLA. HOPS-91 and HOPS-256 were the only non-detected Class 0 systems that
were also observed in A-configuration with the VLA. The others were non-detections in 
C-configuration and removed from the A-array sample. The remaining 8 non-detections associated with Class 0s
for the VLA are wide companions ($>$1000~au separations) associated with Class 0 systems; the companions
were detected by ALMA and not the VLA. Thus, the number of complete systems classified as Class 0 
that do not have detections with the VLA and ALMA are 12 and 8, respectively.

Considering each protostellar system as a whole, we detected both 0.87~mm and 9~mm continuum 
toward 76 Class 0 protostars, 35 Class I, and 16 Flat Spectrum, 1 Class II and 1 unclassified source (likely Class II).
Note that for these statistics we did not subdivide the systems that are small clusters in 
and of themselves. The systems HOPS-108, HOPS-361, and HOPS-384 had
 many continuum sources detected toward them, but these regions are confused at near- to mid-infrared
wavelengths, preventing individual classification. In total, there
are 175 continuum sources detected at both 0.87~mm and 9~mm; 106 are associated with Class 0 protostars, 
41 with Class I protostars, 23 with Flat Spectrum sources, 1 Class II source, 
and 4 unclassified sources that are likely YSOs. Our continuum depth at 
0.87~mm was not extremely sensitive; therefore we do not we expect a significant number of extragalactic detections.

\subsection{Continuum Emission at 0.87~mm and 9~mm}

 We show ALMA and VLA images toward a representative subset of protostars
in Figures \ref{continuum-demo1} and \ref{continuum-demo2}, while images of the full
complement of detected sources are shown in Appendix C. The ALMA 0.87~mm
images show extended dust emission that appears well-resolved and disk-like for many protostars,
while many others show marginally-resolved and/or point-like emission. Our observations zoom in on the
innermost regions of the protostars, resolving the scales on which disks are expected to be present
\citep{tobin2012,seguracox2016, andrews2009, hennebelle2016}. Thus, for simplicity we
refer to the resolved and unresolved continuum structures observed toward these protostars
as disks, despite their Keplerian nature not being characterized in these observations.
Seven Class 0 protostars may contain a large contribution from an envelope; we will 
discuss these in Section 4.5.

Some protostars in the sample exhibit close multiplicity on scales less than 1\farcs25 (500~au), and
many of these close multiple systems can be seen in the individual panels shown
in Figure \ref{continuum-demo1} and Appendix C; some of these systems contain
multiple resolved disks in a single system. Other protostars in the sample exhibit multiplicity on 
scales greater than 1\farcs25 (500~au), and those systems are shown in images with a larger field of view in 
Figure \ref{continuum-demo2}. We only show neighboring sources for separations less than 11\arcsec,
such that they are detectable within the ALMA field of view at 0.87~mm. The multiplicity properties
of the protostars, such as the distribution of separations and multiplicity frequencies, are
not discussed further here and will be published in a forthcoming paper. 
Throughout the paper, it is useful to separate the sample into the full sample and
non-multiple sample. Non-multiples refer to any system
that does not have an ALMA- or VLA-detected companion within 10,000 au.

We consider each detected source, whether it is part of a multiple system
 or not, individually for the measurement of flux densities, computation of 
mass estimates, and radii measurements from Gaussian fitting. There are many cases where the
companion protostars are close enough that they were not resolved in previous
infrared observations from the HOPS program \citep{furlan2016} and \textit{Spitzer}
surveys of the region \citep{megeath2012}. In those instances, we assume that the measurements
of \lbol\ and \tbol\ apply to both components of the protostar system because they are
embedded within a common protostellar envelope and there is no way to reliably determine
the luminosity ratio of the presumed individual protostars associated with the 
compact dust emission from their disks \citep{murillo2016}. Tables 2 and 4
document the observed fields and protostars associated with them, along with
their corresponding \lbol, \tbol, and distance measurements for
ALMA and the VLA, respectively. 
Tables 6 and 7 list the source positions, fields, flux densities, and orientation parameters
derived from Gaussian fitting from the ALMA and VLA data, respectively. The derived 
properties of each source from the ALMA and VLA flux densities and sizes determined from
Gaussian fitting are given in Table 8. We followed the data analysis procedures outlined 
in Section 2.4 to translate our flux densities and source sizes into protostellar dust disk masses
and radii. We also provide the spectral indices from 0.87~mm to 9~mm and the in-band spectral indices
determined from the VLA data alone.

The comparable resolution of both the ALMA and VLA images enables us to 
compare the structure observed at a factor of 10 difference in wavelength. In many instances,
the ALMA images appear significantly more extended than the VLA images, as shown 
in Figures \ref{continuum-demo1} and \ref{continuum-demo2}, and Appendix C. This 
may be indicative of structure whose emission has a wavelength dependence.
The VLA observations at 9~mm are typically dominated by dust emission \citep{tychoniec2018},
but there are instances where free-free emission from jets \citep{anglada1998} 
can contribute significantly to the flux density at 9~mm.
This emission can be compact and point-like, or it may be extended in the jet direction (see an
example in Figures \ref{continuum-demo1} and \ref{continuum-demo2}). 
The emission at 9~mm can be characterized by the spectral index calculated 
within the Ka-band. Values greater than 2 likely reflect a dominant component 
of dust emission, while values less than 2 
require free-free emission to explain the observed flux density.

We show the flux densities for the ALMA and VLA data plotted together
in Figure \ref{fluxdensity}. There is a strong correlation between the 0.87~mm and 9~mm
flux densities that in log-log space is fit with a constant spectral index ($\alpha$) of 2.24$\pm$0.03 
using \textit{scipy}. This indicates that the emission at the two wavelengths is tracing
a similar process, likely dominated by dust emission. Deviations from
the relationship are evident; excess emission at 9~mm indicates 
a large contribution from free-free emission (or high optical depth at 0.87~mm),
and excess emission at 0.87~mm indicates
that there is less flux at 9.1~mm than expected from the same emission process.

The observed flux densities at 0.87~mm and 9~mm are compared to \lbol\ and \tbol\
of each protostellar system in Figures \ref{lum0.87-blt} and \ref{lum9-blt}. Due to the differences
in estimated distance toward each protostellar system, we multiply the flux densities by the square of the
distance in kpc, yielding a luminosity at the observed wavelengths. The 0.87~mm flux densities span three orders of magnitude
independent of class, and the 9~mm flux densities span about 2 orders of magnitude. There are far 
fewer Class I/Flat Spectrum points at 9~mm due to the selection applied for the VLA observations.
It is clear that only a weak trend exists with respect to the observed flux densities and \tbol; Pearson's 
R is $\sim$-0.28 for the 0.87~mm flux densities and $\sim$-0.17 for the 9~mm flux densities. This  indicates 
a modest correlation for 0.87~mm flux densities, but a very weak correlation for the 9~mm flux densities.
Upper limits were ignored in determining these correlations.

The flux densities at 0.87~mm and 9~mm show clear correlations with \lbol\ in Figures \ref{lum0.87-blt} and \ref{lum9-blt}.
We separately plot the full sample including all protostars and the non-multiple sample.
We find that the 0.87~mm flux densities are proportional to \lbol$^{0.41\pm0.04}$ and \lbol$^{0.61\pm0.05}$ for the 
full sample and non-multiple sample, respectively, with Pearson's R
coefficients of 0.50 and 0.64. Similarly, the 9~mm flux densities are proportional to \lbol$^{0.20\pm0.3}$ and \lbol$^{0.38\pm0.07}$ for the 
full sample and non-multiple sample, respectively, with Pearson's R
coefficients of 0.39 and 0.51. The strong correlations with \lbol\ for both 0.87~mm and 9~mm
are not surprising since higher luminosity will result in warmer dust, which will result in higher 
flux densities for a given dust mass. The plots only showing the non-multiple sources exhibit cleaner correlations 
and are likely more robust than the correlations for the full sample. This is because the same bolometric
luminosity is adopted for all members of the multiple systems due to a lack of independent luminosity measurements.
Analysis of the flux densities as they relate to the underlying dust masses toward the
protostellar systems continues in the next subsection.

\subsection{Distribution of Protostellar Dust Disk Masses}

The integrated flux densities measured with ALMA and the VLA enable the dust disk masses to be estimated
under the assumption of an average dust temperature and optically thin dust emission (see Section 2.4
for a more detailed discussion of our methods and assumptions). 
Note that throughout this section and the rest of the paper, disk masses are given in dust mass 
(not scaled by an estimate of the dust to gas mass ratio)
unless specifically stated otherwise.
In the absence of detailed radiative transfer modeling
for all the sources \citep[e.g.,][]{sheehan2017b}, 
the dust disk masses measured from 
integrated flux densities are the most feasible
to compute for a large sample such as the protostars in Orion. The fact that all
the protostars in Orion have had their SEDs, \lbol\ and
\tbol\ characterized enables us to examine trends
in the protostellar dust disk masses in the context of these properties. 
We also note that the dust disk masses we refer to are calculated from the 
ALMA 0.87~mm continuum, unless specifically stated otherwise; however,
we do provide dust disk masses calculated from the VLA 9~mm flux densities in Table 8 for completeness.
It is possible that some of the detected emission is from
an inner envelope. Also, the continuum mass does not reflect the mass already
incorporated into the central protostellar object itself. We consider distributions with
multiple sources included (all or full sample) and excluded (non-multiple sample) to isolate the effect(s)
of multiplicity on the observed dust disk mass distributions.

We examine the protostellar dust disk masses with respect to \tbol\ and \lbol\
in Figure \ref{mass-blt}. We see in Figure \ref{mass-blt} that there
is significant scatter in the dust disk masses as a function of \tbol\ for the
full sample and also non-multiples. 
Given the scatter and lack of clear relation between \tbol\ and protostellar dust disk mass,
we calculated the median dust disk masses for Class 0, Class I, and Flat spectrum 
sources. For the full sample, we find 
median dust disk masses of 25.7, 15.6, and 13.8~M$_{\earth}$, respectively,
calculated from sample sizes of 133, 150, and 132 systems in each class, respectively. If we only 
consider non-multiple sources, then we find median 
dust disk masses of 52.5, 15.2, and 22.0~M$_{\earth}$, respectively,
calculated from sample sizes of 69, 110, and 79 
systems in each class, respectively. The median 
dust disk masses for all and non-multiple protostars include
upper limits in the calculation. The median masses for the different classes
are also listed in Table 9.
While there is a trend of lower dust disk mass with evolution, the amplitude of this trend is 
much smaller that the two orders of magnitude spread in 
dust disk masses for a given class \citep[see also][]{seguracox2018}.
We examine the dust disk mass trends with respect to protostellar class in more detail in the following paragraphs.

The relationship between dust disk mass and \lbol\ is shown in Figure \ref{mass-blt}.
 Such a dependence for protostars could be analogous to the M$_*$~-~M$_{disk}$
relationship for Class II YSOs where M$_{disk}$ $\propto$ M$_{*}^{1.8}$ \citep[e.g.,][]{ansdell2016}.
\lbol\ is the closest proxy for protostar mass available, but this is a relatively poor proxy 
due to a substantial (and unknown) fraction of luminosity coming from accretion \citep{dunham2014}.
We fit a linear slope in log-log space to the M$_{disk}$ versus \lbol\ plot for the
sample including all sources and find that 
M$_{disk}$~$\propto$~\lbol$^{0.11\pm 0.04}$, with a Pearson's R correlation
coefficient of 0.16, indicating a very weak correlation \citep{wall1996}. For a sample limited
to non-multiple sources, we find M$_{disk}$~$\propto$~\lbol$^{0.31\pm0.05}$ and calculate a Pearson's R correlation
coefficient of 0.34, indicating a moderate correlation. We note, however,
that by scaling the average dust temperature by \lbol$^{0.25}$ we have removed much 
of the apparent luminosity dependence on the dust disk mass
(see previous section for relations with flux densities only),
and the remaining correlation
could still be affected by the adopted dust temperatures.

The dust disk mass distributions can be more clearly examined as cumulative distributions
shown in Figure \ref{mass-cumulat}. The plots were constructed using
survival analysis and the Kaplan-Meier estimator as implemented in the 
Python package \textit{lifelines} \citep{davidson-pilon2019}.
We make use of the \textit{left censored fitting} functions that
account for upper limits derived from the non-detections. The width of
the cumulative distributions plotted represents the 1$\sigma$ uncertainty of the distribution.
The larger median masses of Class 0 disks both for the full sample and non-multiple sample are
evident in Figure \ref{mass-cumulat}.

To statistically compare the distributions, we used the log rank test as implemented in
\textit{lifelines}. We found that the distribution of Class 0 masses is 
inconsistent with being drawn from the same distribution as the Class I and Flat
spectrum sources at $>$99\% confidence (p$<$0.01) for both the samples considering all sources
and non-multiples. However, the differences between the Class I and Flat spectrum
sources are not statistically significant.
Thus, the Class I and Flat Spectrum
mass distributions are consistent with being drawn from the same sample. Note that we
obtained consistent results from the Anderson-Darling test\footnote{The Anderson-Darling test is 
similar to the Kolmogorov-Smirnoff (KS) test, but is 
more statistically robust. This because the KS-test uses the maximum deviation to calculate the probability
and is not as sensitive when deviations are at the ends of the distribution or when there are 
small but significant deviations throughout the distribution. 
https://asaip.psu.edu/Articles/beware-the-kolmogorov-smirnov-test} \citep{scholz1987}
on the cumulative distributions alone without considering the upper limits. We list the
p-values from the sample comparisons in Table 10 and also provide the p-values from the
Anderson-Darling tests when conducted.

These cumulative dust disk mass distributions can also be approximated as a log-normal cumulative distribution
function (CDF), which can be directly translated to a Gaussian probability density function (PDF),
as has been demonstrated by \citet{williams2019}.
To determine the mean and standard deviation of the Gaussian PDF, we fit
the cumulative distributions derived from \textit{lifelines} with the survival function (defined
as 1 - Gaussian CDF) using the \textit{curve\_fit} function within \textit{scipy}. 
To calculate the 1$\sigma$ uncertainties, instead of adopting the standard error from the
fit, we performed the same fit on the 1$\sigma$ upper and lower bounds of the 
cumulative dust disk mass distributions from the survival analysis
and adopted the relative values of these parameters as the uncertainties.
We note, however, that the
observed distributions are not precisely Gaussian, so the parameters derived from these
fits may not be completely accurate, nor their uncertainties.

The mean dust disk masses for Class 0, I, and Flat Spectrum systems are  
25.9$^{+7.7}_{-4.0}$, 14.9$^{+3.8}_{-2.2}$, 11.6$^{+3.5}_{-1.9}$~\mearth, respectively, for the full
distributions considering all systems. Limiting the sample to non-multiple
systems we find mean dust disk masses of 38.1$^{+18.9}_{-8.4}$, 
13.4$^{+4.6}_{-2.4}$, and 14.3$^{+6.5}_{-3.0}$~\mearth, respectively. These mean values of the 
distributions are quite comparable to the median dust disk masses for the same distributions, and
the uncertainties on the means further demonstrate that the Class 0 dust disk masses are systematically
larger than those of Class I and Flat Spectrum and differ beyond the 1$\sigma$ uncertainties of the
mean masses. The mean masses of the Class I and Flat Spectrum protostars are consistent within
the uncertainties, a further indication that there is not a significant difference between the disk
masses in these two classes.

\subsection{Distribution of Protostellar Dust Disk Radii}

We utilize the deconvolved Gaussian 2$\sigma$ radius from the fits to 
the continuum images as a
proxy for the radius of the continuum sources, enabling us to characterize the 
disk radii in a homogeneous manner (see Section 2.4). These values are provided in Table 8 for both
the ALMA and VLA measurements. However, we only make use of the ALMA measurements in this
analysis due to the 0.87~mm continuum emission having a greater spatial extent that can more
accurately reflect the full radius of the disk. The VLA continuum emission is often compact
and point-like, even toward protostars with apparent resolved disks at 0.87~mm; see Figures 3 and 4
as well as \citet{seguracox2016,seguracox2018}.

Visual inspection of the Gaussian fits reveals that there are often residuals outside the Gaussian model. This 
affects the larger disks (R~$>$~50~au) more than the compact ones, and our measured radii will
be systematically underestimated in some cases. 
The determination of deconvolved Gaussian 2$\sigma$ radii
can also be subject to some systematics. If the S/N is high enough, then a source
smaller than the beam can be deconvolved from it under the assumption that the
underlying source structure is also Gaussian. \citet{trapman2019} showed that
if the peak S/N of dusty disk emission was $>$10, the disk radius could be
recovered reasonably well. Those authors, however, were using the curve of growth
method rather than Gaussian fitting. Our sample typically has
modest S/N, between 20 to 100, and we regard deconvolved radii significantly smaller
than half size of the synthesized beam (0\farcs05, 20 au) as being possibly unreliable.
In our analysis, we only include sources with strong enough emission such that an estimate of the 
deconvolved size could be made. Weak sources that required their major axis, minor axis, and
position angle to be fixed to the synthesized beam are not included in these plots. We do, however,
tabulate the fits that indicate a deconvolved radius smaller than 10~au even though these
may be too small to be reliable.

We compare the disk radii to \tbol\ in the top panels of Figure \ref{radii-blt}.
The median radii for the Class 0, Class I, and Flat Spectrum sources are
48, 38, and 31 au, respectively, 
for the full sample, and
55, 37, and 38 au, respectively, 
for the non-multiple sample;
we also list these values in Table 9. The amplitude
of this trend is small given the order of magnitude scatter within each class.
Thus, the Class 0 sources have a tendency for larger radii compared to the Class I
and Flat Spectrum sources, and non-multiple sources also tend to have larger disks. The
sensitivity of the current observations may not be sufficient for detecting 
circumbinary emission (disks), however. We do include the deconvolved radii calculated for
the unresolved and marginally resolved disks, which may artificially inflate median
measured disk radii if the disks are significantly smaller than their upper limits.
Even if a disk radius measured from the deconvolved Gaussian is below our 
expected measurement limit of 10~au, we do not plot it as an upper limit and leave it 
at its measured value. Furthermore, we do not account for non-detections when 
calculating the median radii. We also compare the distribution
of radii to \lbol; again, there is no clear trend, and both high- and low-luminosity
sources can have large and small radii. However, the non-multiple sources with 
luminosity greater than 100~\lsun\ have radii of $\sim$120~au, but only for a sample
of 2.

The disk radii distributions are also examined as cumulative distributions 
using survival analysis and the Kaplan-Meier estimator as implemented in the 
Python package \textit{lifelines} and shown 
in Figure \ref{radii-cumulat}. Here the systematically larger sizes of Class 0
disks are evident by eye. To establish the statistical significance of these differences, we compare these distributions 
quantitatively using a log rank test, similar to
how we compared the distributions of dust disk masses. Considering all
sources (multiple and non-multiple), we compared Class 0 vs.\ Class I,
Class 0 vs.\ Flat Spectrum, and Class I vs.\ Flat Spectrum, and the likelihood
that these samples are drawn from the same distribution are 0.63, 0.0002, and 0.003,
respectively. Thus, there is no statistical evidence that the Class 0 and Class I 
radii distributions are drawn from different distributions. However, the distributions of
Class 0 and Flat Spectrum and Class I and Flat Spectrum disk radii are
inconsistent with being drawn from the same parent distribution from the log rank test.
A summary of the sample comparison probabilities is provided in Table 10. 

We then compared the radii distributions for the non-multiple sources, and we
conducted the same comparisons as in the previous paragraph. These were
Class 0 vs.\ Class I, Class 0 vs.\ Flat Spectrum, and Class I vs.\ Flat spectrum,
which have likelihoods of being drawn from the same parent distribution of 
0.59, 0.04, and 0.13, respectively.
Thus, at the 99\% confidence level the distributions of disk radii are all consistent
with having been drawn from the same sample.
However, looking at Figure \ref{radii-cumulat} we would expect the Class 0 sample
to not be consistent with having been drawn from the same distribution as
the Class I and Flat Spectrum samples. This counter-intuitive result could
be caused by the inaccuracy of the log rank test when the cumulative
distributions cross \citep{davidson-pilon2019}, as the Class 0 sample does with
the Class I and Flat Spectrum samples. Furthermore, the uncertainty width
shown in Figure \ref{radii-cumulat} is the 1$\sigma$ width, and within 2$\sigma$ the
distributions would overlap significantly more. As an additional test we performed an 
Anderson-Darling test on the distributions, finding results consistent with the log
rank test.

To characterize the distributions of disk radii further, we fit a Gaussian CDF
to the cumulative distributions of disk radii from the survival analysis in the same
manner as we fitted the Gaussian CDF to the dust disk mass distributions. This enabled us to derive mean radii and
widths of the log-normal distributions with associated uncertainties. 
The mean radii for the full sample of Class 0, Class I, and Flat Spectrum protostars are
44.9$^{+5.8}_{-3.4}$, 37.0$^{+4.9}_{-3.0}$, and 28.5$^{+3.7}_{-2.3}$ au, respectively, and the
mean radii for non-multiple sample are 53.7$^{+8.4}_{-4.2}$, 35.4$^{+6.1}_{-3.5}$,  and 36.0$^{+5.9}_{-3.2}$ au, respectively.
These properties
of the distributions are listed in Table 9. 
The consistency (or lack thereof) of the mean
radii when compared between classes are in line with the results from the log rank tests, 
except for the Class 0 to Class I disk radii for the non-multiple sample, 
where the log-rank test indicates that they are consistent with being drawn from the
same parent distribution.

The mean radii from the Gaussian PDFs 
indicate that the distributions of disk radii are not extremely different
between Class 0, Class I, and Flat Spectrum. 
For the full sample, only the Class 0 and Class I distributions of disk radii 
are consistent with being drawn from the same sample; the Class 0 and Flat 
Spectrum and Class I and Flat Spectrum 
distributions are inconsistent with being drawn from the same sample. 
The radii distributions for the non-multiple samples, however, are all consistent
with having been drawn from the same samples (see Table 10).

The distributions in Figure \ref{radii-cumulat} also
clearly show that disks substantially larger
than the median radii exist for protostars of all classes. However,
taking 50~au as a fiducial number to define the qualitative distinction between
large and small disks, 
$\sim$46\% (N=61) of Class 0, $\sim$38\% of Class I (N=57), and $\sim$26\% (N=35) of
Flat Spectrum disks have radii larger than 50~au. These percentages are 
calculated from (N(R$\ge$50~au)/(N(continuum sources)+N(non-detections)) (Table 5). 
If only non-multiples from each class are considered, the percentage of dust disks with radii
$>$ 50~au are 54\% (N=37), 38\% (N=42), and 37\% (N=29) for Class 0, Class I, and Flat Spectrum, respectively. These percentages are calculated from 
(N(R$\ge$50~au, non-multiple)/(N(non-multiple systems)+N(non-detections)).

\subsection{Distribution of Protostellar Dust Disk Masses versus Radius and Inclinations}

Lastly, we examine the relationship between dust disk mass and radius
for both the full sample and non-multiple systems.
Figure \ref{radii-mass} shows that below 
$\sim$30~\mearth, there is no apparent relation
between dust disk mass and disk radius. In both panels of Figure \ref{radii-mass}, the
radii are clustered around 35~au for dust disk masses less than 
$\sim$30~\mearth, and 
there is a large spread in radius for a given dust disk mass. There is also 
not a clear distinction between the classes with
all three spanning the same range of parameter space
in Figure \ref{radii-mass}.

We do find that for masses greater than 30~\mearth,
there is an apparent trend of increasing radius with mass. We fit the correlation between disk radii
and mass using \textit{scipy}. If we include all the masses and radii in the fit, 
we find that R~$\propto$~M$_{disk}$$^{0.3 \pm 0.03}$ (Pearson's R = 0.54); 
if we only fit masses greater
than 67~\mearth, then R~$\propto$~M$_{disk}^{0.34 \pm 0.09}$ (Pearson's R = 0.37).
These fits are plotted in Figure \ref{radii-mass} as dotted and dashed lines, respectively. 
If we instead limit the sample to non-multiple systems, then we find that 
R~$\propto$~M$_{disk}$$^{0.25 \pm 0.03}$  (Pearson's R = 0.49) and 
R~$\propto$~M$_{disk}$$^{0.26 \pm 0.1}$ (Pearson's R = 0.27)
for the same ranges of dust disk masses used for the full sample, respectively. As a limiting case,
a sample of optically thick disks with a variety of radii would have a disk radius that increases with the square-root of
the dust disk mass. For both fits, the relationship is more shallow than this simple case;
this indicates that the disks we observe should not be optically thick at all radii.

We also estimate the inclination of the protostellar systems, under the assumption
of circular symmetry using the measurements of the protostellar disk radii from 
the deconvolved semi-major and semi-minor
axis of the Gaussian fits. We show the histogram of inclinations in Figure \ref{inclinations} both
in terms of cos($i$) and degrees; an inclination of 90\degr\ refers to viewing a disk edge-on, while 0\degr\
refers to viewing a disk face-on. A completely random distribution of inclinations should have
a flat histogram with equal numbers in each bin of cos($i$). However, we can see that the histogram
of cos($i$) declines at low values which corresponds to high inclinations (near edge-on). The 
histogram of inclinations in degrees is shown in the right panel of Figure \ref{inclinations}, and for
a flat distribution of cos($i$), reflecting a random distribution of inclinations, 
the average value should be 60\degr. The median
and mean values of cos($i$) are 0.596 and 0.601, respectively, corresponding to 53.4\degr\ and
53.1\degr. This average value is less than 60\degr\ due to the lack of 
sources computed to have high inclinations. However, we do not think that this difference 
is significant because looking at the continuum
images in Figure 3 and 5 and Appendix C, there are sources that appear to be oriented near edge-on.
The reason their inclinations do not compute to edge-on is because this requires deconvolved
minor to major axis ratios very near to zero. Furthermore, the disks are known to have
a finite thickness to their dust emission \citep{lee2017}; this, combined with finite 
resolution, will lead to the distribution being biased against
edge-on sources.

\subsection{Regional Comparison of Disk Properties}

\begin{figure}
\includegraphics[scale=1.0]{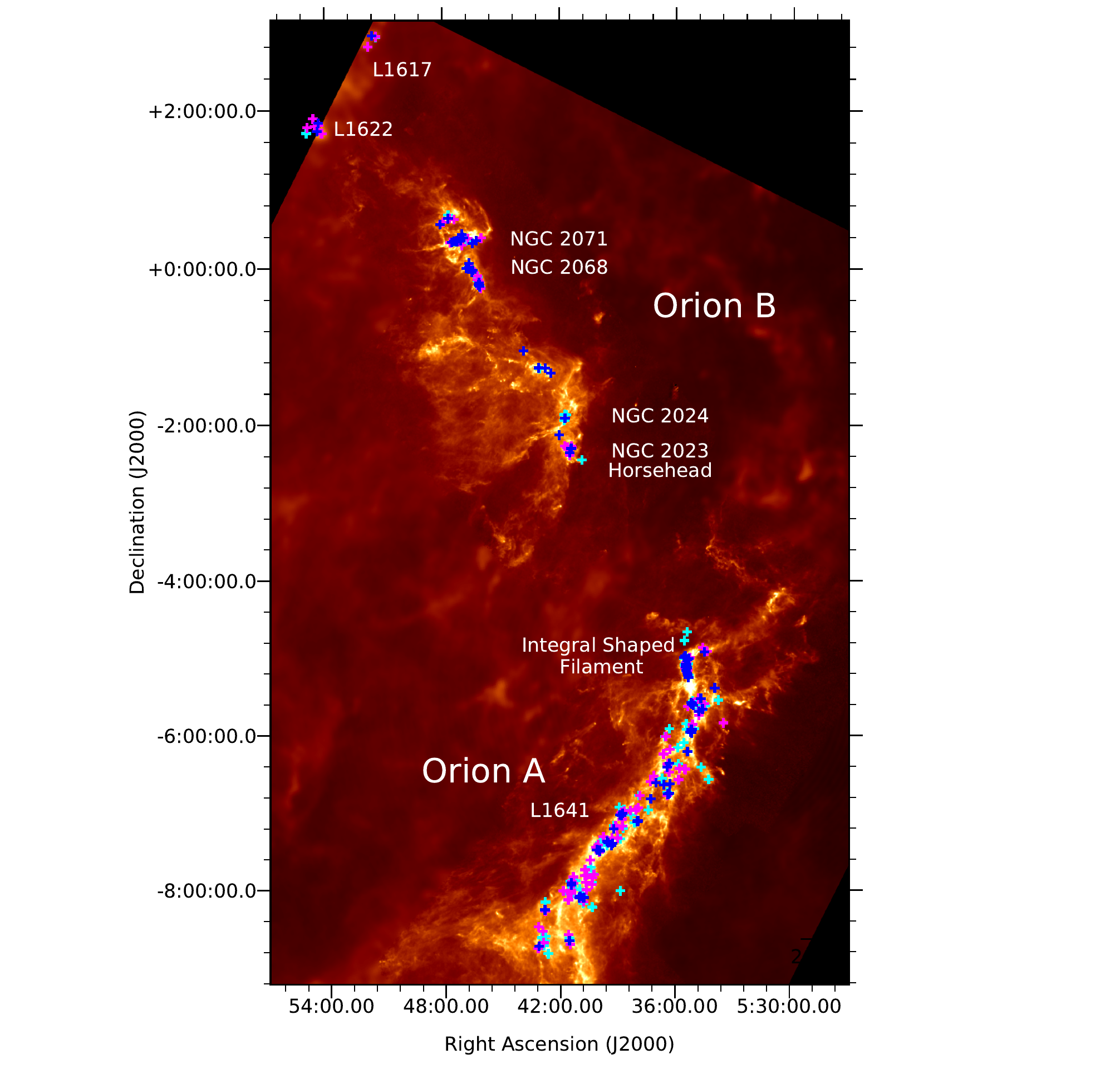}
\caption{Overview of the Orion region and protostellar targets for the ALMA and VLA surveys (crosses).
The image is the \textit{Herschel} and \textit{Planck}
column density map from \citet{lombardi2014} displayed on a square-root color scale. 
Major sub-regions within the clouds are highlighted with
the sub-region labels located adjacent to their positions. The blue crosses are Class 0 protostars, 
magenta crosses are Class I protostars, and cyan crosses are Flat Spectrum protostars.
}
\label{overview}
\end{figure}

\begin{figure}
\includegraphics[scale=0.5]{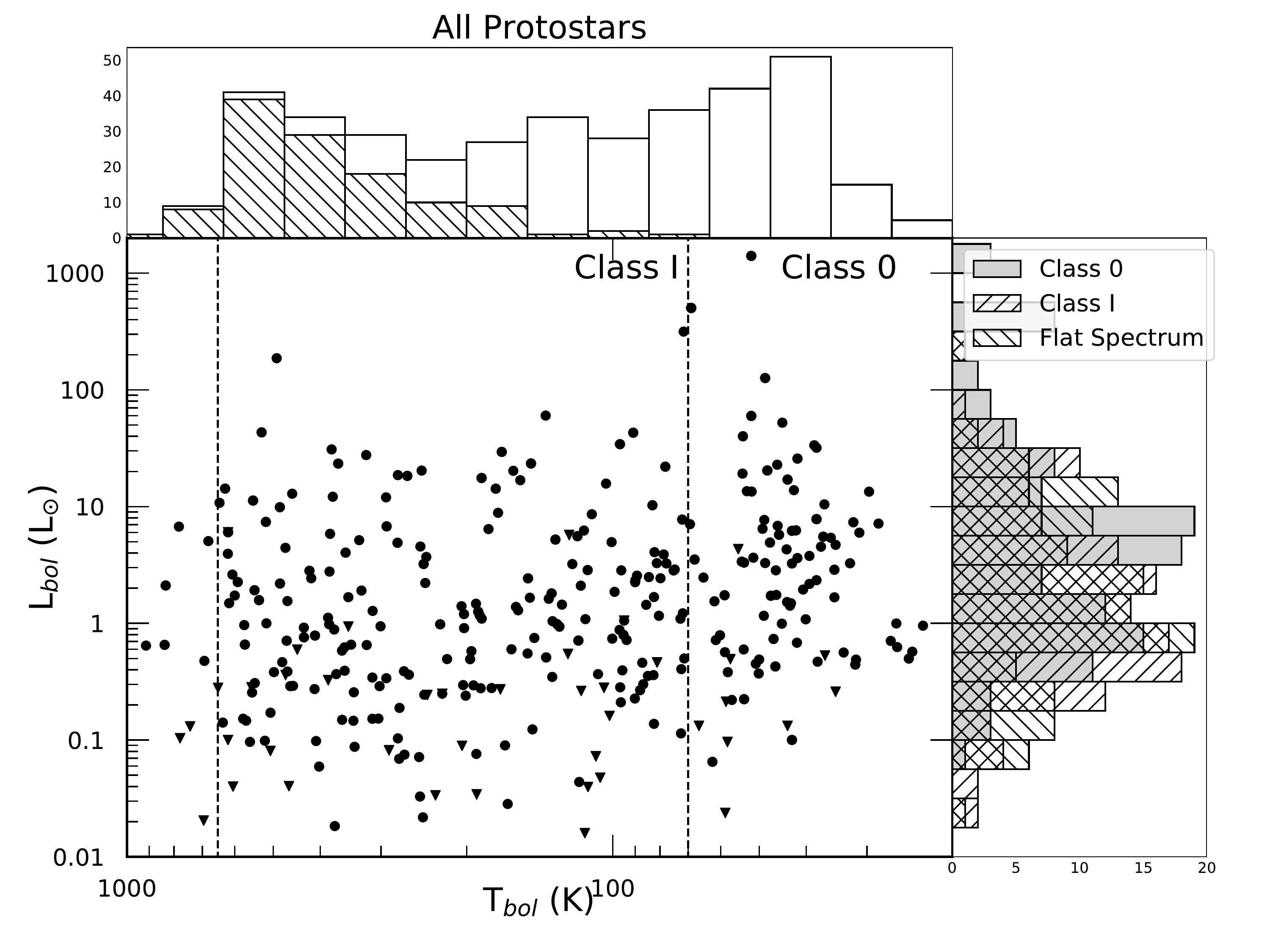}
\caption{Bolometric luminosity (\lbol) versus bolometric temperature (\tbol) 
for the sample of 328 protostars surveyed by ALMA
and the VLA with accompanying histograms. The histogram along the x-axis shows the full number in each
bin, and the hatched region shows the number of Flat Spectrum sources, given that they overlap in this
parameter space. The histogram along the y-axis shows the distribution bolometric luminosities for each
protostellar class in the survey. The distributions are similar, but as shown in \citet{fischer2017} the
Class 0 protostars have slightly higher luminosities on the whole than the Class I and Flat Spectrum protostars.
}
\label{lbol-tbol}
\end{figure}

\begin{figure}
\includegraphics[scale=0.45]{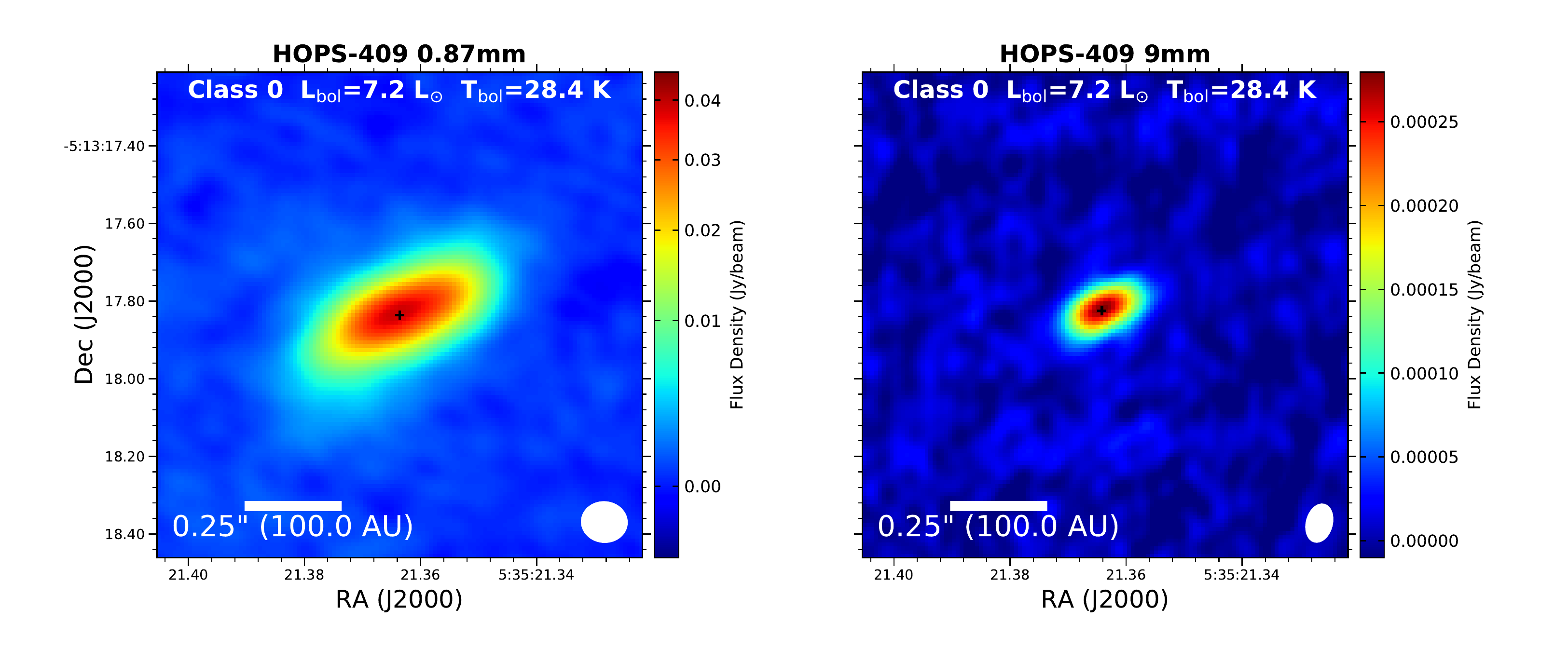}
\includegraphics[scale=0.45]{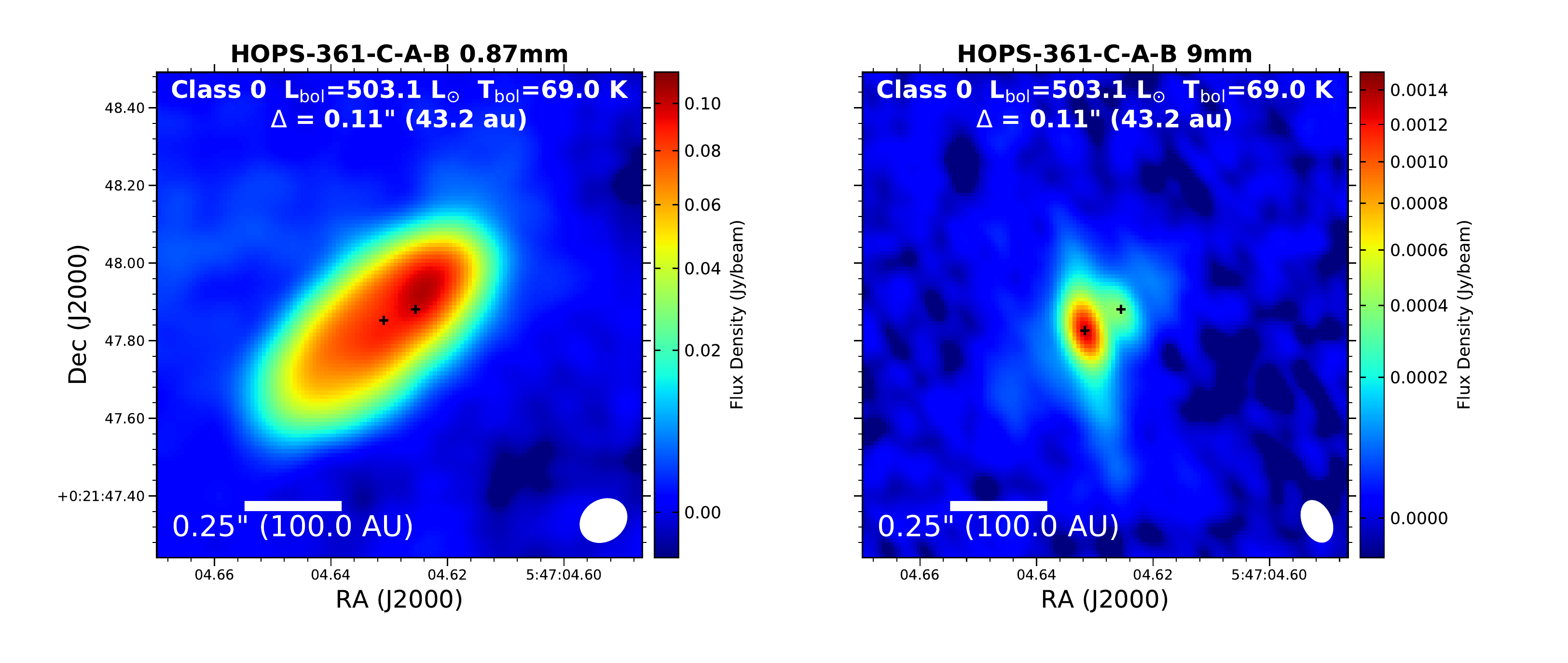}
\includegraphics[scale=0.45]{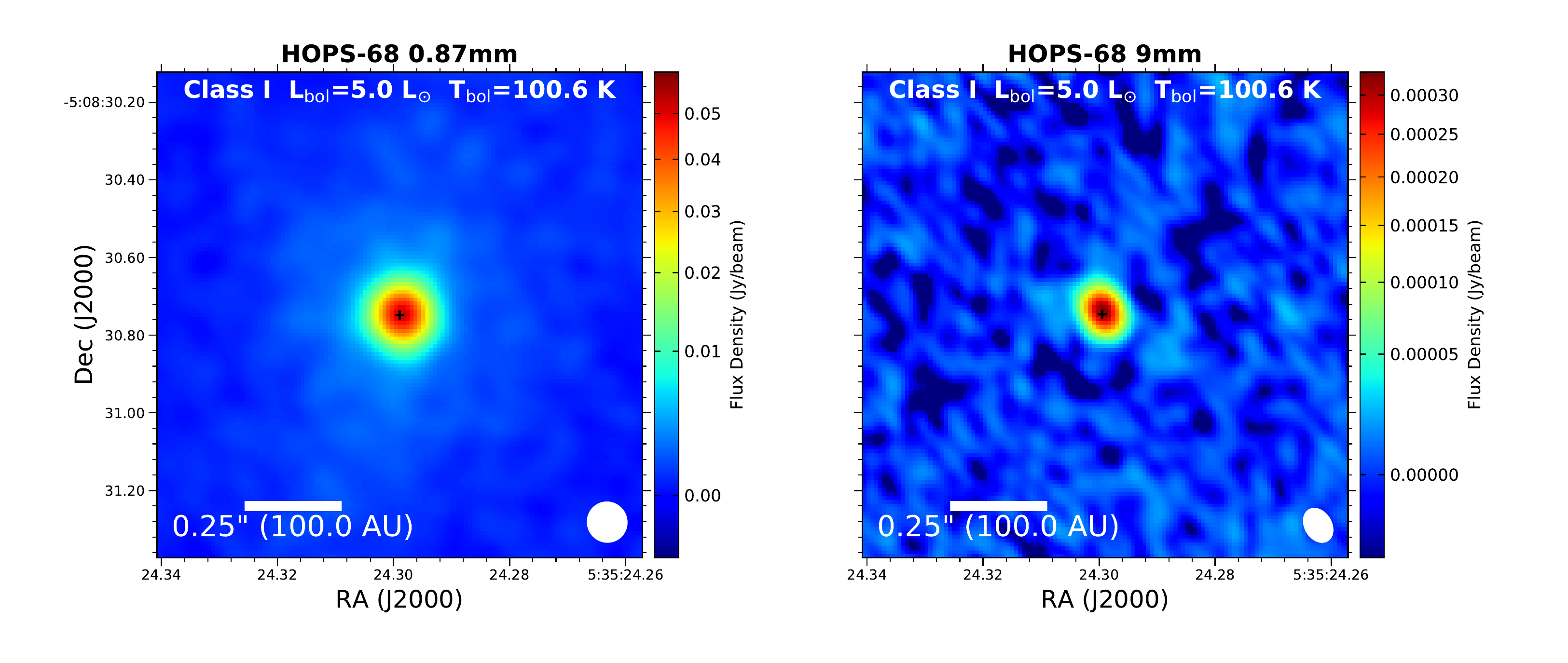}
\caption{Example images from ALMA (left) at 0.87~mm and the VLA (right) at 9~mm toward
selected sources. The top row shows HOPS-409, a Class 0 protostar with an apparent resolved disk at 0.87~mm
and 9~mm. HOPS-361-C, a high-luminosity Class 0 source, is shown in the middle row and appears 
disk-like at 0.87~mm. However, HOPS-361-C is resolved
into a close binary system by the VLA at 9~mm. The brighter source also exhibits an extended
free-free jet at this wavelength. HOPS-68, a Class I source, is shown in the bottom row and
appears compact and only marginally resolved at both 0.87~mm and 9~mm. The synthesized
beams are drawn in the lower right; the typical ALMA synthesized beam is 0\farcs1 and the typical
VLA beam is 0\farcs08.
Images for the remaining protostars are shown in Appendix C. A blank panel for the ALMA 0.87~mm
or VLA 9~mm panel means that observations were not taken toward that particular protostar with
ALMA or the VLA.
\label{continuum-demo1}
}
\end{figure}

The Orion star-forming region, as highlighted in Figure \ref{overview}, encompasses much more than 
just the region around the Orion Nebula. There are two giant molecular clouds 
in Orion denoted A and B. The Orion A molecular cloud encompasses the molecular 
emission south of $\sim$-4.5\degr\ declination, and we consider two regions within Orion A with
distinct properties: the northern half of the Integral-Shaped Filament (ISF) and L1641. We consider
protostars between -4.5\degr\ and -5.5\degr\ declination as part of the northern ISF 
and protostars south of -5.5\degr\ as part of the southern ISF and L1641. The ISF extends to $\sim$-6\degr, 
and the southern ISF between -5.5\degr\ and -6\degr\ has a YSO density similar to L1641, 
so we consider them together.

The northern half of
the ISF is located between the Trapezium and NGC 1977 and has a high spatial density of protostars and
high-density molecular gas \citep{peterson2008,megeath2012,stutz2015,stutz2016}. This region is also referred to as 
Orion Molecular Cloud 2/3 (OMC2/3) and has its protostellar content 
well-characterized \citep[e.g.,][D\'\i az-Rodr\'\i guez in prep.]{furlan2016,tobin2019}.
The central portion of the ISF is located behind the Orion Nebula, where
the SEDs of YSOs do not extend beyond 8~\micron\ due to saturation at longer
wavelengths. In contrast to
the northern ISF, the southern ISF and L1641 have a much lower spatial density of 
protostars \citep{allen2008,megeath2012}.

We then consider protostars located north of -4.5\degr\ as part of Orion B, which itself contains several
sub-regions that we consider together: the Horsehead, NGC 2023, 
NGC 2024, NGC 2068, NGC 2071, L1622, and L1617 \citep{megeath2012}. 
Note that we do not have sources in our sample between declinations of -02:21:17 and -4:55:30, 
so the exact boundary in declination between Orion A and Orion B is not important (Figure 1).
To sample a variety of environments with a reasonably large number of protostars in each sub-sample,
we compared L1641 and southern ISF (low spatial density), to the northern ISF (high spatial density), 
and Orion B (low spatial density). We note that both L1641 and Orion B contain regions
 of high protostellar density, but compared to the northern ISF they have low overall 
spatial density of protostars \citep{megeath2016}.

We examined the dust disk masses within L1641 and Southern ISF, the Northern ISF, and Orion B, finding median 
values of 14.7, 15.2, and 24.3~\mearth, 
respectively, for the full sample with respective sample sizes of 235, 76, and 113 protostars, within each 
region. Limiting the analysis to the
non-multiple sources, the median dust disk masses are 
23.7, 18.7, and 28.3~\msun, 
respectively, with respective sample sizes of 165, 39, 63 protostars (see also Table 9).
The calculations of median mass measurements for each
region include non-detections. While we find the differences
in median masses between all sources and non-multiples reported earlier, there is no significant variation
between the regions. A log rank test performed on the distributions reveals that the
mass distributions for non-multiple sources are consistent with having been drawn from 
the same sample. 
cumulative distributions with a Gaussian CDF in the same manner as described for the full
dust disk mass distributions. When comparing the means of these distributions, only 
 Orion B vs.\ L1641 for the full sample differs by more than 1$\sigma$ (but less than 2$\sigma$); these 
values are listed in Table 9.
A summary of the statistical tests and sample sizes are given 
in Table 10.

We also examined the disk radii within L1641 (N=181), the ISF (N=60), and Orion B (N=93), finding median
radii of 45.6, 38.0, and 39.7 au,
respectively, for the full sample. Limiting the analysis
to the non-multiple sources, we find median radii of 
54.5, 40.4, and 48.2 au,
respectively, with respective
sample sizes of 127, 31, and 47 protostars. The number
of protostellar disks included in each region is different with respect to the number used for mass calculations, 
because we excluded non-detections and the low S/N sources that required Gaussian parameters 
to be set equivalent to the synthesized beam.
The trend of larger disk radii in non-multiple systems is
again evident in these median values, but there are not significant differences between regions. We
confirmed that the radii distributions between the different regions were consistent using a log rank
test for all sources and non-multiple sources. The distributions are consistent
with being drawn from the same sample (see Table 10). 
Moreover, the mean disk radii for these regions, derived from fitting 
a Gaussian to the cumulative distributions, are consistent 
within their 1$\sigma$ uncertainties.

This analysis demonstrates that within the limits of our
dust disk mass and radius measurements, the properties
of protostellar disks do not show statistically significant differences between sub-regions within
the Orion molecular clouds. We do not draw a direct comparison to the disks within the Trapezium
in this section because we targeted very few protostars located within the Orion Nebula itself, under
the influence of the ionizing radiation from the massive stars there. This is due in part to these 
sources not being targeted by the \textit{Herschel} Orion Protostar survey because of the bright emission 
from the nebula in the mid- to far-IR, and hence the sample of protostars toward the Orion 
Nebula is potentially highly incomplete and poorly characterized.  
The Class II disks within the Orion Nebula Cluster, on the other hand, have been studied
with ALMA by \citet{mann2014} and \citet{eisner2018}.

\section{Discussion}

The large sample of protostellar disks detected and resolved in our
survey toward the Orion protostars enables an unprecedented comparison of
protostellar disk properties to SED-derived protostellar properties. The observed
relation of dust disk masses and radii to evolutionary diagnostics such as \lbol\
and \tbol\ enables a better understanding of how disk evolution is coupled to 
protostellar evolution. While the disk radii and masses do not strongly depend
on any evolutionary diagnostic, the protostars overall have lower dust disk masses 
and smaller dust disk radii with increased evolution. The large amount of scatter
in the relations may point toward differences in the initial conditions of star
formation (core mass, turbulence, magnetic fields, net angular momentum, etc.). 
It is important to emphasize that the protostellar classification 
schemes are imprecise tracers of evolution due to the viewing angle dependence
of \tbol\ and the SED slope, but the scatter within a protostellar class
is much too
large to be attributed to classification uncertainty alone \citep[e.g., see Figure 7 of ][]{fischer2017}.
Furthermore, we still lack specific knowledge of the most important
protostellar property, the current mass of the central protostar. Bolometric
luminosity can be used as a proxy for stellar
mass, but it is a very poor proxy with limited relation to the underlying
protostellar mass \citep{dunham2014,fischer2017}. We explore these relationships 
in greater detail in the following section and compare them to predictions of models.

\subsection{Protostellar Dust Disk Masses}

To better understand the evolution of dust disk masses from the protostellar to the Class II phase,
it is essential to compare them to the measured distributions of dust disk masses for both other
protostellar samples and Class II disk samples. We first compare the
distribution of Orion protostellar dust disk masses to those of the Perseus protostellar
disk sample from \citet{tychoniec2018}, the Ophiuchus sample from \citet{williams2019}, and a sample
of Taurus Class I disks from \citet{sheehan2017b}. It is clear from Figure \ref{masses-ori-protos} 
that the Orion protostellar disks lie directly between the Perseus and Ophiuchus disk
mass distributions. 
The Taurus protostellar disks are reasonably consistent with Orion,
despite the smaller sample size and the masses derived from radiative transfer modeling
and different dust opacities.
The mean and median dust disk masses for the various samples are provided in Table 9.

The protostars within the Perseus sample may be similar in protostellar content to 
the Orion sample, since it was also an unbiased survey of the entire region, just with a smaller sample
and lacking as many high-luminosity sources. 
However, the median dust disk mass is 25$\times$ larger than the median for Orion 
(or $\sim$5$\times$ for T=20~K and $\kappa$=3.45 cm$^{2}$~g$^{-1}$), but 
\citet{tychoniec2018} used the VLA 9~mm data for Perseus, corrected for free-free emission
using 4.1 and 6.4~cm data, to calculate their masses. 
The difference in wavelength
and adopted dust opacity introduces a high likelihood of introducing
systematic differences to the distribution of the Perseus dust disk masses. 
They used the
\citet{ossenkopf1994} dust mass opacity at 1.3~mm (0.899 cm$^2$~g$^{-1}$) extrapolated to 
8~mm by assuming a dust opacity spectral index of 1 and a constant average dust temperature
of 30~K. Prior to plotting the Perseus dust disk mass distributions in Figure \ref{masses-ori-protos} 
we adjusted the masses 
to account for the revised distance of 300~pc to the region \citep{ortiz-leon2018}, and we
scaled the dust temperature using \lbol\ and the same temperature normalization 
that was used for the Orion protostars (Section 2.4 and Appendix B).
However, the dust temperature scaling did not significantly alter
the distribution of
Perseus dust disk masses. Another study of Perseus dust disk masses was carried out by \citet{andersen2019}
using Submillimeter Array (SMA) data from the Mass Assembly of Stellar Systems and their Evolution with the SMA (MASSES)
Survey \citep[e.g.,][]{lee2016} using lower resolution data ($\sim$3\arcsec) to estimate
dust disk masses by removing an estimated envelope contribution. We compare the VANDAM 
Perseus dust disk masses with those from \citet{andersen2019} in Appendix D, but they 
similarly find systematically higher dust disk masses with respect to Orion.

Thus, it is unclear if the Perseus protostars
really have systematically more massive dust disks. The adoption of a different dust opacity slope could
easily bring the distributions into closer agreement. While we do not have complementary
observations to longer centimeter wavelengths to enable a more rigorous determination of
free-free contamination to the VLA Orion data, we compare our 9~mm dust disk mass 
measurements to the VANDAM Perseus dust disk mass measurements, 0.87~mm ALMA dust 
disk mass measurements, and the distributions of 9~mm
flux densities in Appendix D. The results indicate Orion at 9~mm is comparable to Perseus at 9~mm, thus
pointing to the adopted dust opacity leading to overestimated dust disk masses. A 
systematic study of the Perseus protostars
using ALMA at a comparable spatial resolution, wavelength, and sensitivity to the Orion 
survey will be necessary to better compare these regions.

In contrast to Perseus, the Orion protostars of all classes have systematically higher dust disk masses than those
in Ophiuchus \citep{cieza2018}, despite similar observing and analysis strategies (Figure \ref{masses-ori-protos}). 
\citet{williams2019} made several different assumptions about dust
temperature (adopting a uniform 20~K) and a larger dust opacity (2.25 cm$^2$~g$^{-1}$ at 225 GHz) 
assuming $\kappa$ = ($\nu$/100 GHz)~cm$^2$~g$^{-1}$. If we make the same assumptions as \citet{williams2019} to
calculate dust masses, the Orion median masses only increase and are still inconsistent
with Ophiuchus (see the bottom panels of Figure \ref{masses-ori-protos}). 
This is because the adoption of a 2$\times$ higher dust opacity does lower the 
masses, but the uniform 20~K temperature cancels out the effect of a higher dust opacity and can significantly
raise the dust disk mass for some protostars.
In fact, the way to bring the distributions into as close as possible agreement is to adopt the 
higher mass opacity, but keep higher temperatures that are adjusted for luminosity. But
even with this adjustment, the distributions of Class I and Flat Spectrum protostars 
are still in disagreement by about a factor of 2. 

Since the dust disk masses between Ophiuchus and Orion cannot be reconciled by adopting the same set of
assumptions, either the protostellar disk properties in Ophiuchus are different
from Orion or there is sample contamination in Ophiuchus. \citet{cieza2018}
selected the sample used in \citet{williams2019} from the \textit{Spitzer} \textit{Cores to Disks} 
Legacy program \citep{evans2009}, and the YSOs were classified according to their 
SEDs. \citet{williams2019} adopted 26 protostars as Class I and 50 as Flat Spectrum. 
However, \citet{mcclure2010} analyzed the region with \textit{Spitzer} IRS spectroscopy, finding
that the 2 to 24~\micron\ spectral slope used by \citet{cieza2018} performs
poorly in Ophiuchus due to the heavy foreground extinction. Thus, \citet{mcclure2010} found that out
of 26 sources classified as Class I protostars from their 2 to 24~\micron\ spectral slope,
only 10 remained consistent with protostars embedded within envelopes
when classified using the IRS spectral slope from 5 to 12~\micron, which they regarded as more robust because
it is less affected by foreground extinction. In addition, \citet{vankempen2009} examined
dense gas tracers toward sources classified as Class I and Flat Spectrum in Ophiuchus, finding that
only 17 had envelopes with emission in dense gas tracers. Thus, it is possible that some of 
the Class I and Flat Spectrum protostars in the \citet{cieza2018} and \citet{williams2019} samples
are actually highly extincted Class II sources. 

However, the dust disk mass distributions as shown indicate that accounting for contamination
in Ophiuchus alone will not fully reconcile
the disagreement with Orion because the high-mass end of the Ophiuchus dust disk mass distribution 
is still inconsistent with Orion. This may signify that there is 
an overall difference in the typical protostellar dust disk masses
in Ophiuchus and Orion.
One possibility is that the Class I and Flat Spectrum Sources in Ophiuchus could be systematically
older than those found in Orion. This is plausible given the relatively small number of Class 0 
protostars in Ophiuchus \citep{enoch2009} relative to Class I and Flat Spectrum. However,
it is also possible that differences in the initial conditions of formation in Ophiuchus
versus Orion could result in a different distribution of dust disk masses.
What is clear from this comparison of different regions is that 
it is essential to compare dust disk mass distributions that utilize data at
comparable wavelengths and resolution
to minimize biases due to the adopted dust opacities, spatial resolution, and 
differences in the methods used to extract the disk properties.

\subsection{Protostellar Dust Disk Masses versus Class II Dust Disk Masses}

To further understand disk evolution past the protostellar phases, it is
essential to compare
protostellar disk samples to the more evolved Class II disk samples.
The distribution of Orion protostellar dust disk masses is shown in Figure \ref{masses-ori-ppds} 
alongside the mass distributions
of Class II disks from surveys of different star-forming regions. The Class II
disk surveys are reasonably complete and representative in their samples: the Lupus
disks are from \citet{ansdell2016}, the Chamaeleon disks are from \citet{pascucci2016}, the
Upper Sco disks are from \citet{barenfeld2016} and the Taurus disks are from \citet{tripathi2017}.
We also list the median dust disk masses from the various surveys in 
Table 9.
There are more surveys \citep[even in Orion itself e.g.,][]{eisner2018,vanterwisga2019}, but to avoid
making the comparison plots overly complex, we limited
our comparison to some of the most complete surveys. 

The Class II dust disk mass distributions shown in Figure \ref{masses-ori-ppds} have systematically
lower masses with increasing age of the stellar population, with Upper Sco being the oldest. 
We note that these other surveys
typically adopt a uniform temperature of 20~K and a dust mass opacity law of $\kappa$=($\nu$/100 GHz)$^{\beta}$, where
$\beta$ = 1. Our main results are formulated using a different dust opacity and the assumption of 
average dust temperatures based on \lbol; however, for comparison we show
plots in Figure \ref{masses-ori-ppds} using the same assumptions as the Class II disk studies.

It is clear that the distribution of protostellar dust disk masses in Orion is
systematically higher than the distributions found for Class II disks
by a factor greater than 4 (Figure \ref{masses-ori-ppds}), depending on
which samples are being compared to Orion. The lower dust masses for Class II disks are particularly
important for establishing the feasibility of giant planet formation
in the context of the core accretion model \citep[e.g.,][]{pollack1996}.
The formation of a 5 to 10 \mearth\ planetary cores built up from the dusty solid material within a disk
is required before the gas can be accreted from the disk, enabling the formation of a giant planet
\citep[e.g.,][]{hubicky2005,piso2015}. 
Thus, most Class II disks may not have the requisite raw material within their disks (that can be detected by ALMA)
to build up such large solid
bodies from scratch, while many of the protostellar disks in Orion (and other regions)
have sufficient raw material to form
many giant planets within a single system. However, it is possible (perhaps likely) that 
solids have already grown beyond millimeter sizes
in Class II disks, limiting the ability of millimeter/submillimeter observations
to detect their emission. Thus, planetesimal formation and perhaps the cores of giant 
planets may have already formed within the protostellar disks prior to their
evolution into Class II disks. Moreover, the
disks around pre-main-sequence stars are frequently found to have substructure within them in the
form of rings, gaps, cavities, and asymmetries \citep[e.g.,][]{andrews2018,vandermarel2019}.
There have even been indications of such substructures within protostellar disks 
\citep[][Segura-Cox et al. in prep.; Sheehan et al. in prep.]{sheehan2017a,
sheehan2018}. While there are multiple theoretical explanations for these structures, the most tantalizing
is planet formation.

This means that protostellar disks may better represent
the initial conditions for planet formation because they are more likely to have pristine environments 
where significant dust evolution is just 
beginning \citep[e.g.,][]{birnstiel2010}. Models of dust evolution indicate that dust
grains can grow to cm sizes in the protostellar phase within a few 100,000 yr. Thus, no matter if the
full protostellar phase lasts $\sim$500~kyr \citep{dunham2014} or the sum of the half-lives ($\sim$222~kyr)
\citep{kristensen2018}, it is possible
that Class II disks have already been formed with large dust particles and perhaps even
produced planetesimals or planets. Therefore, the dust disk masses around the protostars may provide a more accurate 
measurement of the amount of raw material available for planet formation.   

It is important to highlight that not all samples of Class II dust disk masses follow the trend of systematically
decreasing dust disk mass with the stellar population age.
Both Ophiuchus and Corona Australis have Class II populations with ages comparable to Lupus and Taurus,
but their dust disk mass distributions are systematically lower than Lupus and Taurus. Their
mass distributions are similar to that of Upper Sco \citep{williams2019,cazzoletti2019}. 
Moreover, the Class I and Flat Spectrum dust disk masses in Ophiuchus are lower than those
in Orion, possibly pointing to a global phenomenon resulting in lower dust disk masses for all 
YSO classes in Ophiuchus.

\subsection{Protostellar Disk Radii and Their Evolution}

Prior to the inclusion of non-ideal MHD processes in numerical simulations,
protostellar collapse simulations with flux-frozen magnetic fields prevented
the formation of disks around nascent protostars \citep{allen2003, hennebelle2008,mellon2008}.
On the other hand, hydrodynamic models that neglected the possible removal
of angular momentum by magnetic fields as well as radiative feedback
frequently produced large, gravitationally
unstable disks that were prone to fragmentation \citep[e.g.,][]{yorke1999,
bate2005,stamatellos2009,kratter2010,bate2012,bate2018}.
Non-ideal MHD processes (i.e., Ohmic dissipation, the Hall effect, 
and ambipolar diffusion) are now regularly included in numerical codes and
enable the formation of disks during the protostellar phase \citep[e.g.,][]{dapp2010,
machida2011,li2011,masson2016,hennebelle2016}. The Hall effect
depends on the magnetic field polarity; one polarity will encourage the formation of
a disk, while the opposite polarity may inhibit it. 
Furthermore, non-idealized
initial conditions that include turbulence or magnetic fields that are misaligned
with respect to the rotation axis can also facilitate the formation of disks
\citep[e.g.,][]{hennebelle2009,seifried2012,joos2012}.

One of the principal evolutionary differences between disks that form in the hydrodynamic case
and those whose formation is enabled by the dissipation of magnetic flux from non-ideal MHD processes
is that the former can rapidly grow with time to
100s of au in size while the latter can have their sizes limited to a few 10s of au. 
Hence, protostellar disks may grow as rapidly as R$_C$~$\propto$~t$^3$ in a rotating, infalling envelope
without consideration of magnetic fields
\citep{ulrich1976,cassen1981,tsc1984}. This is because, in the
context of inside-out collapse of an envelope with solid-body rotation, material from larger
radii with more angular momentum will become incorporated into the disk at later times. Even
though rotationally-supported disks are able to form readily in non-ideal MHD simulations,
their growth is not as rapid as those disks in the pure hydrodynamic simulations \citep{dapp2010,
masson2016}. Moreover, if
magnetic braking plays a role in keeping the disk radius small initially, its
efficiency should also be reduced with evolution (decreasing envelope mass), because angular momentum is no longer carried away as
efficiently \citep{machida2010,li2014} from the inner to outer envelope.
This means that disk growth at later times is expected from both
simple analytic models and numerical simulations that include magnetic
fields.

The trend exhibited within the Orion sample, however, is that
that the protostellar disk radii decrease (or are at least constant) from Class 0 to Flat Spectrum sources.
This result is seemingly at odds 
with the predictions of non-magnetized models invoking 
inside-out collapse with initially solid-body
rotation.
It is also at odds with the predictions of disk growth with time in MHD models as
the envelope dissipates and the
prediction of a bimodal distribution of protostellar disk radii from the Hall effect. 
However, it is important to stress that our observations are providing constraints on dust
disk radii and not the gas disks. This distinction is important because the dust disks
can appear smaller than the gas disks due to radial drift of large dust 
particles \citep{weiden1977}. Simulations have shown that rapid dust growth
and radial drift are possible in protostellar disks \citep{birnstiel2010}, subsequently affecting the 
apparent radius measured from dust emission. Under the assumption
that the protostellar classes reflect time evolution, we expect more radial drift to have occurred for 
the more evolved systems (i.e., Class I, Flat Spectrum, and Class II disks).
Different grain sizes are expected to have different drift rates; thus, it is useful to examine the disk sizes at
two wavelengths (e.g., 0.87~mm and 9~mm). There is some evidence for protostellar dust disks to appear more
compact at longer wavelengths \citep[e.g.,][]{seguracox2016,seguracox2018}; however, the surface brightness/dust mass
sensitivity difference between ALMA at 0.87~mm and the VLA at 9~mm makes a simple comparison difficult, and so this
line of investigation will be pursued in future work. For the moment, 
 assuming that the dust disk radii at 0.87~mm reflect the gas disk radii and ignoring possible radial drift, 
the observed distributions are inconsistent
with the expectations from theory and simulations. Thus, observations of the gas disk 
radii are necessary to fully test the expectations from theory and simulations.

Another process that could be at work and keep the gas disk radii small is
angular momentum removal by disk winds and outflows \citep{bai2016}.
This process may also cause disk radii to get smaller with time.
However, the expected impact on disk radii in the protostellar phase is not clear, 
and it may be difficult to disentangle this effect from radial drift if the
gas and dust disks are well-coupled.

To examine the evolution of dust disk radii
beyond the protostellar phase, we compared the 
distributions of disk radii to samples of Class II disk radii from Taurus,
Lupus, Chamaeleon, the ONC, and Upper Scorpius 
\citep{tripathi2017,ansdell2016,pascucci2016,eisner2018,barenfeld2016}, as shown in
Figure \ref{radii-ori-ppds}. We can see that 
the protostellar disk radii in Orion fall between the extremes for the protoplanetary disks. The
Orion distribution does not contain as many disks with radii as large as the disks in Taurus, but
the disks in Orion have radii similar to disks in Lupus and larger than 
those in the ONC and Upper Scorpius.
The environment within
the Orion molecular clouds is not as extreme as it is in the Orion Nebula
Cluster, but also not as low-density as Taurus and Chamaeleon. However, different
methodologies are employed within different studies to measure the disk radius. For
example, \citet{ansdell2016} adopt the radius where 90\% of the flux is enclosed as
an effective radius, other studies adopt Gaussian fitting \citep[e.g.,][]{eisner2018}
similar to our methodology, and others model the disk radii \citep[e.g.,][]{tripathi2017}.
Thus, to better compare disk radii between populations, a common set of methods needs to be used.

It is necessary to understand the implications of declining disk radii with evolution
for the protostars in Orion with respect to the typically larger disk radii for Class II
disks (e.g., in Taurus). The apparent decrease
of dust disk radii from Class 0 to Flat Spectrum 
could in part be a systematic bias introduced by measuring the radii from
Gaussian fitting. Low intensity outer disks can often be left as residuals from 
fitting a single Gaussian component; however, this is probably not the case for the
majority of the Class I and Flat Spectrum sources. The idea
of disk radii increasing through the protostellar phase via infall of higher
angular momentum material may be too simplistic for more dynamic star-forming environments.
Furthermore,
\citet{offner2014} find that the outflow from the protostar can remove a significant
amount of gas from the envelope within $\sim$100,000~yr after protostar formation. 
In the context
of an isolated, inside-out collapsing core, this could prevent high angular momentum
material from being incorporated into the disk and limit the growth of the disk radius
in the protostellar phase. 

The different regions within Orion
that reflect different environments (e.g., isolated vs. clustered) do not have 
statistically significant differences in their disk radii. But, we do find that non-multiples
have distributions of disk radii (and masses) that are 
systematically larger and more massive disks with respect to the full sample. 
This may indicate that the formation of both wide and close multiple
star systems (and their evolution) affects the observable disk radii and masses.

Following the protostellar phase, there are other mechanisms that can enable the disks to
grow to larger radii. Viscous disks are expected to spread in radius due to angular
momentum transfer \citep{ss1973} via the accretion process, and the protoplanetary 
disks with their typical
ages of 2~Myr \citep{hernandez2008,dunham2014} have sufficient time after the
protostellar phase to spread to larger radii. Whether the length of the
protostellar phase is the sum of their half-lives (222~kyr for Class 0, Class I, and
Flat Spectrum combined) as recently suggested by \citet[][see Section 1]{kristensen2018} or
$\sim$500,000~yr \citep{dunham2014}, the protostellar disks have had less
time to viscously spread. Thus, the systematically smaller radii of protostellar disks
is not in tension with the larger disks that can be found in some Class II
disk samples. Moreover, only $\sim$2\% of disks within the Lupus survey
have radii greater than 200~au \citep{vanterwisga2018}. Thus, large protoplanetary 
disks appear to be the exception rather than the rule.

It is, however, unclear if viscosity drives the evolution of Class II disks because
observations attempting to characterized turbulence in Class II disks find levels too
low for turbulent viscosity to be important \citep{flaherty2018,teague2018}. Thus,
recent studies posit that angular momentum from the disks is carried
away by disk winds rather than viscous spreading \citep{bai2016}. If this is the case, the
disk winds should cause disks to become smaller in radius with time because
the angular momentum is simply being removed from the system rather than being redistributed.
To minimize the impact of environmental differences in understanding the evolution
of the Orion protostellar disks to Class II disks, 
the disk properties of the Orion Class II population must be characterized
at the same resolution and in a similar environment. Current studies either lack the necessary resolution 
\citep[e.g.,][]{vanterwisga2019} or are probing Class II disk properties in too
extreme of an environment to compare with the protostars \citep{eisner2018}.

Finally, the frequency of protostellar disks with radii greater than 
$\sim$50~au is $\sim$36\% (153/(379+42)) toward the targeted protostars in 
our sample (46\% for Class 0 protostars, 38\% of Class I protostars,
and 27\% of Flat Spectrum sources). This finding
for Class 0 protostars is not in tension with the recent results of 
\citet{maury2018}, where five\footnote{This includes the large circum-multiple disk 
around L1448 IRS3B \citep{tobin2016b} which was excluded from the \citet{maury2018} numbers.}
protostars out of 16 had disks larger than 60 au.
However, from our results we can conclude that protostellar (and specifically Class 0)
disks larger than 50~AU are not rare.

\subsection{Potential for Gravitationally Unstable Disks}

The distribution of dust disk masses found in the study of Orion protostars, many of which
have dust disk masses in excess of $\sim$30~\mearth\ (possibly $\sim$0.01~\msun\ in gas mass),
begs the question of how likely some of these disks are to be
gravitationally unstable. Considering a simplified relationship for Toomre's Q from
\citet{kratter2016}, we can use the distribution of dust disk masses to infer which disks
are most likely to be unstable with a few assumptions. Q can be approximated as
\begin{equation}
\label{eq:qapprox}
Q \approx 2\frac{M_*}{M_d}\frac{H}{R},
\end{equation}
where $M_{*}$ is the mass of the protostar (in solar masses), $M_d$ is the 
total mass of the disk (gas and dust) in solar masses,
H is the vertical
scale height of the disk, and R is the radius of the disk. A disk is considered gravitationally
unstable for values of Q~$<$~1, at which point the disk is susceptible to fragmentation;
in general, Q is not expected to venture far below unity because fragmentation would occur.
The disk is considered marginally
unstable for 2.5~$>$~Q~$\ge$~1, where gravitational instability could still transport angular 
momentum and excite spiral arm formation \citep{kratter2010}. 
Because Toomre's Q requires the total mass of the disk, we multiply the derived dust disk
mass by 100, the gas-to-dust ratio in the interstellar medium \citep{bohlin1978}, and
convert to solar mass units.

If we consider a fiducial
protostellar mass of 0.25~\msun, we can calculate Q for the sample of protostars in Orion.
 We determine
the scale height (H/R) using the adopted fiducial protostar mass, since H = c$_s$/$\Omega$ for a
geometrically thin disk. We calculate c$_s$ at the outer radius of the disk (R), which is determined
from observations, based on the dust temperature. 
We estimate the dust temperature at this radius using both the relationship 
between luminosity and dust temperature and the relationship that T~$\propto$~R$_{disk}^{-0.46}$,
both which were established from radiative transfer models (Appendix B).
We calculate Q for each disk and provide this number in Table 8
to provide a reference for how unstable a disk might be. These should be
regarded with caution and are meant to only serve as rough estimates.
Values for Q significantly below 1 (e.g., $<$0.5) are likely the result of mass 
overestimates that could result from a combination of
temperature underestimates, dust opacity underestimates, 
and our assumption of a constant dust-to-gas mass ratio. Furthermore, 
the assumption of a uniform protostellar mass of 0.25~\msun\ in Orion could be low
for some of the systems with L~$>$~$\sim$10~\lsun, but their higher average 
dust temperatures mitigate gross underestimates of Q.

We find that there are three Class 0 (HOPS-317-B, HOPS-402, HOPS-400-B)
and one Class I (HH111mms-A) systems whose disks have Q $<$1 and thus could be prone to fragmentation. 
Disks can also be marginally unstable (1~$<$~Q~$\le$~2.5) and develop features like spiral arms
without leading to runaway fragmentation of the disk \citep{kratter2010}. We find
7 Class 0 (HOPS-87, HOPS-224, HH212mms, HOPS-403, HOPS-124, HOPS-398, and HOPS-404) and  
4 Class I (HH270mms1-A, HOPS-188, HOPS-140-B, and HOPS-65) 
have 1~$<$~Q~$\le$~2.5. 
These systems for which we calculate Q$\le$2.5 and do not have detected 
multiplicity may be ideal systems to search for 
spiral arms generated by self-gravity.
However, HOPS-317-B, HOPS-402, HOPS-400-B, HOPS-87, HOPS-403, HOPS-398, and HOPS-404 
are likely to have significant envelope contamination (see Section 4.5).  
Also, HOPS-140-B and HOPS-65 have large gaps in their disks (Appendix C, Sheehan et al. in prep.), 
making the Gaussian fit very poor and overestimating the flux density and dust disk mass, such that these
disks are not likely gravitationally unstable.
Thus, after the removal of the likely false positives, the remaining the Class 0 systems HOPS-124, HH212mms, 
and HOPS-224 and the Class I systems HH111mms-A, HH270mms1-A, and HOPS-188 
have distinctly disk-like morphologies in the images shown in Appendix C.
Thus, these protostars are the most likely
systems for which the disks may be self-gravitating,  but
we cannot rule out the possibility that others are self-gravitating (or that some of these
are indeed non-self gravitating) given our simple estimates
of mass, radii, and temperature.

\subsection{How Much Envelope Contamination is Present?}

We have interpreted the continuum emission in our survey as tracing 
protostellar disks because the high resolution effectively resolves out
the large-scale envelope, only leaving compact structure less than 2\arcsec\ in diameter for
 most cases. Indeed, the images for
many sources obviously appear disk-like, and the distribution of inclinations is nearly
consistent with expectations for a random distribution of disk inclinations (Figure \ref{inclinations}). If
much of the resolved emission we trace was coming from a dense inner envelope, we 
expect that it would be distributed in a more symmetric manner and could mimic
a face-on disk.

There are also at least two mechanisms that could produce the appearance 
of flattened structure within envelopes and potentially
masquerade as a disk: 1) 
idealized magnetic collapse could produce a flattened inner envelope \citep[e.g.,][]{galli1993} and 2)
rotational flattening of the infalling envelope \citep[e.g.,][]{ulrich1976,cassen1981,tsc1984}.
Thus far, systems that
were expected to be likely candidates for dense, small-scale flattened envelopes, based on their
large-scale envelope morphology and magnetic fields, do not exhibit such structures mimicking a
disk viewed at intermediate to near-edge-on viewing geometry
\citep[e.g., L1157, NGC 1333 IRAS4A;][]{girart2006,looney2007,stephens2013,tobin2013,cox2015}. Moreover,
systems with observed rotation and extended continuum emission \citep[e.g., HH111MMS, L1527 IRS;][]{lee2014,lee2018,aso2017}
are typically found to actually reflect rotationally-supported disks. Also, the necessary density
structure of the continuum emission from the apparent disks require a significant increase
in density above that of the envelope present outside the disk.

Thus, most of the compact continuum structures that we detect
are likely to reflect emission from the protostellar disks. 
While we cannot exclude that some inner envelope emission may be present around the disks that
we detect, our methods of Gaussian fitting tend to leave residuals at large radii for
the most extended sources. Therefore, we expect that our methods will 
implicitly reduce the envelope contribution to the measured flux densities and disk radii.

Despite these arguments, inspection of the images toward the Class 0 protostars with Q~$<$~2.5 (section 4.4) 
reveals that there is a very bright, extended structure surrounding 
HOPS-317-B, HOPS-402, HOPS-400-B, HOPS-87, HOPS-403, HOPS-398, and HOPS-404 
at 0.87~mm and 9~mm that is not obviously disk-like. In fact, the surface brightnesses toward HOPS-400-B, HOPS-402, 
and HOPS-403 appear quite uniform at 0.87~mm, while there are
peaks evident at 9~mm. This is highly suggestive that the emission at 0.87~mm is optically thick.
It is not likely that 
these are all disks viewed nearly face-on given that the outflows (when detected and resolved)
are extended in the plane of the sky (\citealt{takahashi2012,tobin2016b}; Karnath et al. submitted). 
Thus, for at least some of these most massive, non-disk-like sources, we may be detecting very dense, 
compact inner envelope emission.

It is important to point out that HOPS-398, HOPS-400, HOPS-402, HOPS-403, and HOPS-404
belong to a special subset of Class 0 protostars 
that collectively have very dense envelopes. These are
the PACS Bright Red Sources (PBRS) that were discovered
by \citet{stutz2013} and further characterized by \citet{tobin2015b,tobin2016b}. Moreover,
HOPS-317-B, but it was not well-resolved from HOPS-317-A by \textit{Herschel}, and it has characteristics 
similar to the PBRS. The PBRS have luminosities
between 0.5 and 5.0~\lsun\ and appear to be among the youngest protostars in Orion, and HOPS-87 is
likely an extremely young source as well \citep{takahashi2012}. 

Protostars with 
apparently massive inner envelopes that are not disk-like appear to be
infrequent throughout the entire sample. If we remove the aforementioned sources 
that are likely dense inner envelopes from the statistics 
of disk radii and masses, the Class 0 median disk radii for the full
sample (and non-multiples) are reduced from 
45 au to 42~au (54~au to 52~au), and the Class 0 median dust disk masses for the full sample (and non-multiples) 
are reduced from 26~\mearth\ to 22~\mearth\ (53~\mearth\ to 33~\mearth).
Removing these protostars from the disk radii statistics also lowers the percentage of 
Class 0 disks with radii $>$ 50~au to 41\% for the full sample and 46\% for non-multiples.
The reduction in mean radii is not substantial, but the reduction in median mass for the non-multiple sample is quite significant.
However, the mean of the distribution calculated from the Gaussian PDF only changes from 25.9$^{+7.7}_{-4.0}$~\mearth\
to 22.5$^{+6.5}_{-3.4}$~\mearth\ for the full sample and from 38.1$^{+18.9}_{-8.4}$~\mearth\ to
31.2$^{+15.7}_{-6.9}$~\mearth\ for the non-multiples. These changes are well within the uncertainties of the distributions of radii, and thus
the inclusion of a few protostars with significant envelope contamination does not strongly
alter our conclusions.

\subsection{Comparison to Simulations of Protostellar Disks}

Simulations of large samples of disks with realistic global initial conditions
appropriate for comparison to our survey of Orion are presently very limited.  
Most magneto-hydrodynamic (MHD) simulations of star formation on the molecular
cloud scale use grid-based methods
with adaptive mesh refinement, such as those by \citet{li2018} with 28~au resolution, 
and are thus unable to resolve protostellar disks on the scales that we observe them. Only
the largest protostellar disks would be resolved in those simulations; a disk
with a radius of 50~au would only have $\sim$4 grid cells across its diameter.
However, models using smoothed particle hydrodynamics (SPH) can achieve higher
resolution while still sampling the molecular cloud scales, such as in \citet{bate2009, bate2012}.
Thus, we will focus on comparing our results to those found in the \citet{bate2018} 
simulations and to a more limited set of MHD zoom-in simulations from \citet{kueffmeier2017}.
We also emphasize that the disk properties from the simulations are gas disk masses and gas
disk radii. For the sake of comparison with the simulations, we infer the dust disk masses 
from the simulations by dividing by 100 and we
assume that the gas disk radius is the same as the dust disk radius.

\subsubsection{\citet{bate2018} Cluster Simulation}

\citet{bate2018} performed an analysis of the masses and radii
of protostellar disks formed in the simulation from \citet{bate2012}.
This SPH simulation 
included radiative transfer and followed a 500 M$_\odot$ collapsing cloud, which ultimately produced 183 protostars. 
While the SPH method afforded relatively high resolution, the disks in the simulations
are nonetheless poorly resolved due to the limited number of SPH particles per disk. For example, the lowest
gas disk masses are $\sim$0.01~\msun\ (33~\mearth\ in dust)
and contain only $\sim$700 SPH particles, and the 
scale height is not adequately resolved in any of the disks \citep[e.g.,][]{nelson2006}.
The simulation also did not include protostellar outflows or magnetic fields, 
which impact angular momentum transport and disk size \citep{li2014}.
Finally, although the simulation included radiative transfer, it neglected radiative 
feedback from protostars, which acts to further increase disk stability and stellar masses
\citep{offner2010,krumholz2016,jones2018}.  
Despite these limitations, the \citet{bate2018} study constitutes one of the few numerical 
studies of disk formation within star clusters that includes both a statistically 
significant sample of protostars and disks with masses down to $\sim$0.01~\msun\ (33~\mearth\ in assumed dust mass).

Class 0 protostars are likely the best sample
to compare with the simulations, because 
the simulations run for a total time of $\sim$2.25$\times$10$^5$~yr, and
star formation only occurs for $\sim$95,000~yr. \citet{dunham2014}
calculated that the lifetime of a Class 0 protostar is $\sim$160,000~yr,
but using a different set of assumptions \citet{kristensen2018}
found that the half-life of the Class 0 phase could be 47,000 yr, while
the half-lives of the Class I and Flat Spectrum phases are 88,000 and 87,000~yr,
respectively.

We show the distribution of disk masses and radii 
derived from the simulations and compared to the observations of Orion 
in Figures \ref{mass-cumulat} and \ref{radii-cumulat}. In both cases,
we compare to the cumulative distributions of disk radii and masses
for the full Orion sample as well as for a restricted
sample that excludes multiple systems detected in the observations. For the simulations, this means
comparing with the set of systems that never had another
protostar within 2000~au. 
The distribution of simulated disk radii for the full sample
is in reasonable agreement with the distributions of 
Class 0 and Class I disk radii (see Figure \ref{radii-cumulat}). The distribution from the simulations
does fall below our distributions at $\sim$30~au, which indicates
that we detect disks with
smaller radii than those found in the simulations. Considering only
the non-multiple systems from the observations and non-interacting systems
from the simulations, the observed distributions are again in reasonable agreement with the 
distribution of simulated disk radii, overlapping significantly with the
Class 0 distribution.
We do note, however, that the disk radii are calculated differently
for the simulations and observations.
The disk radius measured by \citet{bate2018} corresponds to the point where 63\% of 
the mass is enclosed, which may be smaller than
the Gaussian 2$\sigma$ used for measuring the size of protostellar disks
from dust emission. However, gas disk radii are being measured directly
from the simulations
and are not post-processed to account for optical depth, observational resolution,
and instrumental effects, as required to make a more detailed comparison.
Thus, the general agreement between the simulations and observations should be regarded with caution.

While the radii seem to be in rough agreement between the simulations
and observations, there
are differences for the dust disk mass distributions.
When the full sample (multiple systems included) is considered (see Figure \ref{mass-cumulat}), the
simulations have higher masses until $\sim$50~\mearth, and
at this point the observed systems have a higher fraction of disks at masses less than
$\sim$50~\mearth\ for the Class 0, Class I, and Flat Spectrum disks. If the
non-interacting systems from the simulations are compared to our non-multiple
sample, the distribution of
simulated masses is systematically larger than the observations.
This difference could be due to opacity limiting our ability to measure the 
masses of disks to the degree of accuracy afforded by the simulations.
Alternatively,
the simulations are very likely to have a deficit of low-mass disks that will also skew their
disk mass distribution. Disks with masses
 below $\sim$33~\mearth\ do not form at all in the simulation, while disks 
with masses close to $\sim$33~\mearth\ suffer from high numerical viscosity, 
which causes the disk gas to rapidly accrete onto the protostar thereby reducing the disk lifetime. 
Higher SPH resolution would likely increase the number of small disks in the 
distribution. However, it is worth noting that higher resolution may not necessarily 
increase agreement with observations. \citet{bate2018} also performed a disk resolution 
study and showed that increasing the SPH resolution also increases the disk mass. Thus, 
the high-mass disks in the simulated sample are likely to be more discrepant with the observed disks.

Some of the disagreement between the mass may also be due to statistical bias. 
In order to increase the statistics, the simulation data include disks sampled from a number 
of different snapshots, and a disk around the same sink particle may 
be included multiple times in the distribution at different 
ages. A number of disks form and dissipate over the course of the simulation, but higher-mass disks are more
likely to persist and thus be counted in more snapshots.
Therefore, masses measured from the simulations may include some bias towards more massive disks and
may not fully capture the evolution of disk mass or radius.

In conclusion, improved comparisons of observations 
to simulations will require taking into account radiative transfer effects 
and reconciling different methodologies. Nevertheless, the current simulations give an initial 
indication that the large-scale SPH simulations may be not have all the requisite physics
to reproduce the distributions of observed disk radii and masses. 
Specifically, the simulations underpredict the number of low mass disks (less than 0.01~\msun in gas or
33~\mearth\ in dust) and they overpredict the number of massive disks. The disk radii,
on the other hand, appear to agree reasonably well.

\subsubsection{MHD Simulations}

Due to the greater computational requirements of MHD simulations, a large
characterization of disks formed with MHD is currently difficult. 
\citet{kueffmeier2017} conducted large-scale ideal MHD simulations with turbulence and zoomed in
on several protostellar systems with 22 levels of refinement to have a best resolution
of 2~au. As such, they could not conduct a large number of zoom-in simulations and were only
able to follow 9 of these small-scale zoom-ins to examine the properties of the forming
disks. Simulations can be conducted with higher resolution of individual systems using
grid-based methods \citep[e.g.,][]{kratter2010, machida2011, li2011, tomida2015}, but the context of the 
global star-forming environment is then lost.
The results from \citet{kueffmeier2017} are mixed with regard to the disk properties; 
some disks grew to 100s of au in radius, while others stayed at a few 10s of au in radius. 
This range of disk radii is broadly consistent with our observations, but the 
simulated sample was not large enough to enable a statistical comparison to our data.

Recognizing the difficulty in building up a large sample of disks formed in 
non-ideal MHD simulations, \citet{hennebelle2016} derived analytic prescriptions
for the radii of disks whose formation is regulated by magnetic fields that depend
on the ambipolar diffusion timescale, magnetic field strength, and the combined 
disk mass and stellar mass. The analytic approximation of disk radius with ambipolar diffusion
enabling formation is given by
\begin{equation}
\rad  \simeq  18~{\rm au}~\times~\delta^{2/9 }\left( {   \eta_\mathrm{AD} \over 0.1 \, {\rm s}   } \right)^{2/9} \left( {B_z \over 0.1\,{\rm G} } \right) ^{-4/9}   \left( { M_\mathrm{d} + M_* \over 0.1 \msun} \right)^{1/3},
\end{equation}
and, in the limit of hydrodynamics only, the relation is
\begin{equation}
\rhd  \simeq  106~{\rm au}~\times~\frac{\beta}{0.02} \left( { M_* \over 0.1 \msun} \right)^{1/3} \left( {   \rho_\mathrm{0} \over 10^{-18} \, {\rm g~cm^{-3}}   } \right)^{-1/3}.
\end{equation}
These equations are given in \citet{hennebelle2016}, where a more generalized form and
derivation is also presented. Within these equations, $\delta$ is the scale factor
of the initial density profile adopted from the density profile of the 
singular isothermal sphere \citep{shu1977},
$\eta_\mathrm{AD}$ is the ambipolar diffusivity, $B_z$ corresponds to the poloidal magnetic field
strength, $M_d$ is the disk gas mass, $M_*$ is the protostar
mass, $\beta$ is the ratio of rotational energy to gravitational potential energy,
 and $\rho_\mathrm{0}$ is the central density of the protostellar cloud. For the sake
of simplicity, we adopt the fiducial values for the variables in the equations (except for disk and protostar mass)
and $\delta$=1. We adopt a fiducial protostar mass of 0.25~\msun, similar to
our calculation of Toomre's Q. We plot these relations with the data in Figure \ref{radii-mass} using the
assumption that the dust mass of the disk is 100 times less than the gas mass.
Variations in protostar mass only serve to shift the relationship
up and down slightly and move the curvature to higher disk masses because
$r_{AD}$~$\propto$~(M$_{disk}$ + M$_{*}$)$^{1/3}$. The power-law dependence of disk
radii on dust disk masses found in Section 3.5 is between M$_{disk}$$^{0.25}$ and M$_{disk}$$^{0.34}$,
depending on the sub-sample selected. This power-law dependence is 
similar to the scaling of $r_{AD}$ at high dust disk masses, but the predicted masses of the 
$r_{AD}$ curve only overlap with the distributions of masses and radii for
dust disk masses less than $\sim$70~\mearth.
Most protostars with dust disk masses $>$ 170~\mearth\ lie above the $r_{AD}$ line with the
fiducial protostar mass of 0.25~\msun, but below the line representing
the disk radius predicted from hydrodynamics only. This can be interpreted
as variations in the initial conditions as well as protostar mass governing the disk
radii in the various systems.

\section{Conclusions}

We have conducted a high-resolution survey toward 328 well-characterized
protostars in the Orion A and B molecular clouds using ALMA at 0.87~mm
and the VLA at 9~mm. The resolution of the observations is $\sim$0\farcs1 (40~au in diameter) 
and $\sim$0\farcs08 (32~au) for ALMA and the VLA, respectively, enabling disks
to be characterized and multiple systems to be detected for the 
entirety of the sample, with unprecedented sensitivity for such a large sample.
We detect 286 out of 328 targeted sources with ALMA; however, the total number of discrete continuum
sources detected is 379 with the inclusion of multiple sources and nearby sources
that fell within the ALMA primary beam. The VLA survey targeted 98 protostar systems 
and 4 fields in the OMC1N region, detecting a total of 232 discrete continuum sources.
Many of the sources detected in addition to the targeted protostars and their companions
were nearby Class I sources, Class II sources, or members of 
small groups that are blended at infrared wavelengths.

Our main results are as follows:

\begin{itemize}

\item The mean dust disk masses for the Class 0, Class I, and
Flat Spectrum protostars are 
25.9$^{+7.7}_{-4.0}$, 14.9$^{+3.8}_{-2.2}$ , and 11.6$^{+3.5}_{-1.9}$~\mearth, 
respectively, for the
full sample of detected protostellar continuum sources (including unresolved disks and non-detections). 
When we exclude multiple systems 
(systems having an ALMA- or VLA-detected companion within 10000 AU), the
mean dust disk masses are 38.1$^{+18.9}_{-8.4}$, 13.4$^{+4.6}_{-2.4}$, 14.3$^{+6.5}_{-3.0}$~\mearth\
for Class 0, Class I, and Flat Spectrum sources, respectively. Class I protostars and 
Flat Spectrum protostars have similar dust disk masses, and the dust disk mass distributions 
for the non-multiple protostars in these classes
 are consistent with being drawn from the same sample.
Overall, the dust disk mass systematically
decreases with evolution for the Orion protostars.

\item The Orion protostellar dust disk mass distributions were compared to other populations
of protostellar disks in Ophiuchus, Perseus, and Taurus. The Perseus disks have systematically
larger dust disk masses, but are measured using 8.1~mm flux densities \citep{tychoniec2018}.
The Orion dust disk masses calculated from the 9~mm flux densities are comparable to the Perseus dust disk masses,
as are the 9~mm flux density distributions. Thus, no true difference of protostellar dust disk masses
can be discerned between Orion and Perseus at present.
The Orion disks have mass distributions comparable to the protostellar disks in Taurus, 
despite the Taurus masses being derived from radiative transfer modeling.
However, the Orion disks
are systematically more massive than those in Ophiuchus 
from the \citet{cieza2018,williams2019} survey, even when 
the mass distributions are constructed with the same set of assumptions for dust
opacity and temperature. Contamination of the Ophiuchus Class I and Flat Spectrum
sample with highly extincted Class II sources could partly account
for this inconsistency \citep{mcclure2010}, 
but would not fully reconcile the difference.

\item The protostellar disks in Orion are
more than four times more massive than the samples of Class II disks
that have been surveyed in other regions. This indicates that
the Orion protostellar disks have more raw material from which planet formation can 
take place. Also, this finding could signify that significant growth 
of solids and perhaps the formation of planetesimals and planetary cores 
happens in the protostellar phase, prior to evolving into Class II disks.
In this case, significant evolution of the solids must happen during the protostellar phases,
and perhaps seeding the Class II disks with large particles and perhaps planetesimals.

\item The mean dust disk radii for the Class 0, Class I, and
Flat Spectrum protostars are 
44.9$^{+5.8}_{-3.4}$, 37.0$^{+4.9}_{-3.0}$, and 28.5$^{+3.7}_{-2.3}$~au,
respectively, for the
full sample of detected protostellar continuum sources. 
When we exclude multiple systems (both wide and close), 
the mean radii are 53.7$^{+8.4}_{-4.2}$, 35.4$^{+6.1}_{-3.5}$, and 36.0$^{+5.9}_{-3.2}$~au
for Class 0, Class I, and Flat Spectrum sources, respectively.
Despite the apparent decrease of mean disk radii, statistical comparisons of the
disk radii distributions indicate that the Class 0 and Class I disk radii distributions
for the full sample are consistent with being drawn from the same sample, while the Class 0 
and Flat Spectrum distributions are inconsistent with being drawn from the same sample, as are the 
Class I and Flat Spectrum distributions. On the other hand, the distributions of disk
radii for the non-multiple samples are all consistent with having been drawn from the same sample.
These findings are seemingly contrary to simple 
predictions for disk formation in rotating, collapsing cores; however, these comparisons are
for dust disk radii and may not reflect the same distribution as that of the gas disks.

\item There are 61 dust continuum sources associated with Class 0 protostars having dust disk radii
inferred to be greater than 50 au, in addition to 57 Class I and 35 Flat
Spectrum sources, corresponding to 46\%, 38\%, and 26\%, respectively, 
of the detected continuum sources and non-detections in each class. If we
only consider the non-multiple protostars, the percentages of protostars
with a disk $>$ 50~au for Class 0, Class I, and Flat Spectrum are 54\%, 38\%, and 37\%, respectively. 
The distributions of the dust disk radii for the Orion Class 0,
Class I, and Flat Spectrum sources show that protostellar disk radii are systematically
smaller than Class II disks in Taurus, appear comparable to Lupus, and are
systematically larger than the samples in Chamaeleon and the Orion
Nebula Cluster. Some of the differences could be due to the measurement techniques, thus it is
currently unclear if the main drivers of disk evolution can be derived from comparing these
samples.

\item The protostellar dust disk masses and radii exhibit no statistically
significant differences between Orion B, the northern ISF, and L1641 and the southern ISF. These regions 
within Orion sample a variety of protostellar and gas densities. The similarity
between these distinct regions within Orion may suggest that there is the potential
for protostellar disk properties as an ensemble to be similar between different star-forming regions.
This is because the apparent differences in physical conditions between these sub-regions
of Orion have not lead to a significant difference in their disk properties.
 However, the differences of Orion compared to Ophiuchus and Perseus remain to be
reconciled and fully understood.

\item When compared to current numerical simulations of star formation that
include molecular cloud scales down to disk scales, we find that 
simulations without magnetic fields have comparable disk radii but larger masses 
as compared to the observations. Simulations with magnetic fields are 
not as extensive but may also compare 
favorably with the observations. Many 
disk radii measured toward the Orion protostars are between
an analytic approximation for disk radii formed in non-ideal MHD simulations
and predictions of disks formed without the influence of magnetic fields. 
Thus, it seems likely that the initial conditions for collapse play a role
in setting the properties of the protostellar disks, but the relative importance (or
lack thereof) of magnetic fields, turbulence, and envelope rotation are still uncertain.

\end{itemize}

The authors wish to thank the referee J. Williams for a constructive report 
and useful suggestions with regard to comparing with the Ophiuchus disk sample.
The authors also thank J. Eisner for supplying the dust disk mass distributions
for the Class II disk samples and M. Bate for useful discussions regarding
his simulations.
J.J.T. acknowledges  support from  NSF AST-1814762  and past support 
from the Homer L. Dodge Endowed Chair.
ZYL is supported in part by NASA 80NSSC18K1095 and NSF AST-1716259, 1815784 and 1910106. 
GA, MO, and AKD-R acknowledge financial support from
the State Agency for Research of the Spanish MCIU through the AYA2017-84390-C2-1-R grant
(co-funded by FEDER) and through the ``Center of Excellence Severo Ochoa'' award for the
Instituto de Astrof\'\i sica de Andaluc\'\i a (SEV-2017-0709). M.L.R.H. acknowledges support 
from a Huygens fellowship from Leiden University.
This paper makes use of the following ALMA data: ADS/JAO.ALMA\#2015.1.00041.S.
ALMA is a partnership of ESO (representing its member states), NSF (USA) and 
NINS (Japan), together with NRC (Canada), NSC and ASIAA (Taiwan), and 
KASI (Republic of Korea), in cooperation with the Republic of Chile. 
The Joint ALMA Observatory is operated by ESO, AUI/NRAO and NAOJ.
The PI acknowledges assistance from Allegro, the European ALMA Regional
Center node in the Netherlands.
The National Radio Astronomy 
Observatory is a facility of the National Science Foundation 
operated under cooperative agreement by Associated Universities, Inc.
This research made use of APLpy, an open-source plotting package for Python 
hosted at http://aplpy.github.com. This research made use of Astropy, 
a community-developed core Python package for 
Astronomy (Astropy Collaboration, 2013) http://www.astropy.org.

 \facility{ALMA, VLA}
\software{Astropy \citep[http://www.astropy.org; ][]{astropy2013,astropy2018}, 
APLpy \citep[http://aplpy.github.com; ][]{aplpy}, CASA \citep[http://casa.nrao.edu; ][]{mcmullin2007}}

\begin{small}
\bibliographystyle{apj}
\bibliography{ms}
\end{small}

\clearpage

\begin{figure}
\includegraphics[scale=0.45]{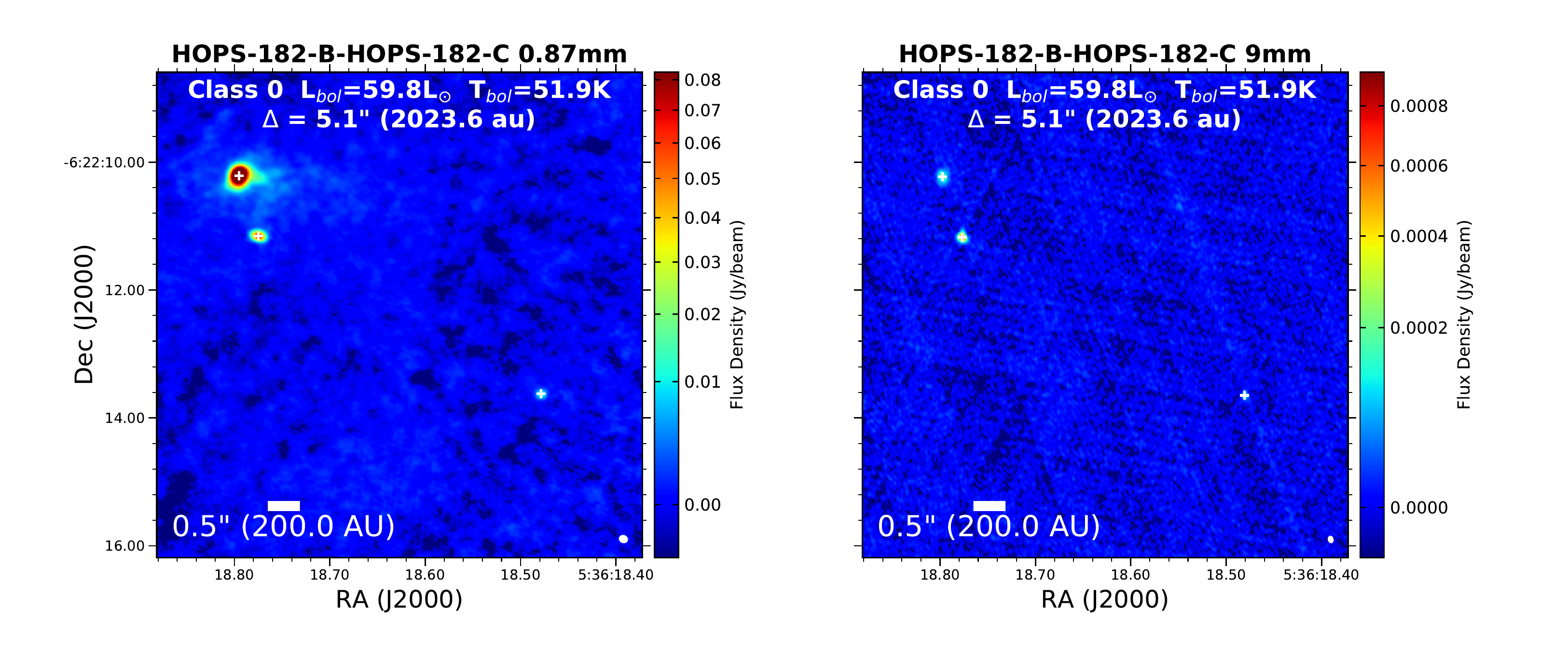}
\includegraphics[scale=0.45]{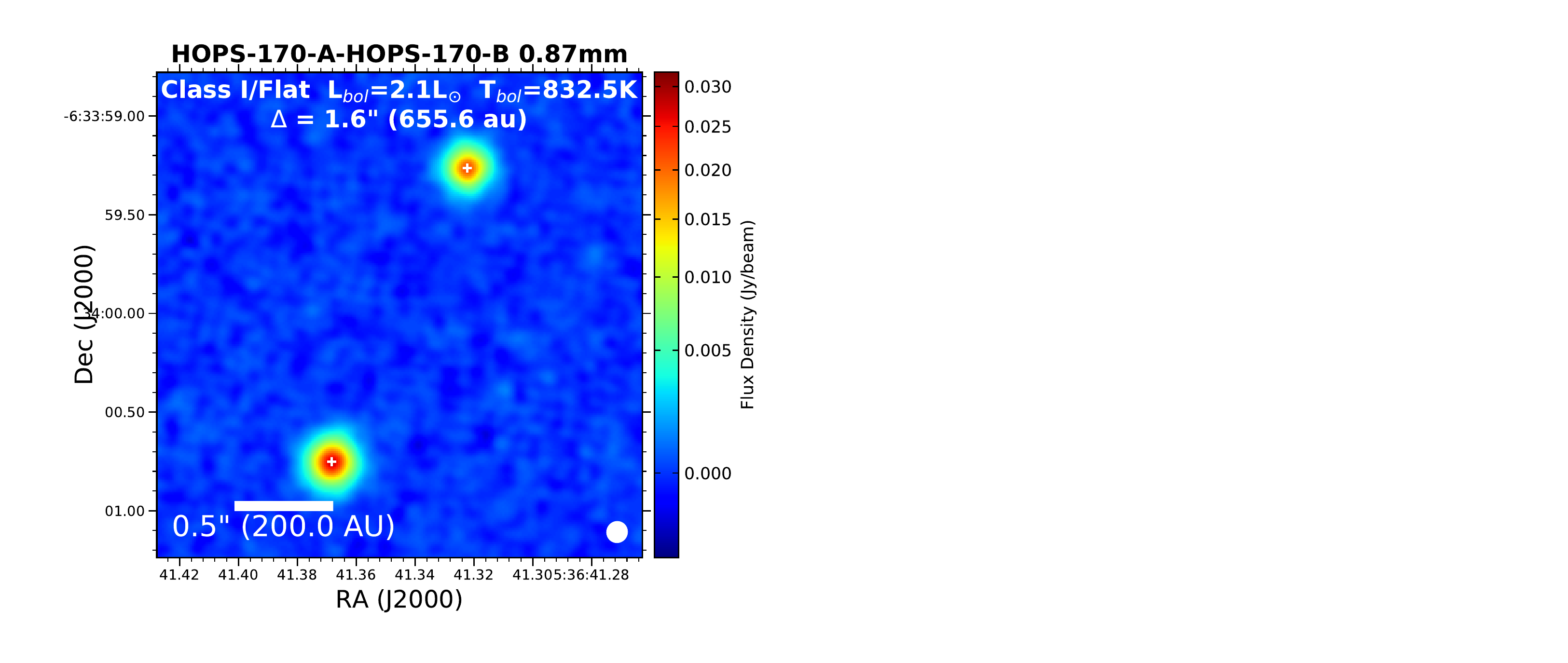}
\includegraphics[scale=0.45]{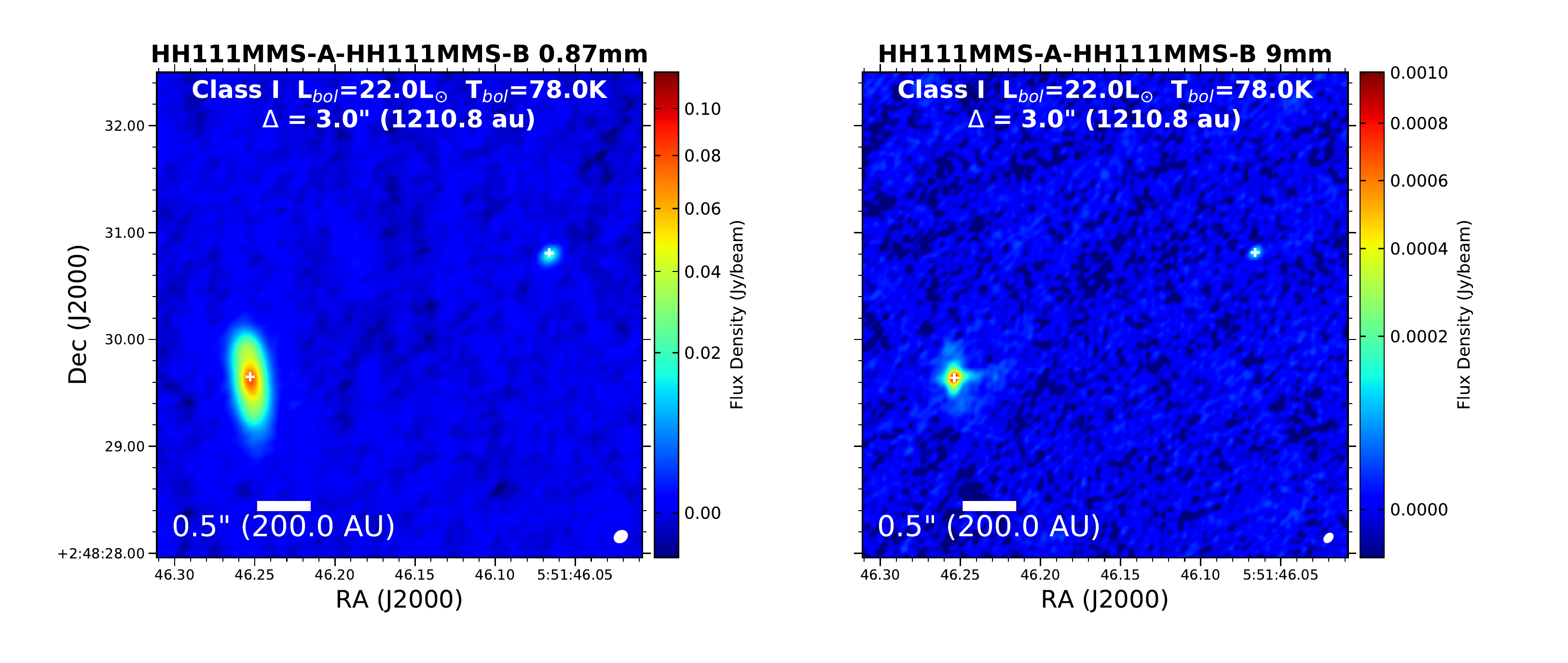}
\caption{Example images from ALMA (left) at 0.87~mm and the VLA (right) at 9~mm toward
selected wide multiple systems. The top row shows HOPS-182 (also known as L1641N),
which is made up of a close and wide multiple system. The brighter sources, HOPS-182-A and -B
both appear resolved at 0.87~mm and 9~mm. A binary Class I source, HOPS-170, is shown 
in the middle panel, these sources both show apparent resolved disks that are likely face-on.
HH111mms is shown in the bottom panel, showing the brighter disk-like source with a fainter
companion separated by 3\arcsec and detected at both wavelengths. Furthermore, the jet from HH111mms
is also detected at 9~mm. The synthesized
beams are drawn in the lower right, the typical ALMA synthesized beam is 0\farcs1 and the typical
VLA beam is 0\farcs08. Images for the remaining sources are shown in Appendix C.
A blank panel for the ALMA 0.87~mm
or VLA 9~mm panel means that observations were not taken toward that particular protostar with
ALMA or the VLA.}
\label{continuum-demo2}
\end{figure}

\begin{figure}
\includegraphics[scale=0.5]{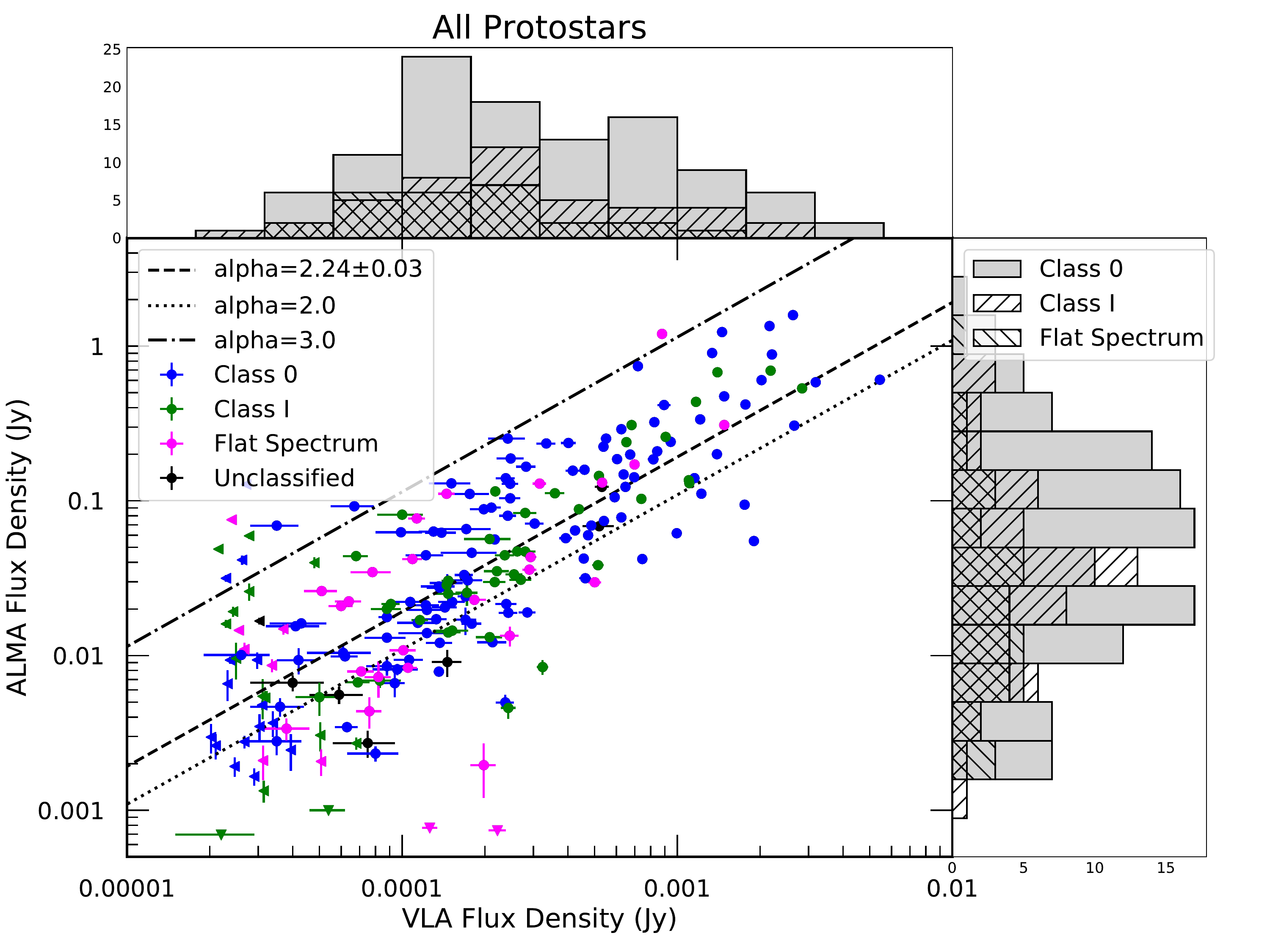}
\caption{Plot of ALMA 0.87~mm flux density versus VLA 9.1~mm flux density in log-log space. The observational
class of the sources are denoted by color where Class 0 protostars are blue, Class I 
protostars are green, and Flat Spectrum sources are
magenta; unclassified sources are black. A strong correlation between the ALMA and VLA flux densities is 
present, and a constant spectral index can fit the correlation with a 
spectral index ($\alpha$) $\alpha$ = 2.24$\pm$0.03 (ignoring upper limits). We also plot 
the line representing $\alpha$ = 2, points below this line require additional free-free emission to 
explain their shallow spectral slopes. Most sources have $\alpha$ between 2 and 3, 
consistent with optically thick to optically
thin dust emission. The left pointing or downward pointing
triangles denote upper limits for the VLA 9~mm or ALMA 0.87~mm observations, respectively.
}
\label{fluxdensity}
\end{figure}

\begin{figure}
\includegraphics[scale=0.31]{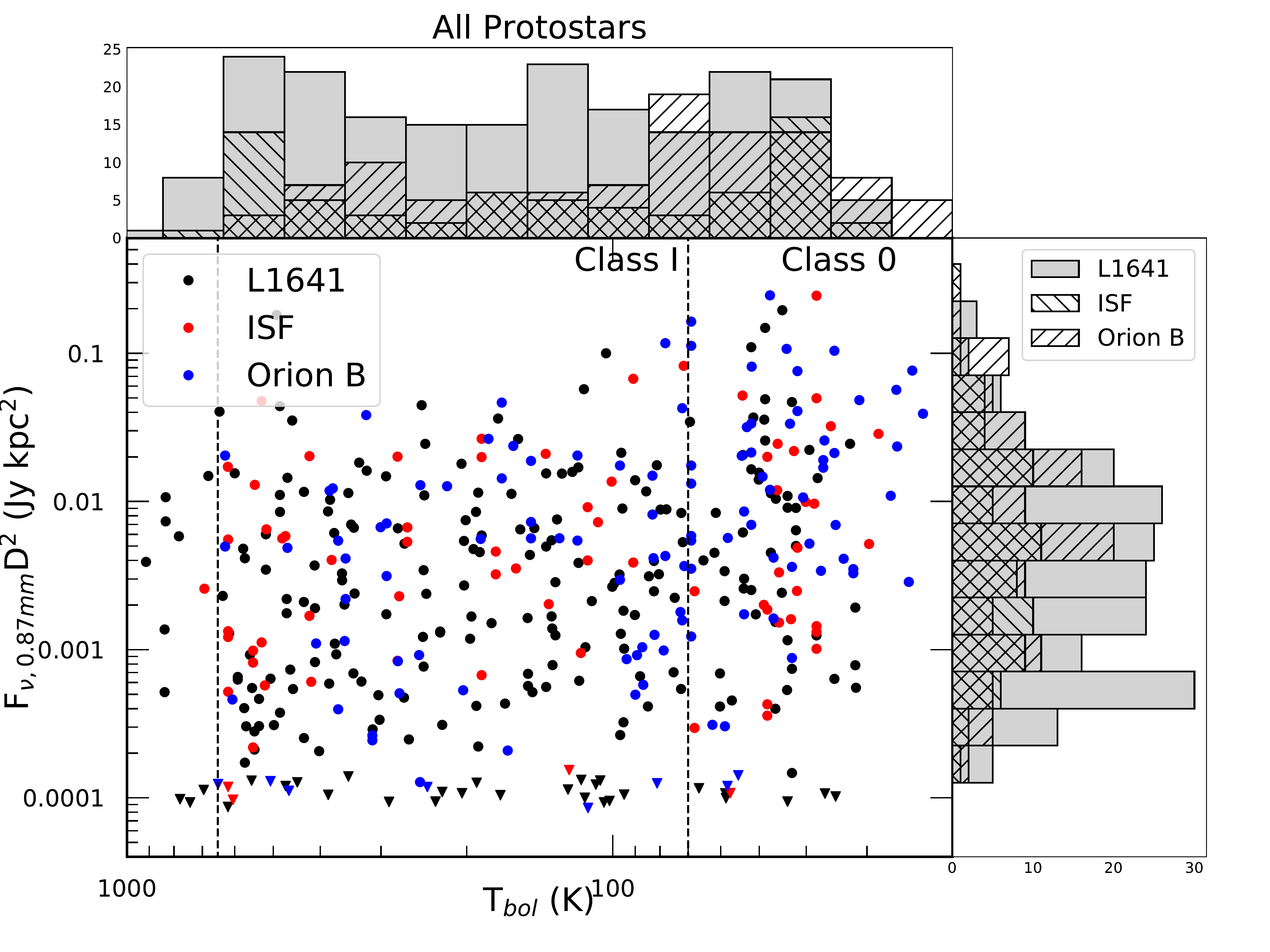}
\includegraphics[scale=0.31]{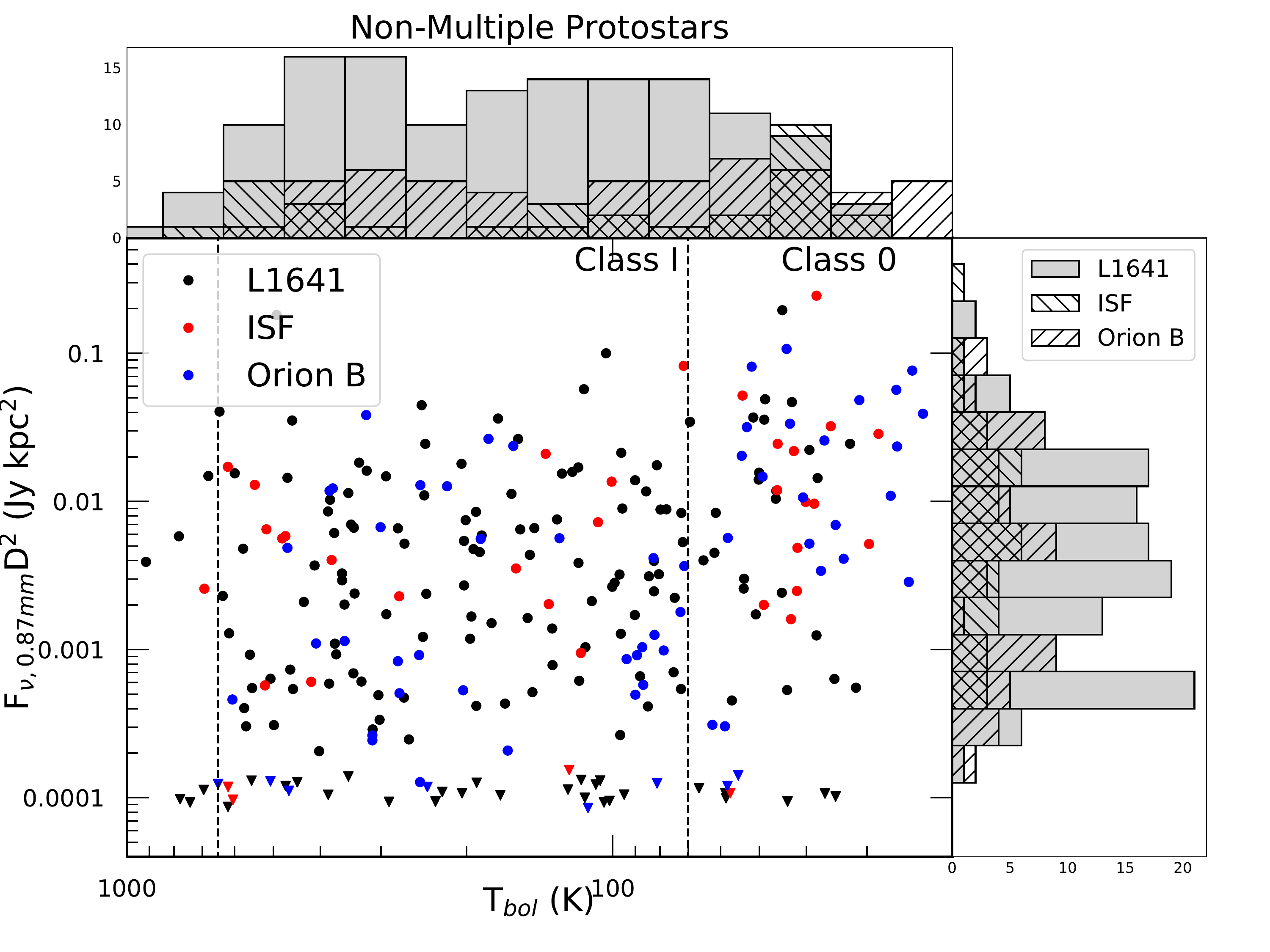}
\includegraphics[scale=0.31]{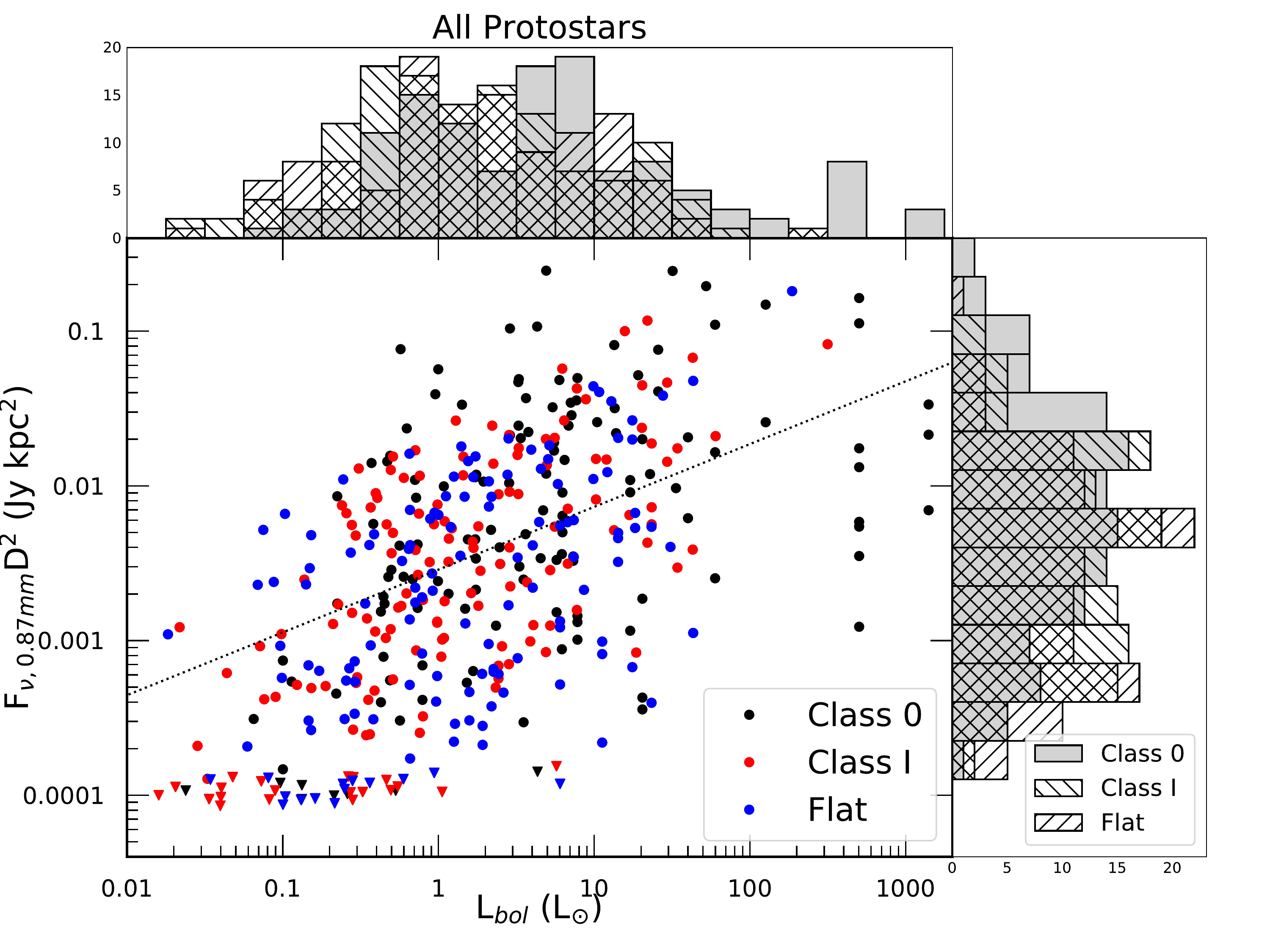}
\includegraphics[scale=0.31]{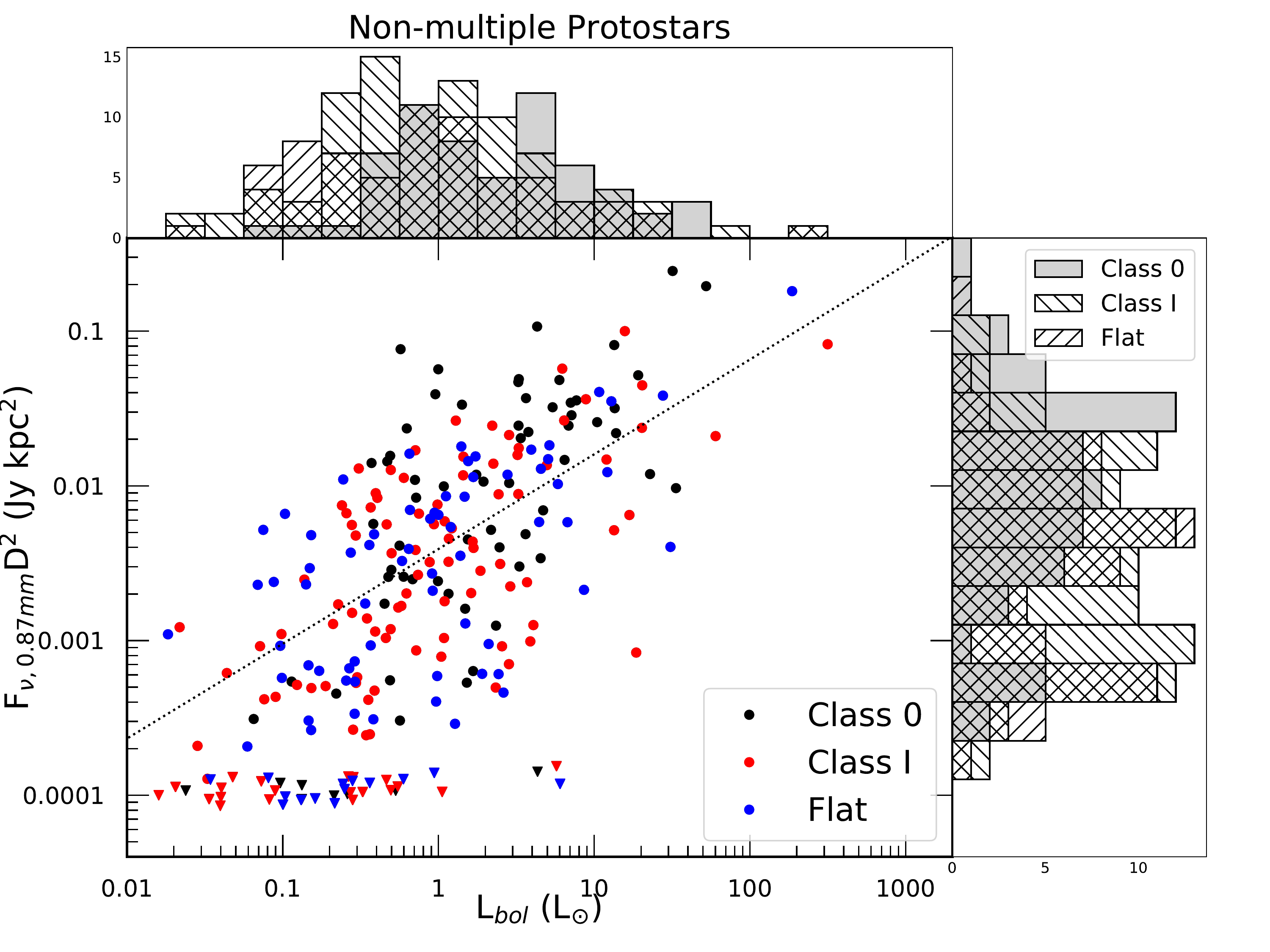}
\caption{
Comparison of 0.87~mm flux densities to \tbol\ (top panels) and \lbol\ (bottom panels); 
the
flux densities are multiplied by the square of their distance in kpc to remove scatter due to 
different distances within the region. The left panels
show the full survey sample with a corresponding measurement
of \lbol\ and \tbol, while the right panels show only the non-multiple sample.
 The colorization and histograms associated with the top panels
separates the sources by their regions (black: L1641 and southern ISF, red: Northern ISF, blue: Orion B).
In the bottom
panels, the colors and associated histograms separate the sources by class;
 black corresponds to Class 0 protostars, red corresponds to Class I protostars,
and blue corresponds to Flat spectrum protostars.
Upper limits are denoted as downward facing triangles.
The lines in the lower panels are the fits to the 0.87~mm flux densities versus \lbol, for the
full sample (left panel) we find $F_{\nu}$D$^2$~$\propto$~L$_{bol}^{0.4\pm0.04}$
 and for the non-multiple sample (right) we find $F_{\nu}$D$^2$~$\propto$~L$_{bol}^{0.61\pm0.05}$. 
The vertical lines in the top panels denote the separations between Class 0 
and Class I (70~K) and Class I and Class II (650~K).
}
\label{lum0.87-blt}
\end{figure}

\begin{figure}
\includegraphics[scale=0.31]{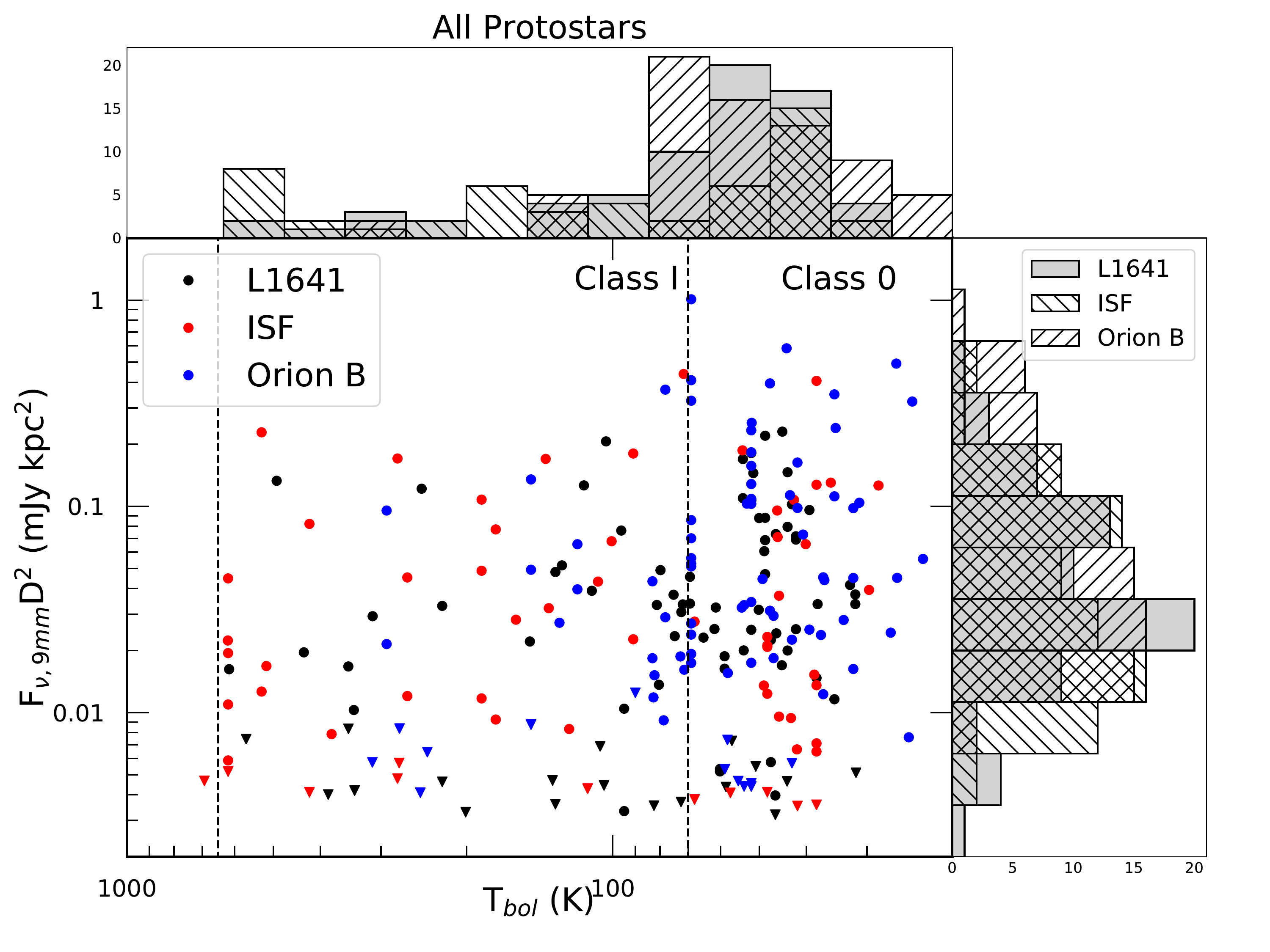}
\includegraphics[scale=0.31]{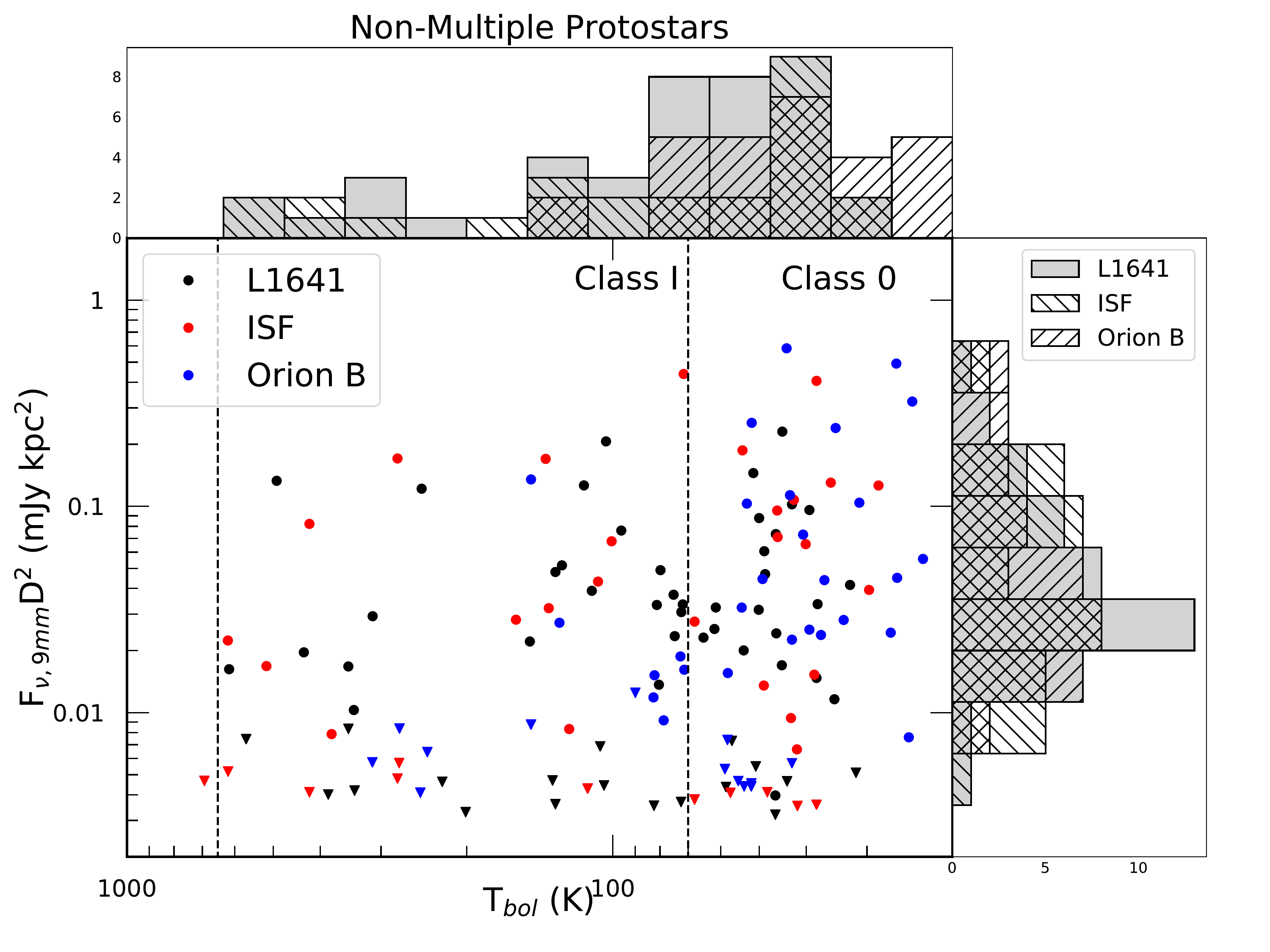}
\includegraphics[scale=0.31]{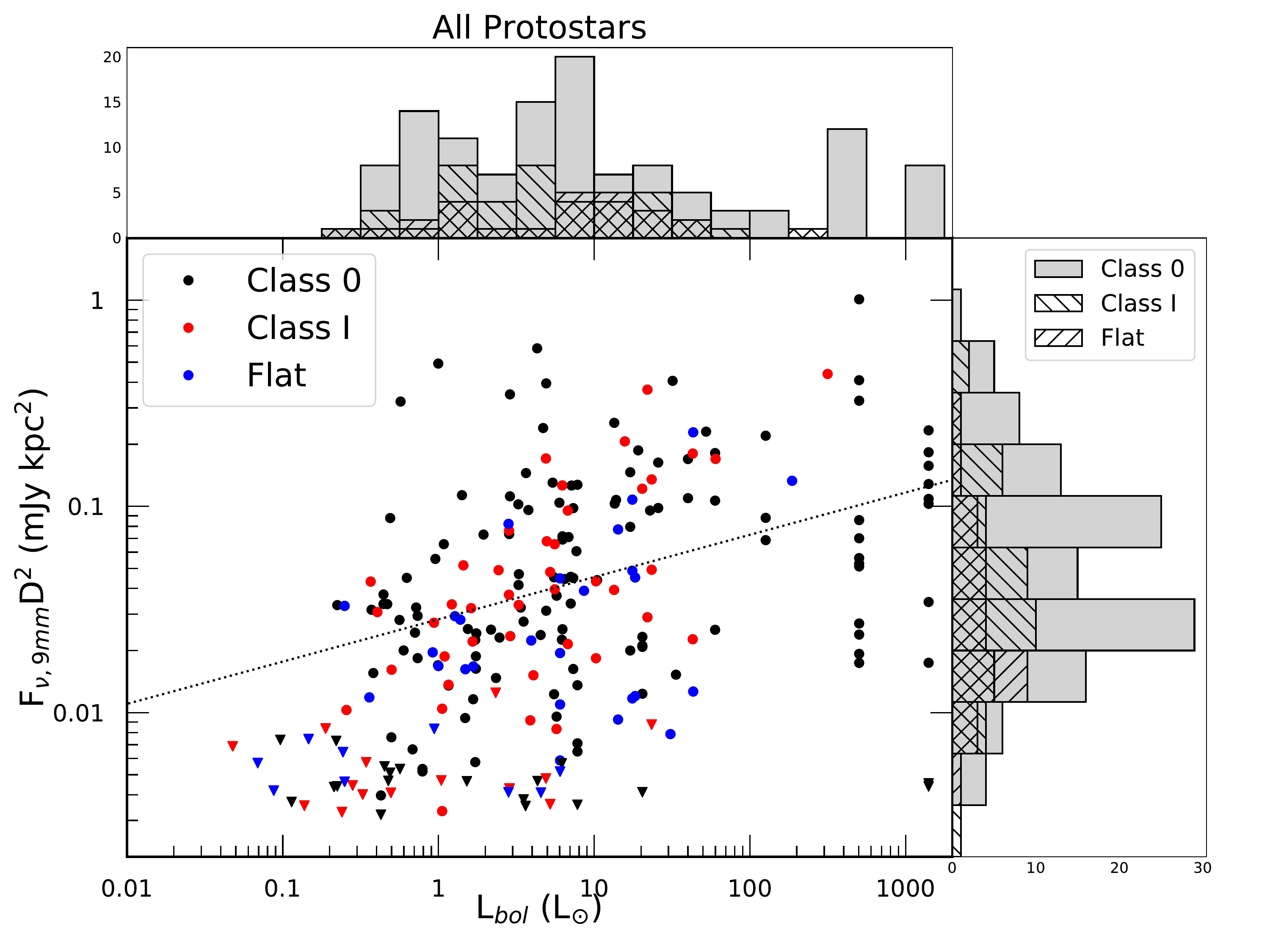}
\includegraphics[scale=0.31]{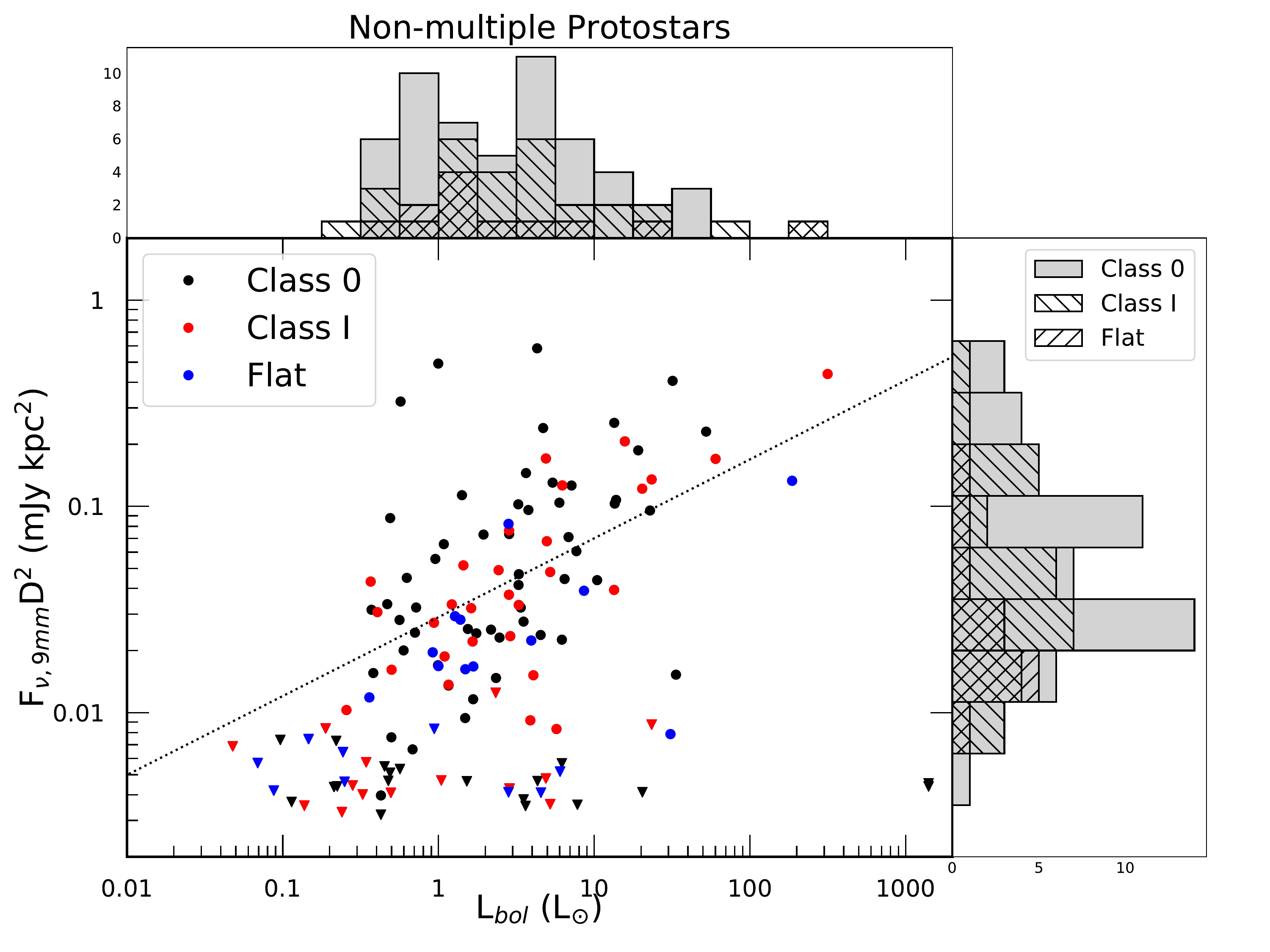}
\caption{
Same as Figure \ref{lum0.87-blt}, but comparing the 9~mm flux densities.
The lines in the lower panels are the fits to the 9~mm flux densities versus \lbol, for the
full sample (left panel) we find $F_{\nu}$D$^2$~$\propto$~L$_{bol}^{0.2\pm0.03}$
 and for the non-multiple sample (right) we find $F_{\nu}$D$^2$~$\propto$~L$_{bol}^{0.38\pm0.07}$.
}
\label{lum9-blt}
\end{figure}

\begin{figure}
\includegraphics[scale=0.31]{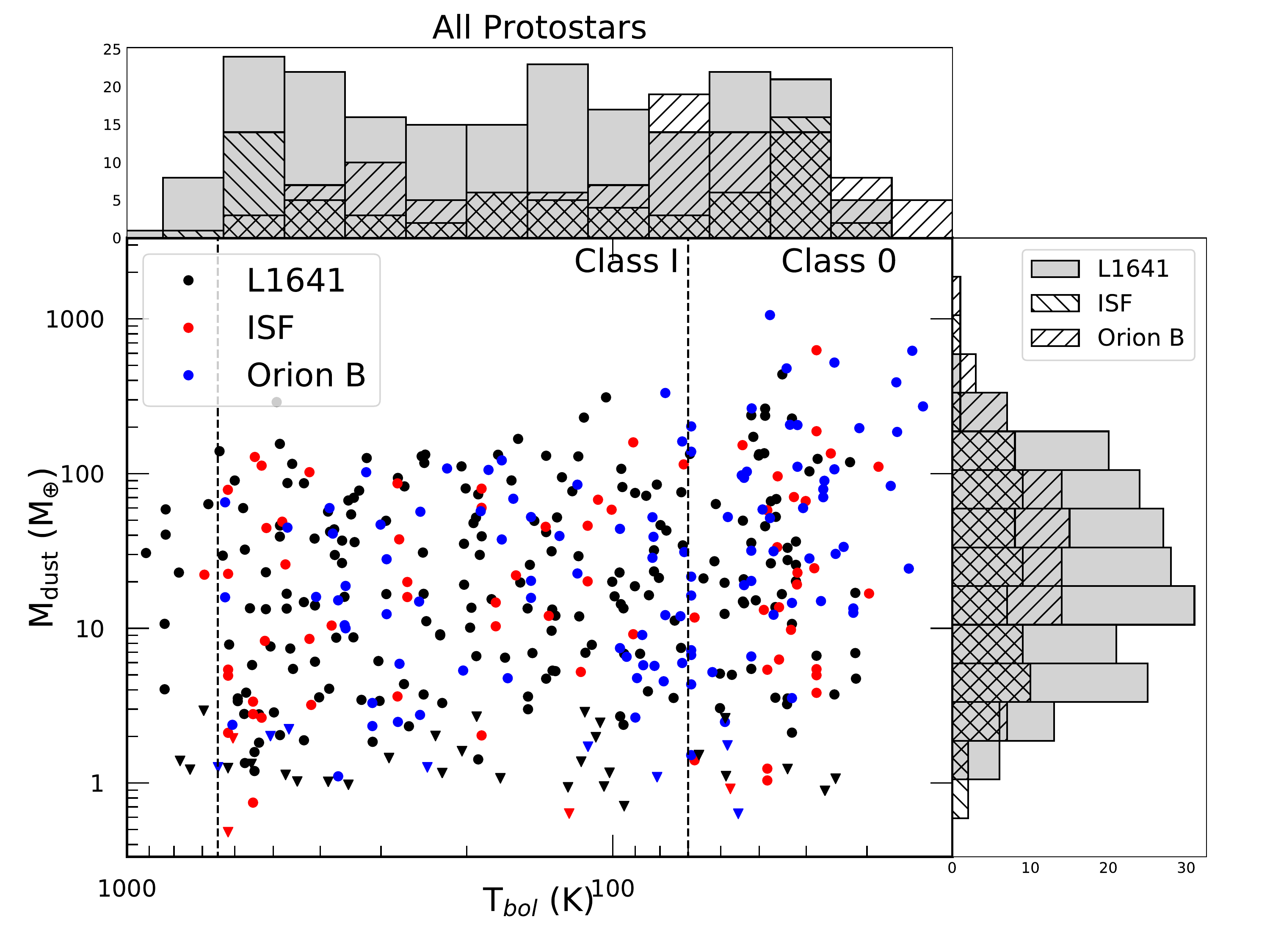}
\includegraphics[scale=0.31]{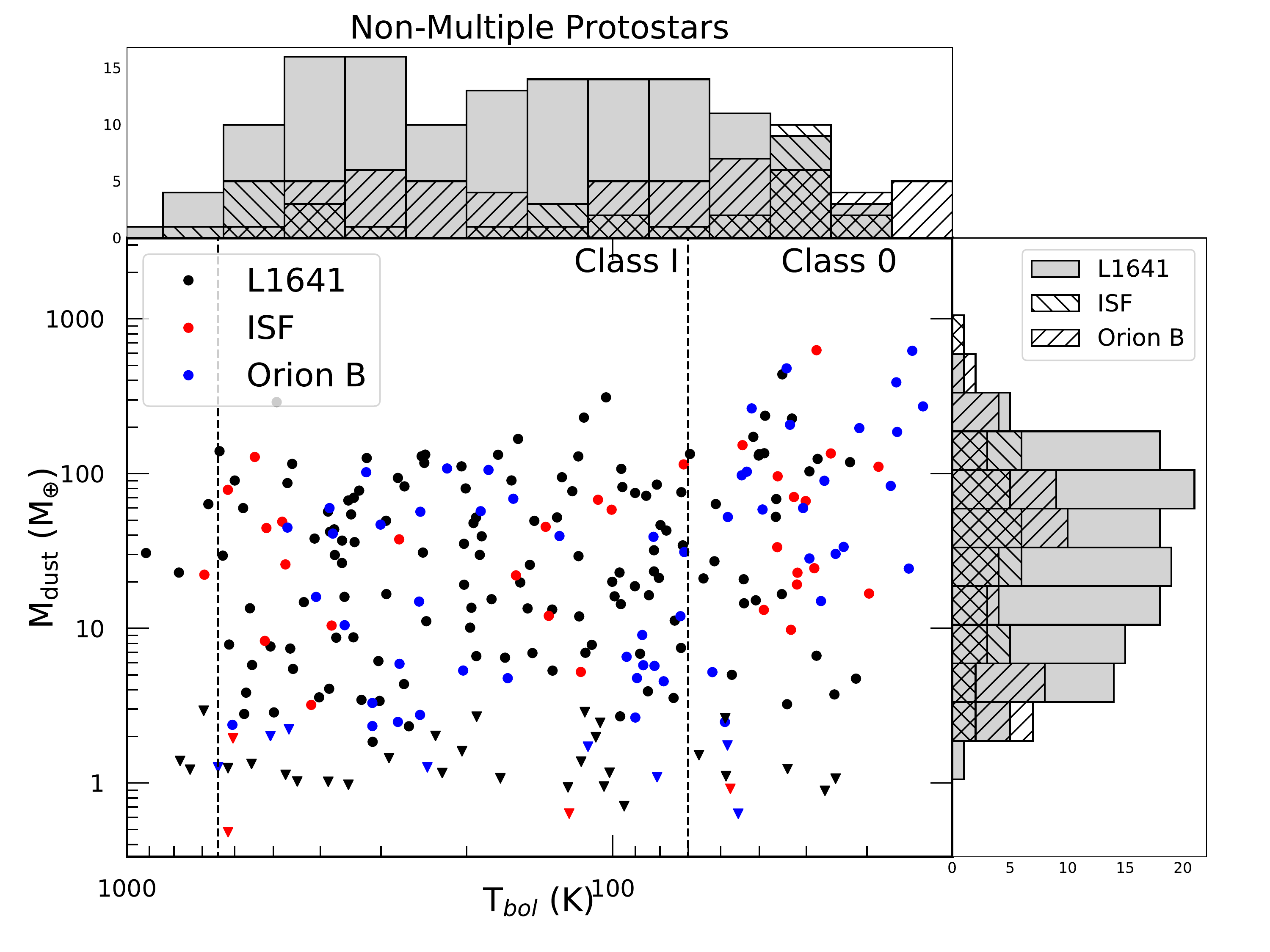}
\includegraphics[scale=0.31]{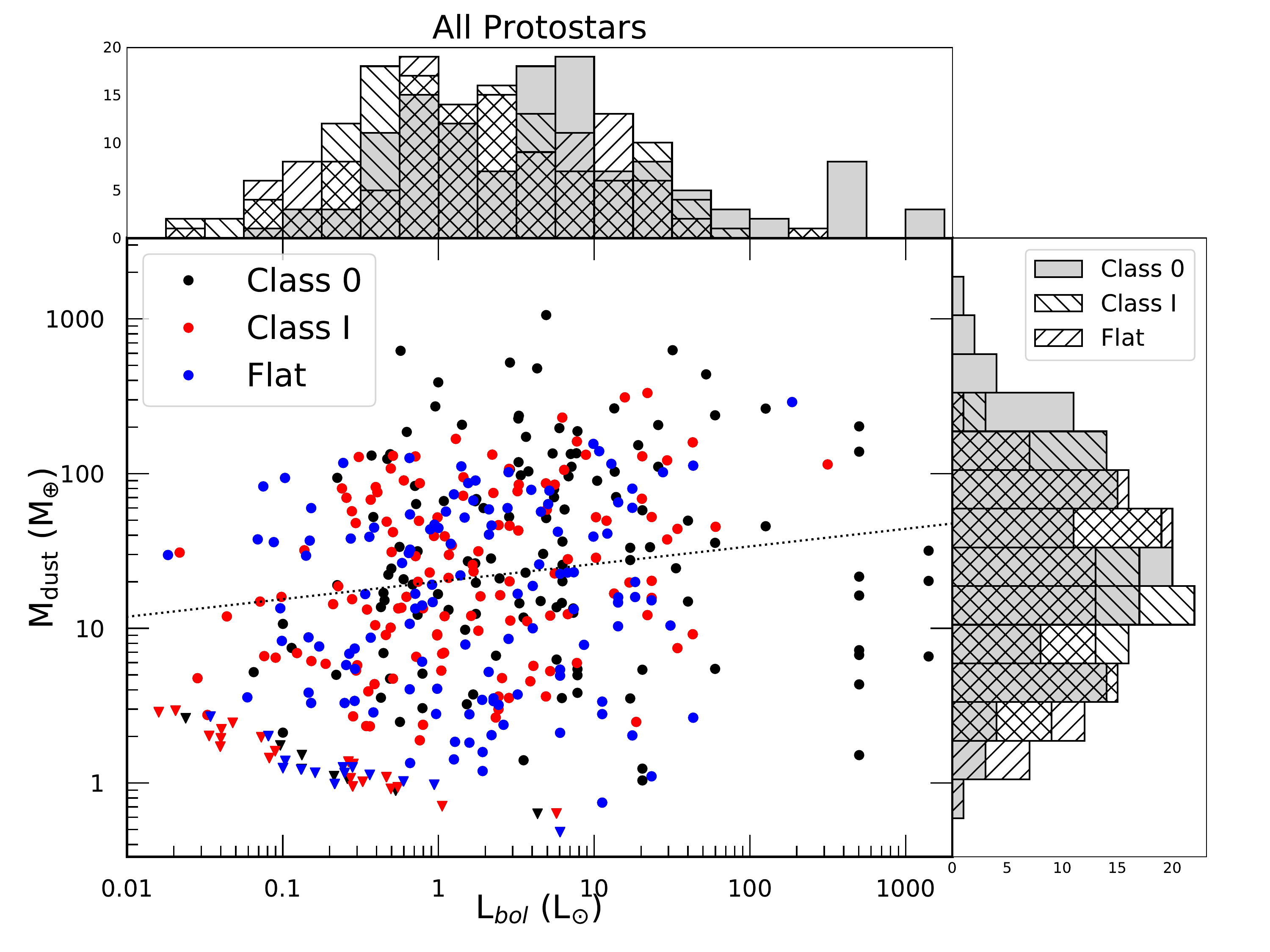}
\includegraphics[scale=0.31]{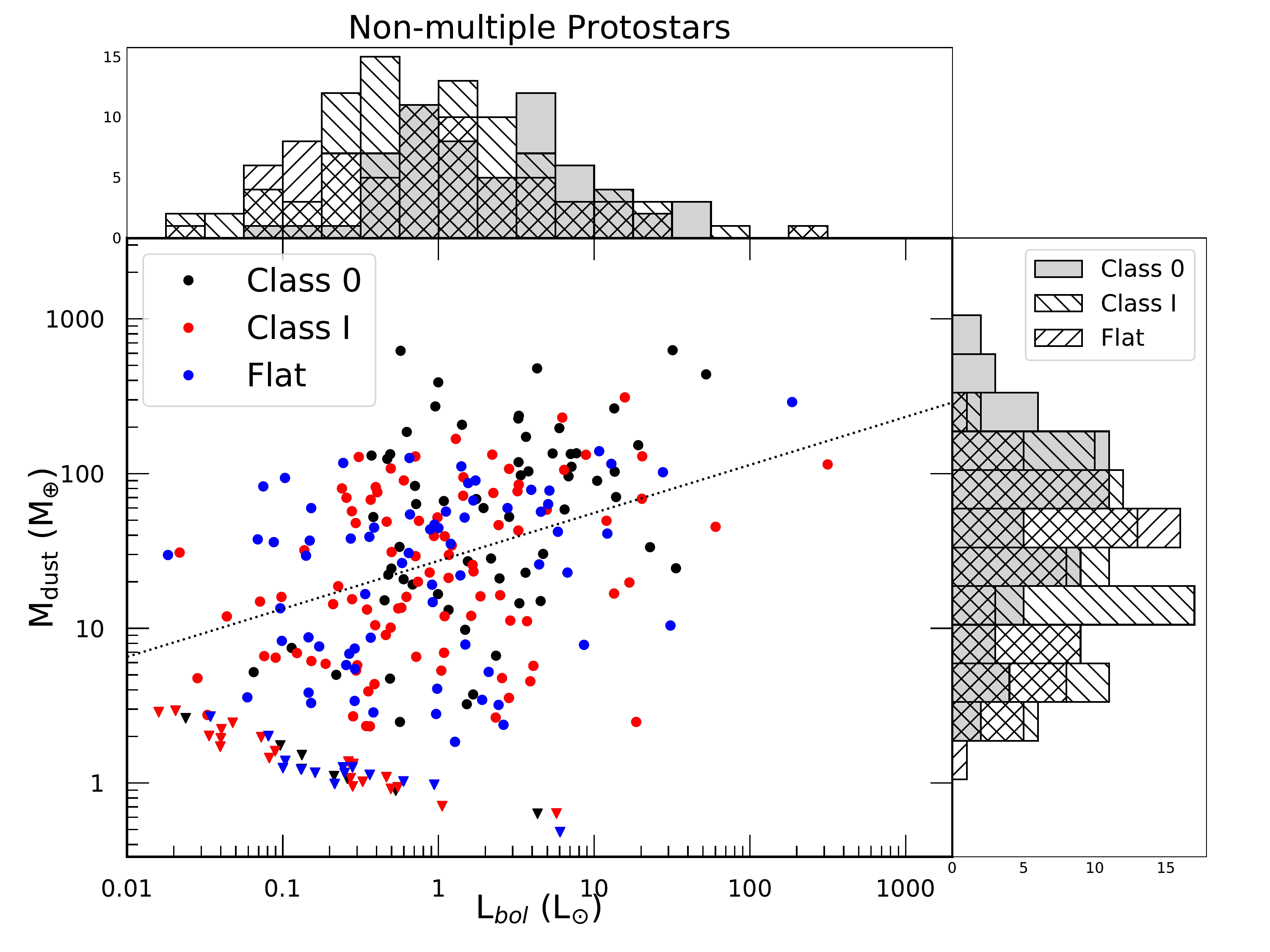}
\caption{
Comparison of dust disk masses to \tbol\ (top panels) and \lbol\ (bottom panels). 
The left panels
include the full sample from the survey with a corresponding measurement
of \lbol\ and \tbol, while the right panels show only the non-multiple sample. 
The colorization and histograms associated with the top panels
separates the sources by their regions (black: L1641 and southern ISF, red: Northern ISF, blue: Orion B).
In the bottom
panels, the colors and associated histograms separate the sources by class;
 black corresponds to Class 0 protostars, red corresponds to Class I protostars,
and blue corresponds to Flat spectrum protostars.
Upper limits are denoted as downward facing triangles.
The lines in the lower panels are the fits to the dust disk masses versus \lbol, for the
full sample (left panel) we find $M_{disk}$~$\propto$~L$_{bol}^{0.11\pm0.04}$
 and for the non-multiple sample (right) we find $M_{disk}$~$\propto$~L$_{bol}^{0.31\pm0.05}$.
 The vertical lines in the top panels denote the separations between Class 0 and Class I (70~K) and Class I and Class II (650~K).
}
\label{mass-blt}
\end{figure}

\begin{figure}
\includegraphics[scale=0.31]{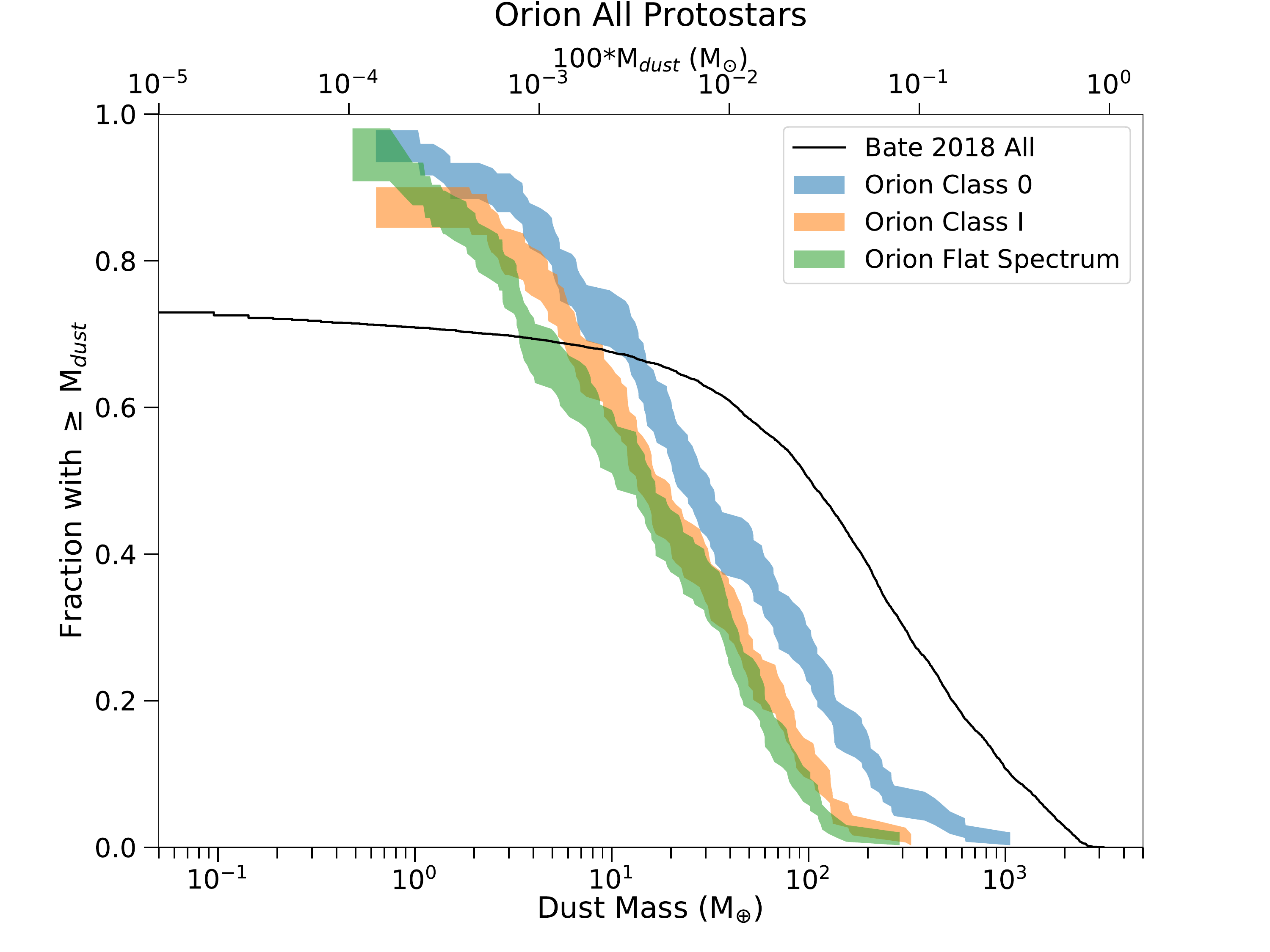}
\includegraphics[scale=0.31]{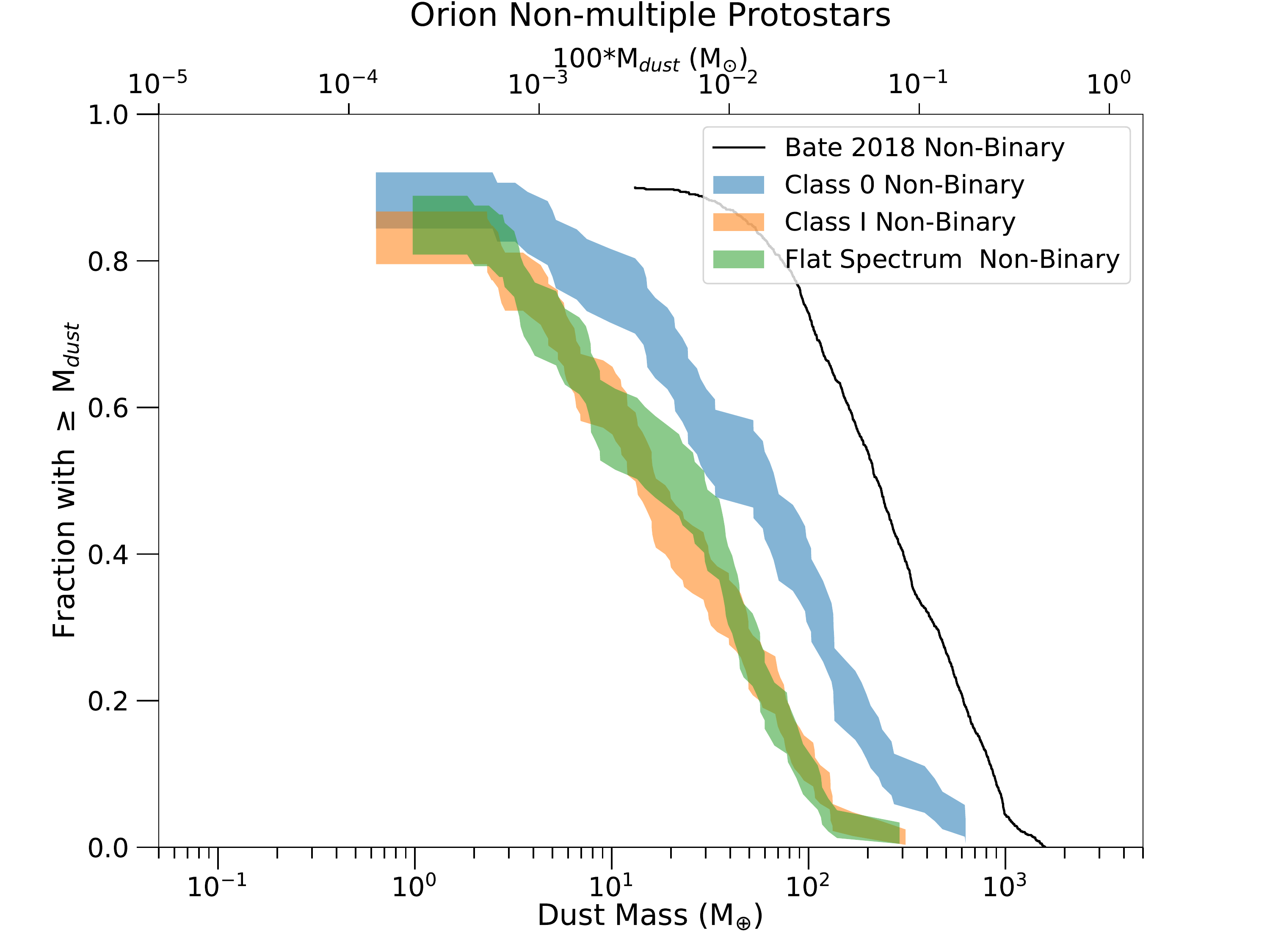}
\includegraphics[scale=0.31]{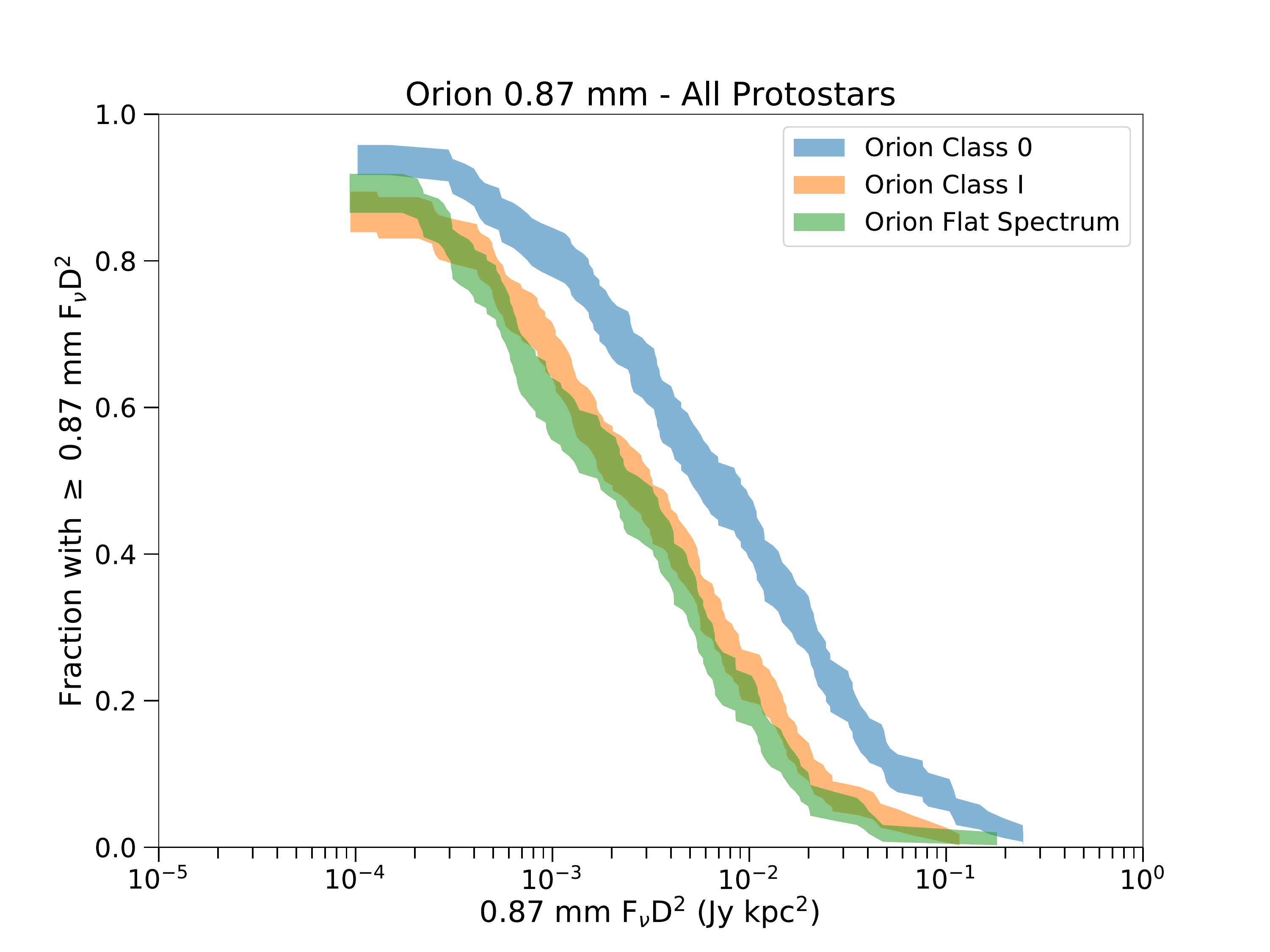}
\includegraphics[scale=0.31]{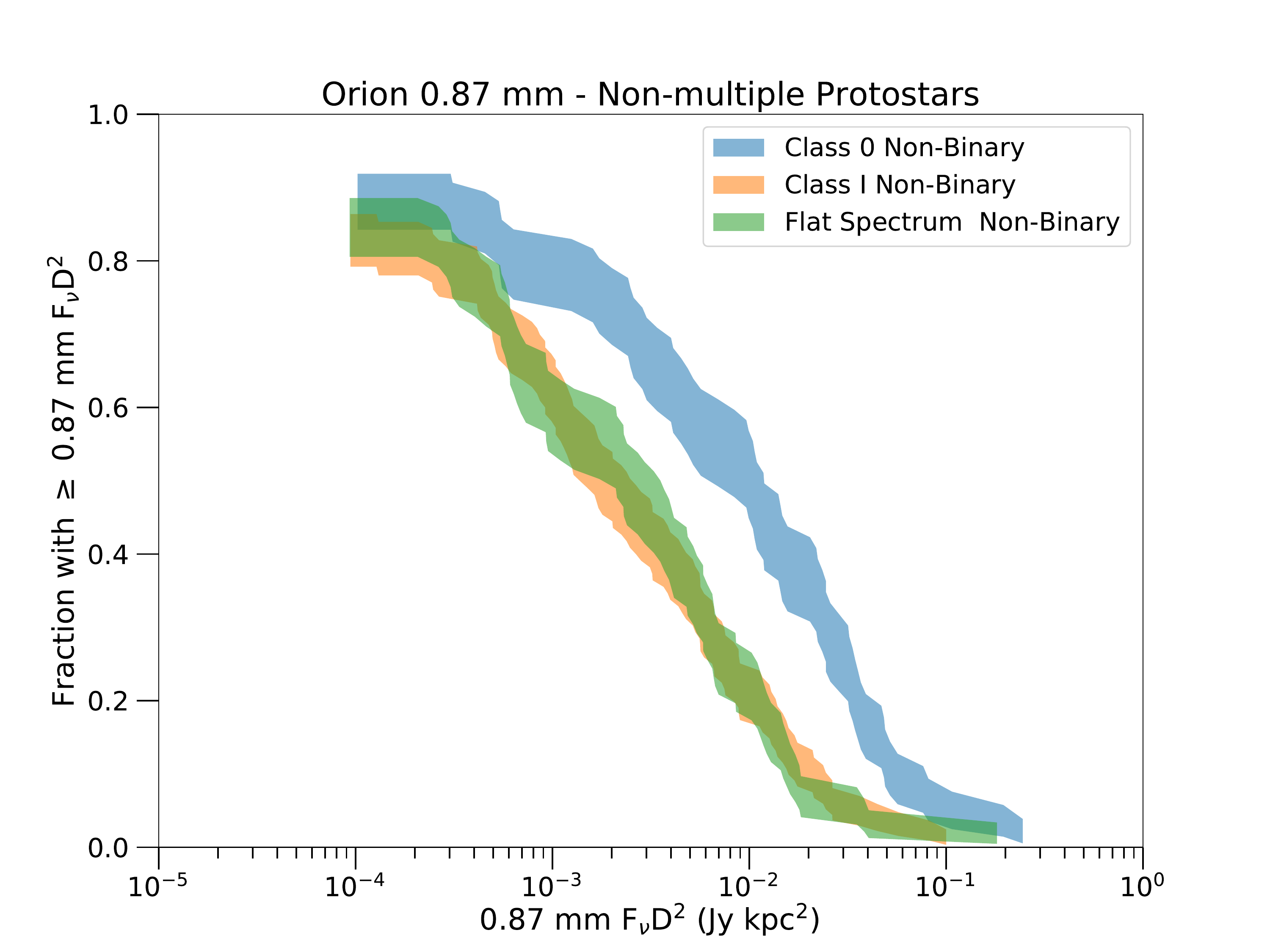}
\caption{
Cumulative distributions of dust disk masses
(top row) and 0.87~mm flux densities multiplied by their distance in kpc squared (bottom row) 
within the Orion sample. The left panels show the full sample, 
while the right panels only show the non-multiple
sample. Class 0 dust disk masses
are drawn with a blue shaded region, Class I dust disk masses are drawn with an orange shaded region, Flat
Spectrum dust disk masses are drawn with a green shaded region.
 The Class 0 dust disk masses clearly have their distribution
shifted toward higher masses, while there is less difference between the Class I and Flat
Spectrum dust disk masses; this is true for the full and non-multiple samples.
The full and non-multiple samples are also compared to the dust disk masses derived from the
large-scale simulations of \citet{bate2018}. The mass distributions from the simulations have
a systematic shift toward higher masses in both the full and non-multiple samples. 
Statistical comparisons between the dust disk masses of each class is discussed in the text.
The distributions of 0.87~mm flux densities, from which the dust disk masses are derived, show the same
overall trend of higher 0.87~mm flux densities for Class 0. The Class I 0.87~mm flux densities are
systematically higher than Flat Spectrum sources for the population of all protostars, but 
have substantial overlap for the non-multiple sample.
}
\label{mass-cumulat}
\end{figure}

\begin{figure}
\includegraphics[scale=0.31]{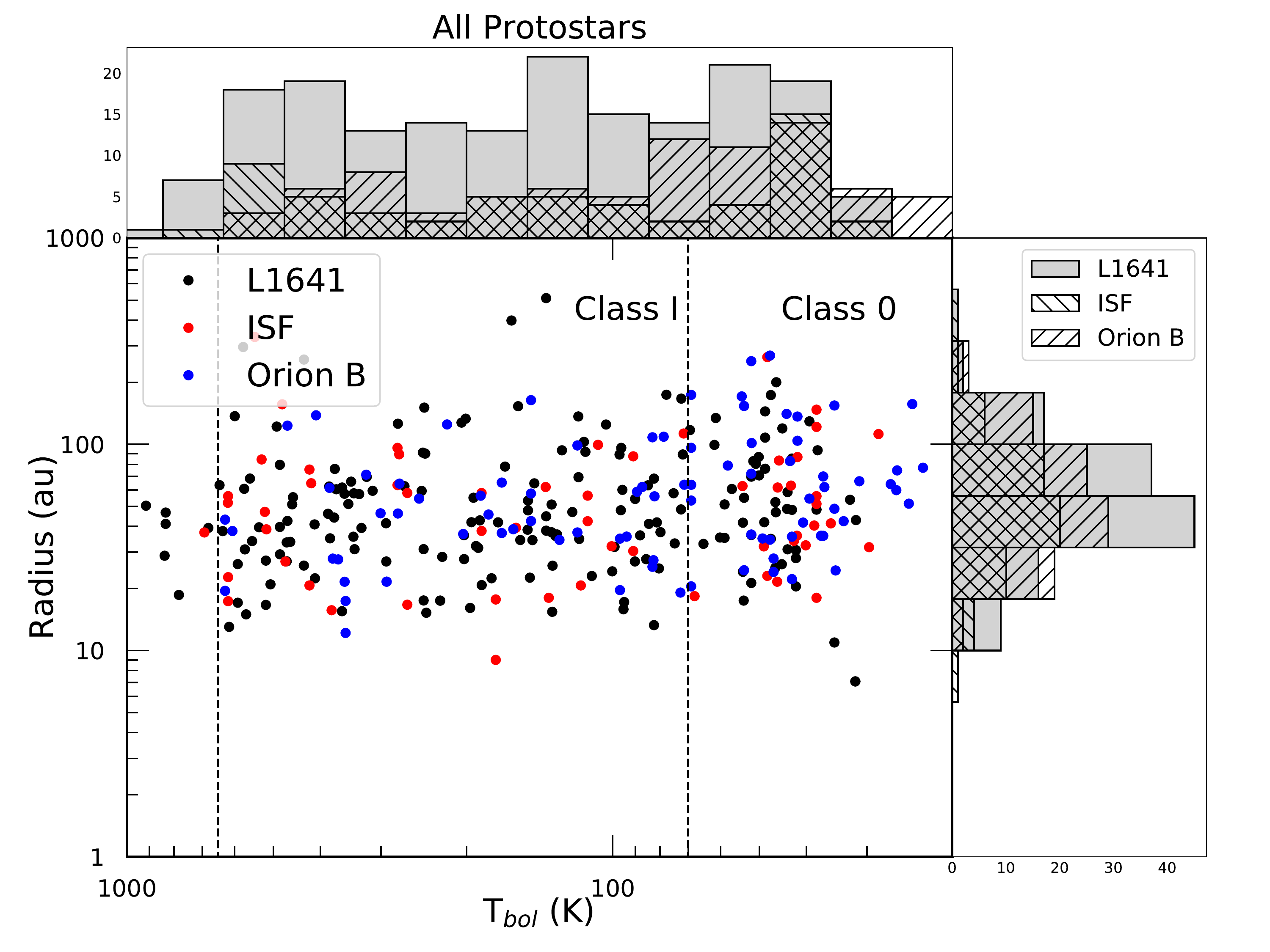}
\includegraphics[scale=0.31]{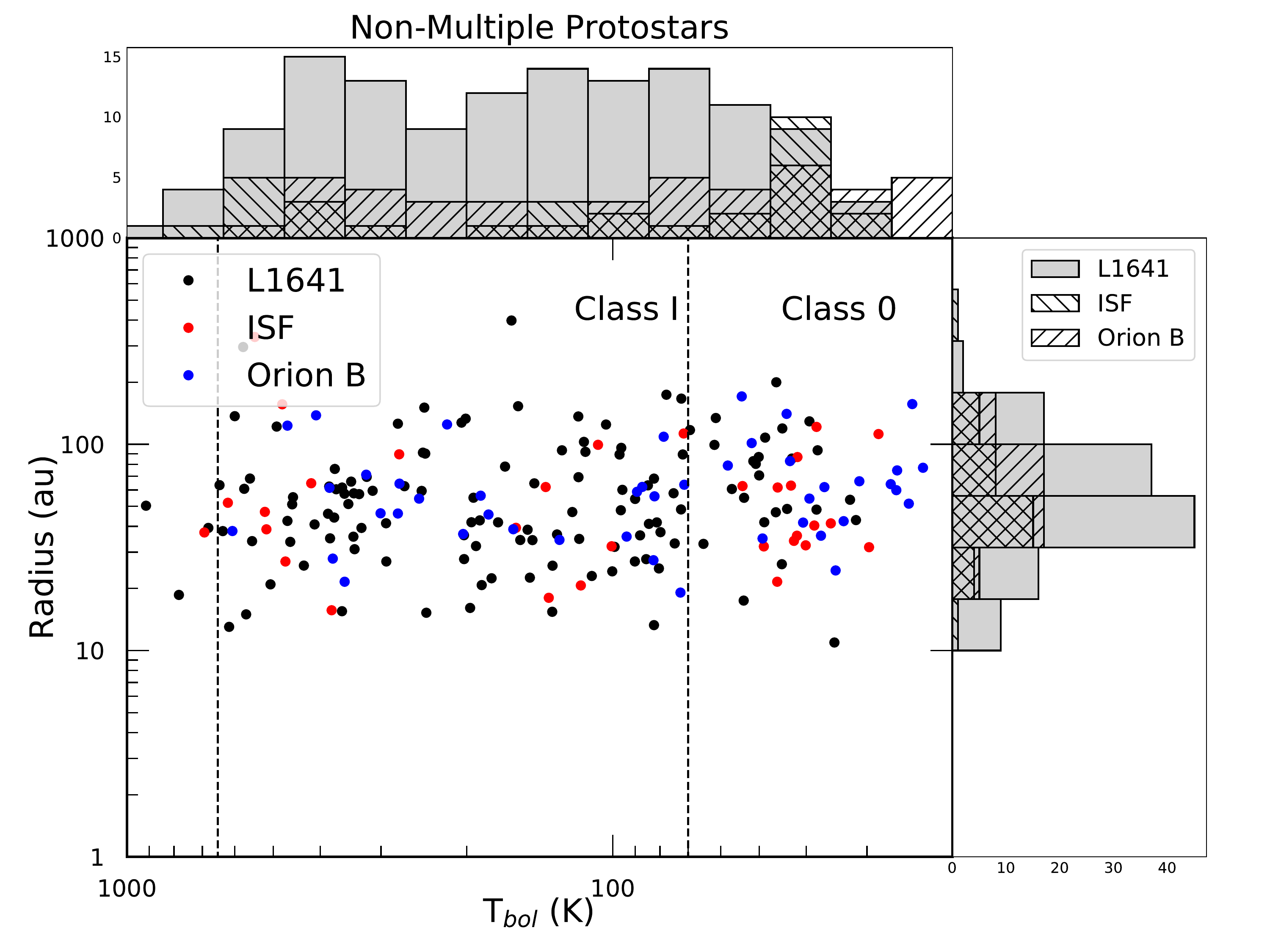}
\includegraphics[scale=0.31]{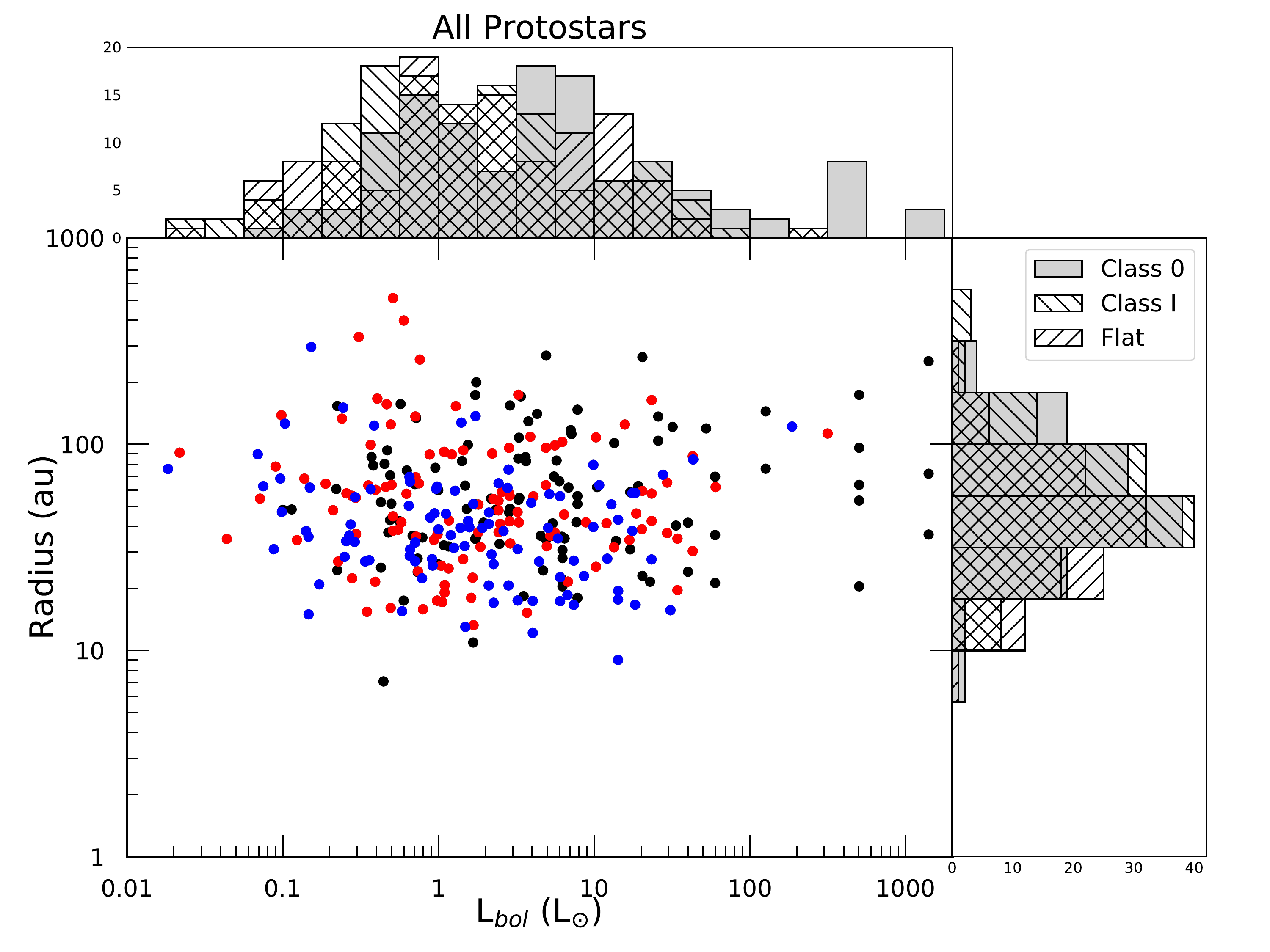}
\includegraphics[scale=0.31]{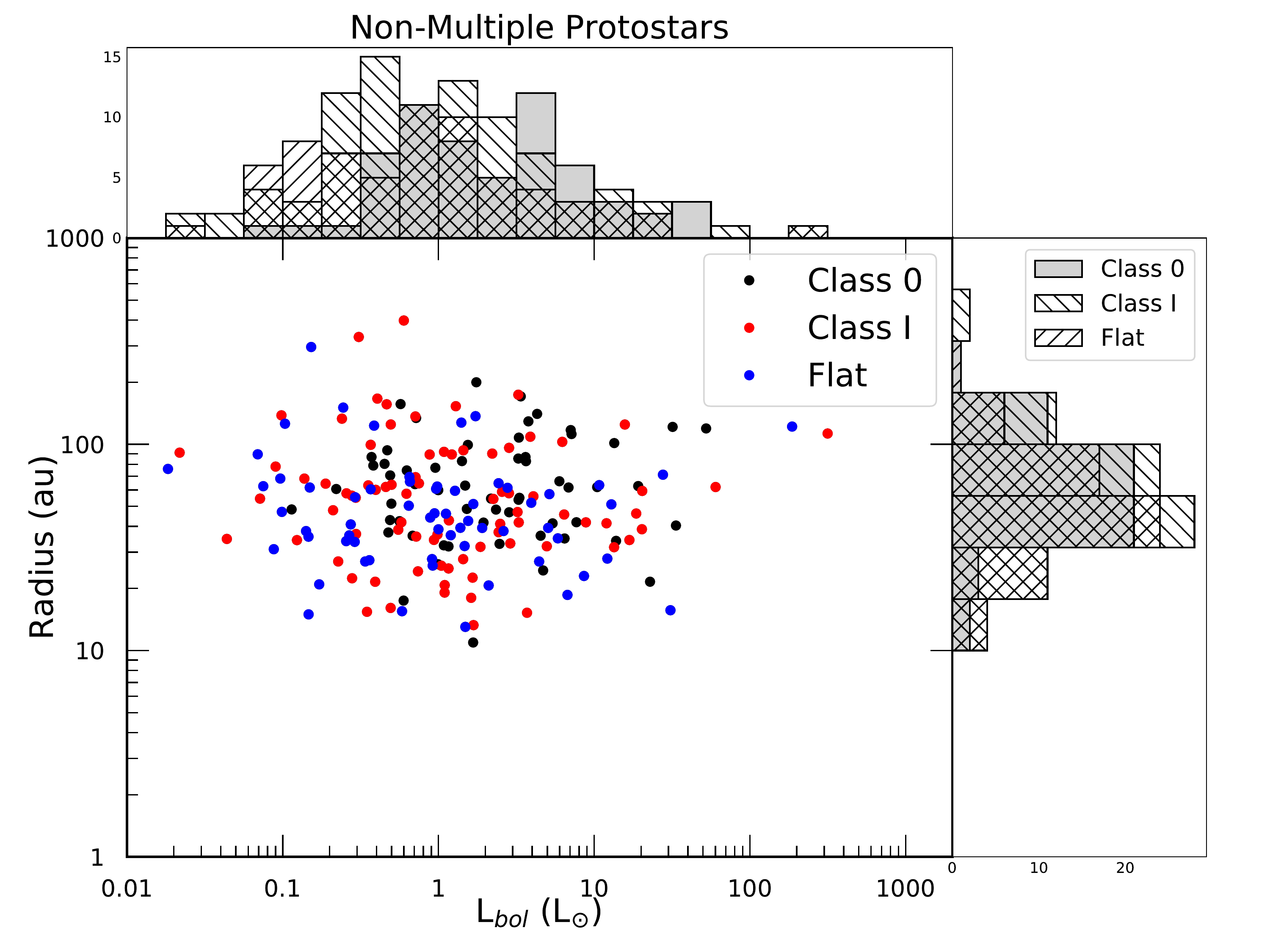}
\caption{
Comparison of protostellar disk radii to \tbol\ (top panels) and \lbol\ (bottom panels). The accompanying histograms
in the top panels separate the sample by their region and in the bottom panels
the sample is separated by class. The colorization and histograms associated with the top panels
separates the sources by their regions (black: L1641, red: integral-shaped filament [ISF], blue: Orion B).
In the bottom
panels, the colors and associated histograms separate the sources by class;
 black corresponds to Class 0 protostars, red corresponds to Class I protostars,
and blue corresponds to Flat spectrum protostars. The left panels
show the full sample with a corresponding measurement
of \lbol\ and \tbol, while the right panels show only non-multiple sample.
There is no strong correlation between disk radii and \tbol\ or \lbol. 
}
\label{radii-blt}
\end{figure}

\begin{figure}
\includegraphics[scale=0.31]{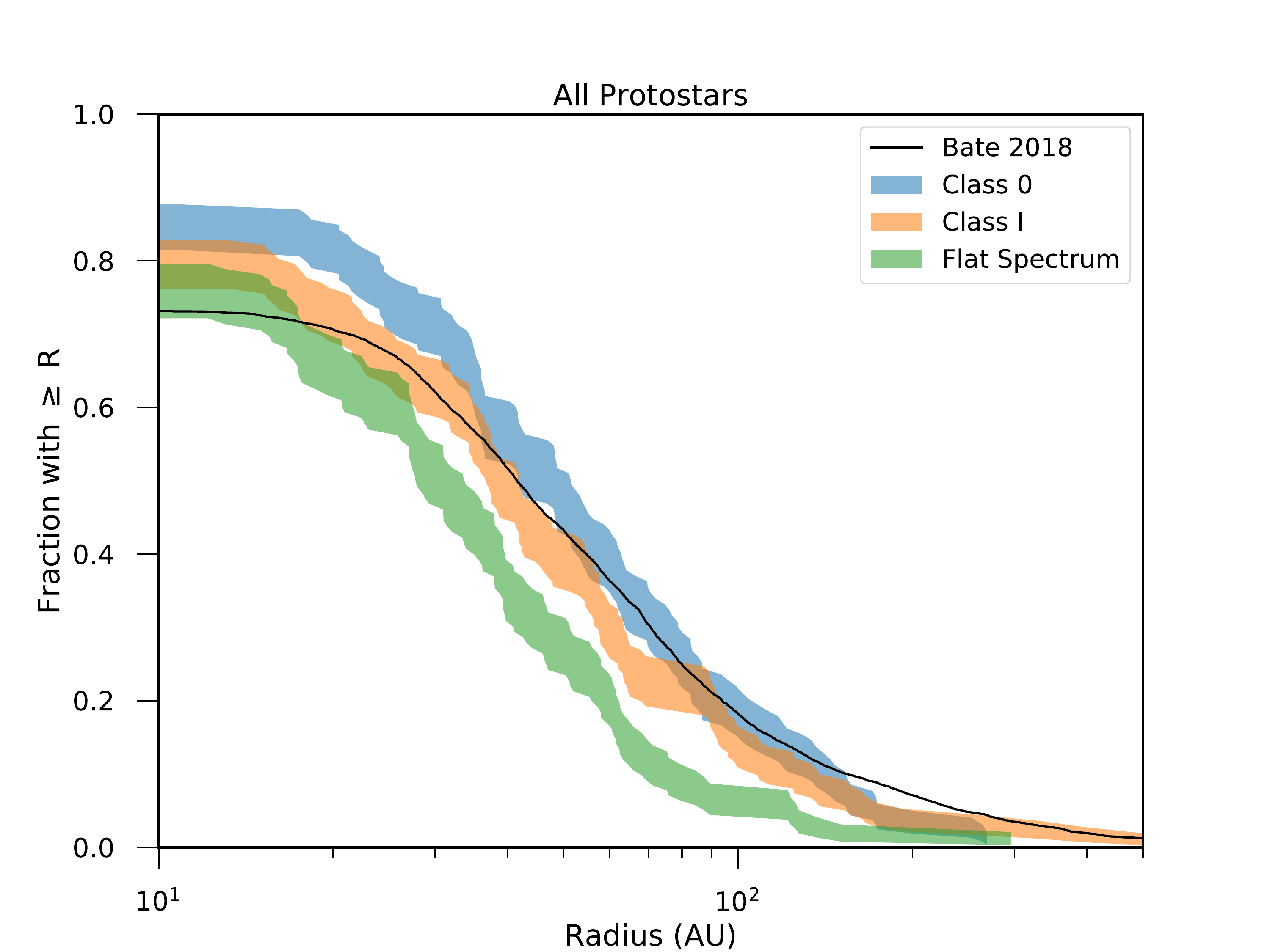}
\includegraphics[scale=0.31]{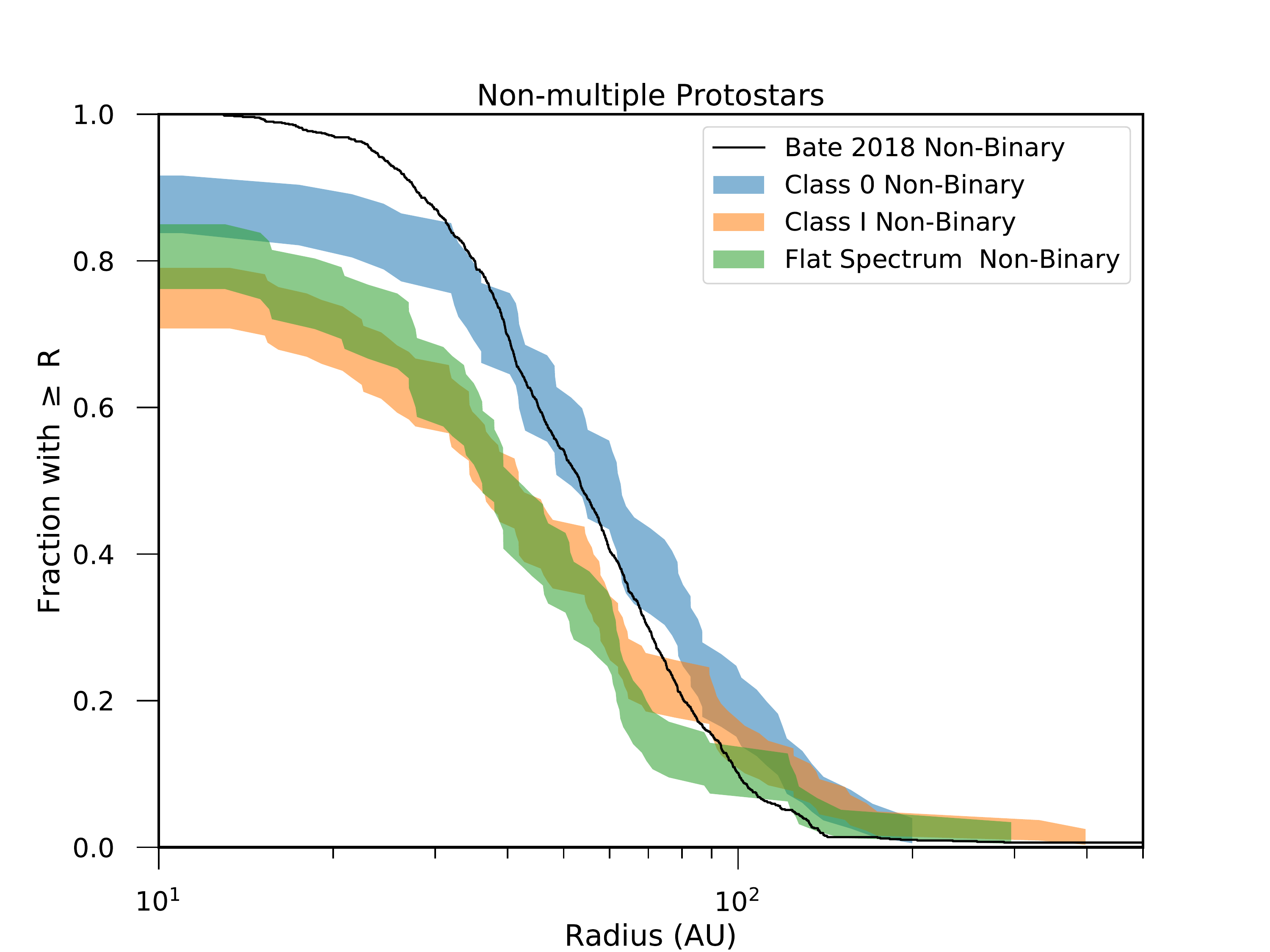}
\caption{
Cumulative distributions of disk radii within the Orion sample. The left panel shows the
full sample, while the right panel only shows the non-multiple
sample. Protostars with a measured dust disk radius that is $\le$~10~au or did not have a measurement due
to low S/N is considered an upper
limit for the purposes of the survival analysis produced by the \textit{lifelines} package. Class 0 protostars
are drawn with a blue shaded region, Class I protostars are drawn with an orange shaded region, Flat
Spectrum sources are drawn with a green shaded region. The Class 0 sources clearly have their distribution
shifted toward larger radii; this is more evident in the case of non-multiple sources.
The distribution of Class I radii is slightly higher than the Flat spectrum radii
for the full sample, but when only considering non-multiple systems, there is very little
difference in the distributions.
The full sample and non-multiple samples are also compared to the disk radii derived from the
large-scale simulations of \citet{bate2018}. The radii distributions from the simulations 
are comparable to the Class 0 and Class I disk radii for both the full sample and non-multiple cases.
}
\label{radii-cumulat}
\end{figure}

\begin{figure}
\includegraphics[scale=0.31]{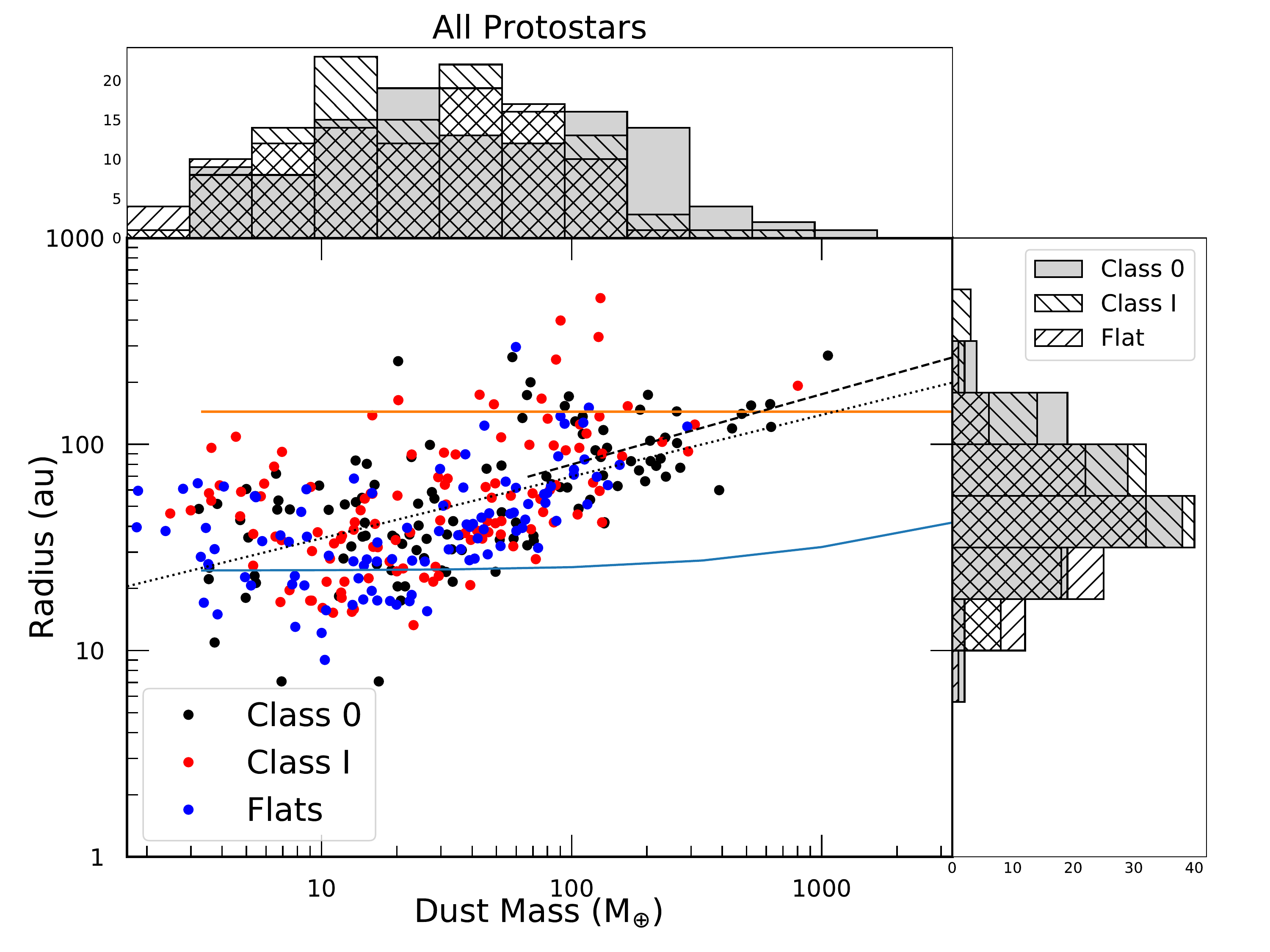}
\includegraphics[scale=0.31]{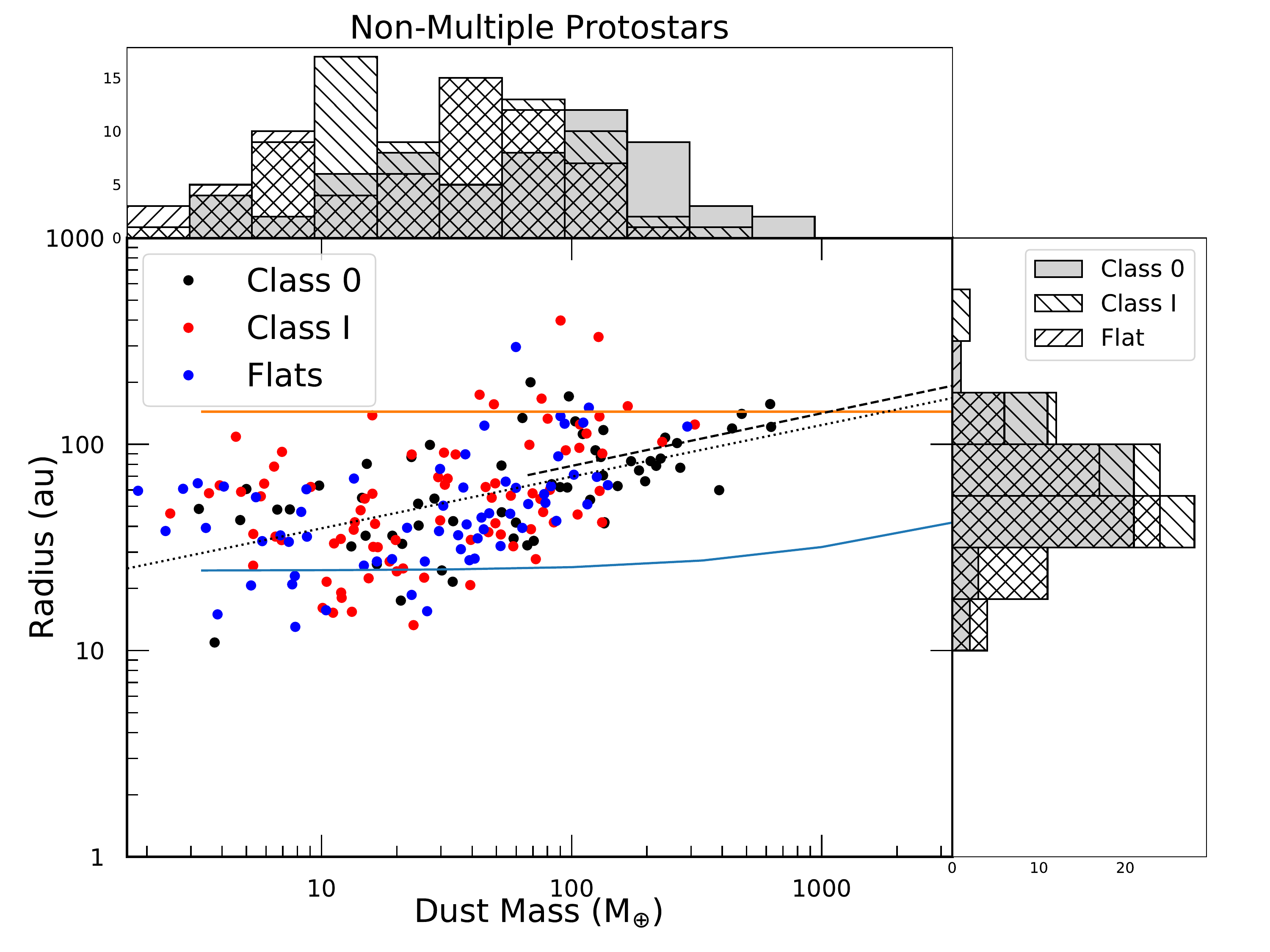}
\caption{
Comparison of disk radii versus dust disk mass for the full sample (left) 
and the non-multiple sample (right). The Class 0 disks are represented
by black points, the Class I disks by red points, and Flat spectrum disks
by blue points. There is a clear correlation in that
the sources with larger disk radii have higher masses.
Smaller disks could have higher masses that cannot be measured due to high optical depth. 
Upper limits are not shown because an upper limit on disk radius is unphysical if there
is not a detection.
The solid lines are the analytic prescriptions for disk radii vs. dust disk mass from \citet{hennebelle2016}.
The lower line is for the case of disks formed in the presence of magnetic fields, while the upper line
is for the case of a disk formed in the hydrodynamic limit; see section 4.6.2.
The dotted and dashed lines in the right and left panel are fits to the correlation between radius and mass;
the dotted lines only consider masses $>$33~\mearth. The relationships are all
comparable to R~$\propto$~M$_{disk}^{0.3}$,
see Section 3.5 and 4.6.2 for further detail.
}
\label{radii-mass}
\end{figure}

\begin{figure}
\includegraphics[scale=0.31]{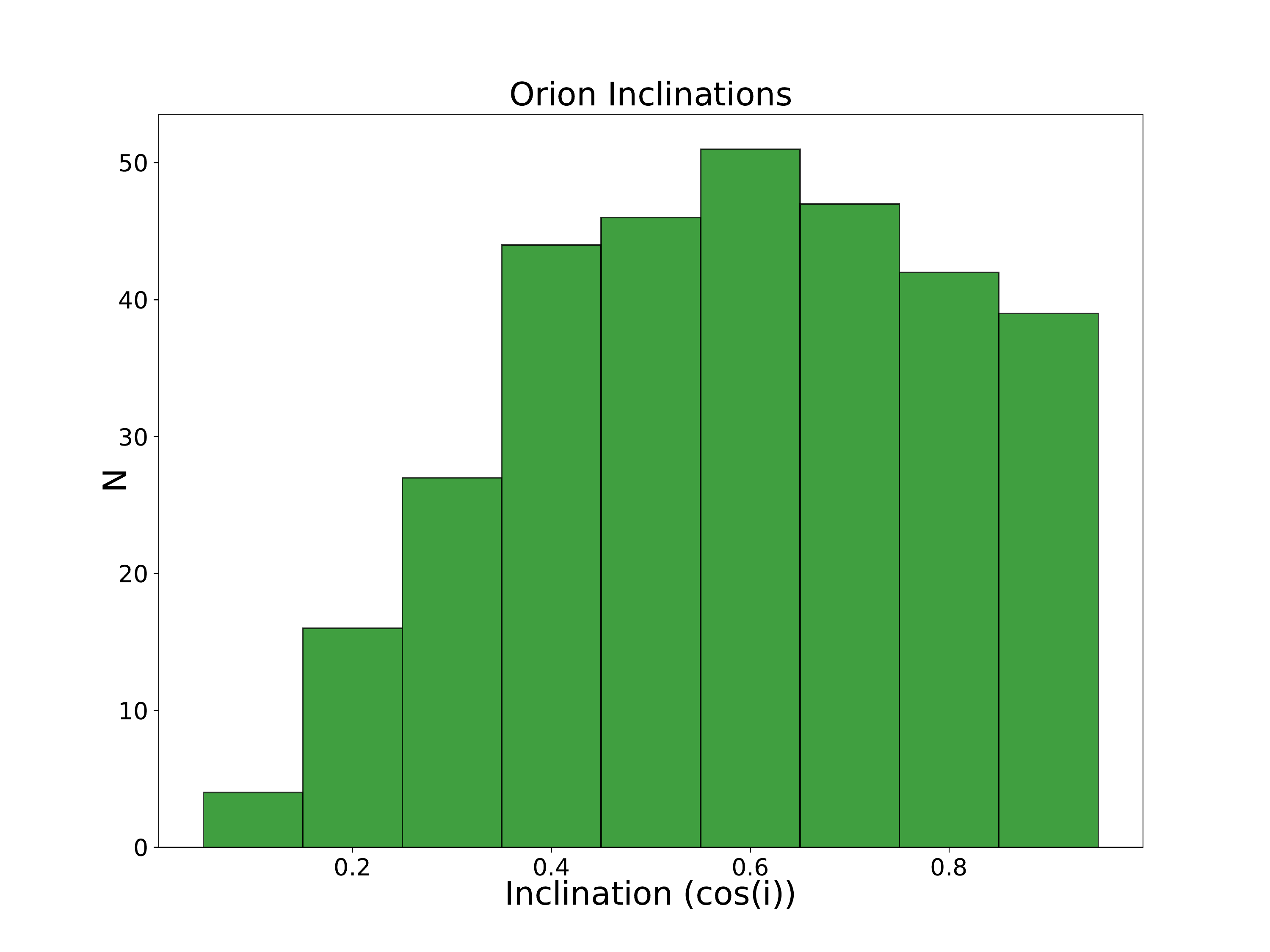}
\includegraphics[scale=0.31]{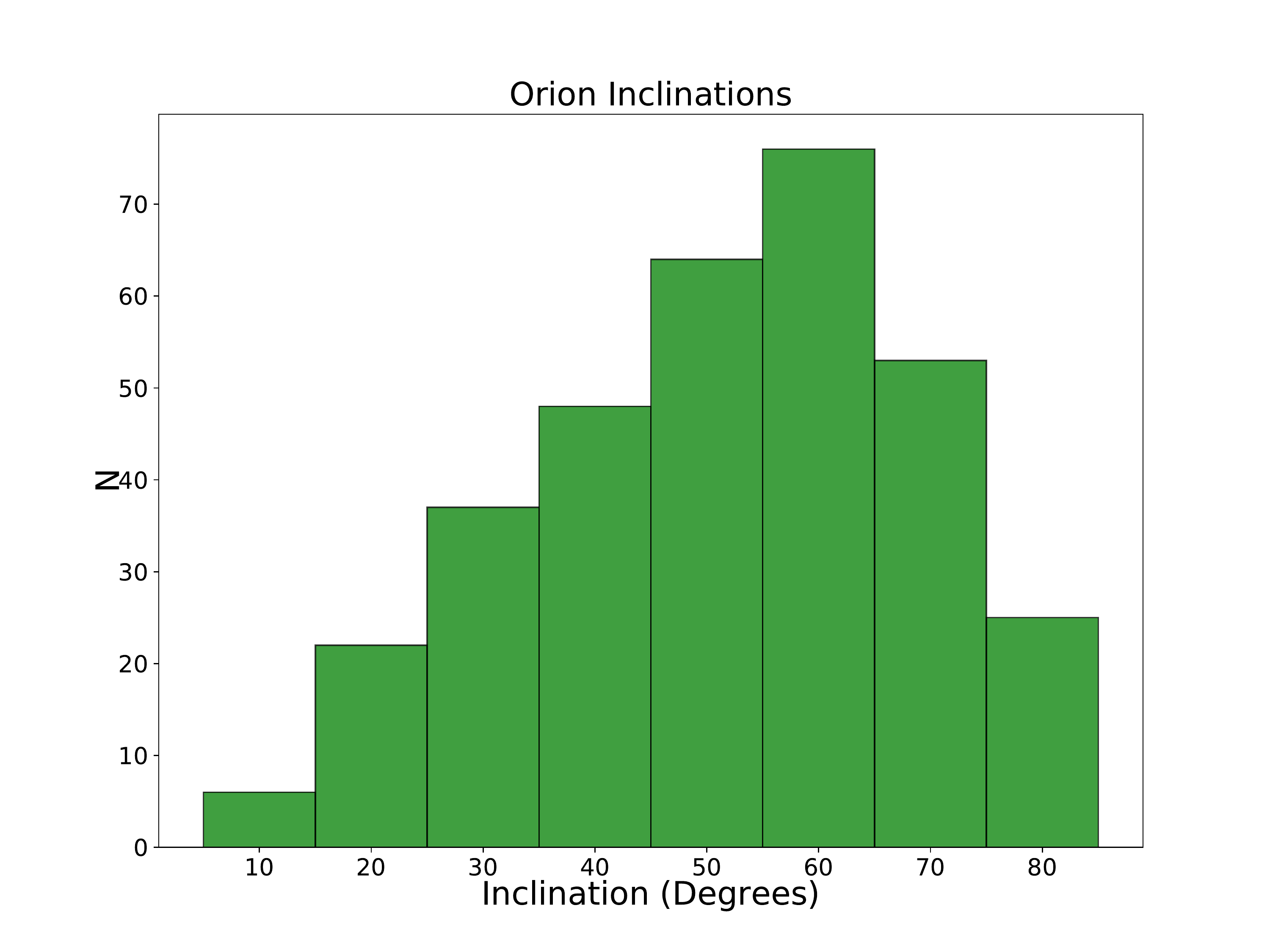}
\caption{
Histogram of inclinations derived from the deconvolved major and minor axes 
from Gaussian fitting in cos($i$) (left panel) and degrees (right panel). 
A completely random distribution would be flat in cos($i$) but peaked at 
60\degr. We see that the distribution in cos($i$) starts to drop 
at values lower than 0.4, which is likely because the deconvolved minor axes 
fit for sources close to edge-on ($\sim$90\degr; cos($i$)= 0) are generally overestimated due to the 
resolution of the survey.
}
\label{inclinations}
\end{figure}

\begin{figure}
\includegraphics[scale=0.31]{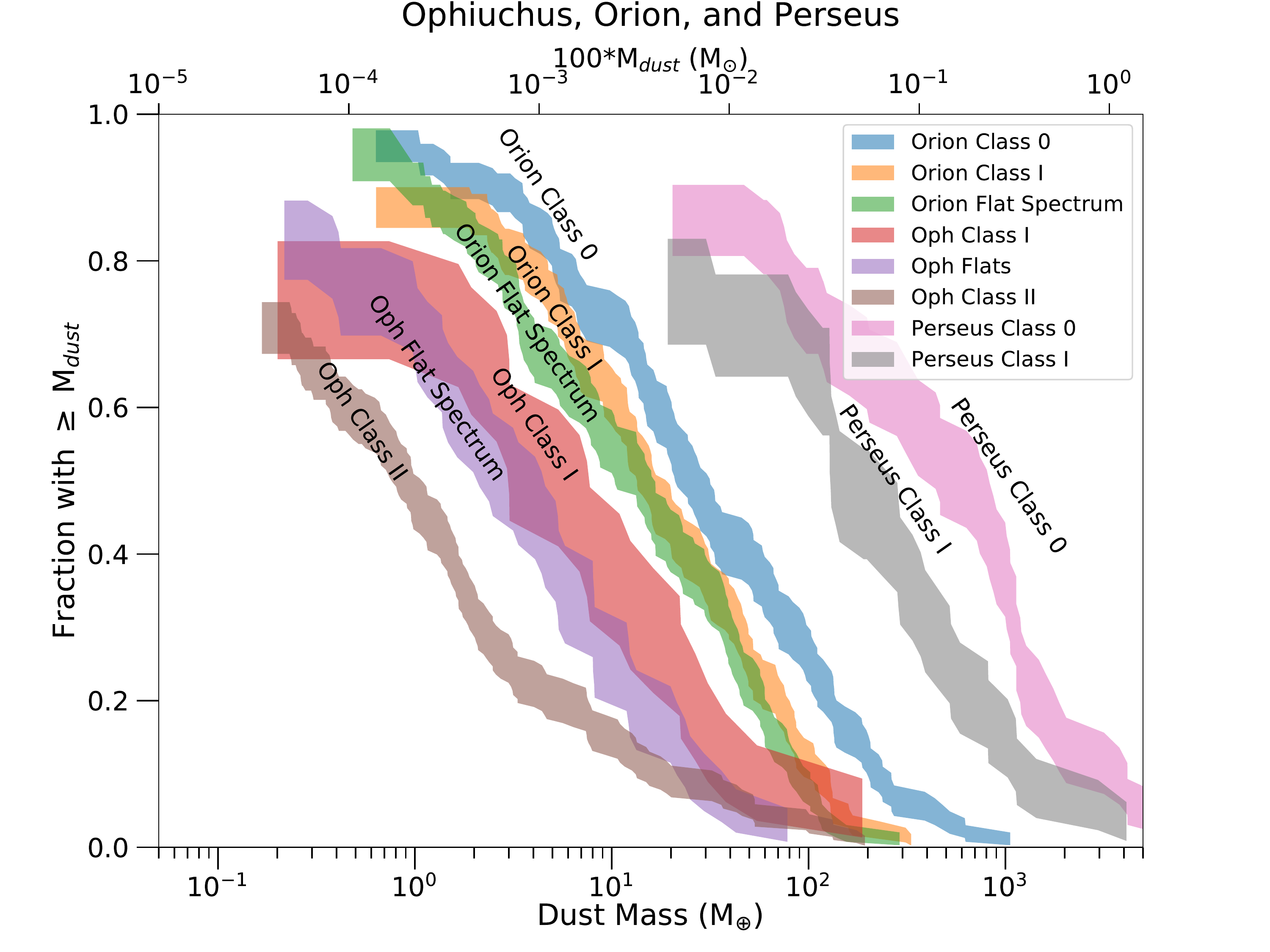}
\includegraphics[scale=0.31]{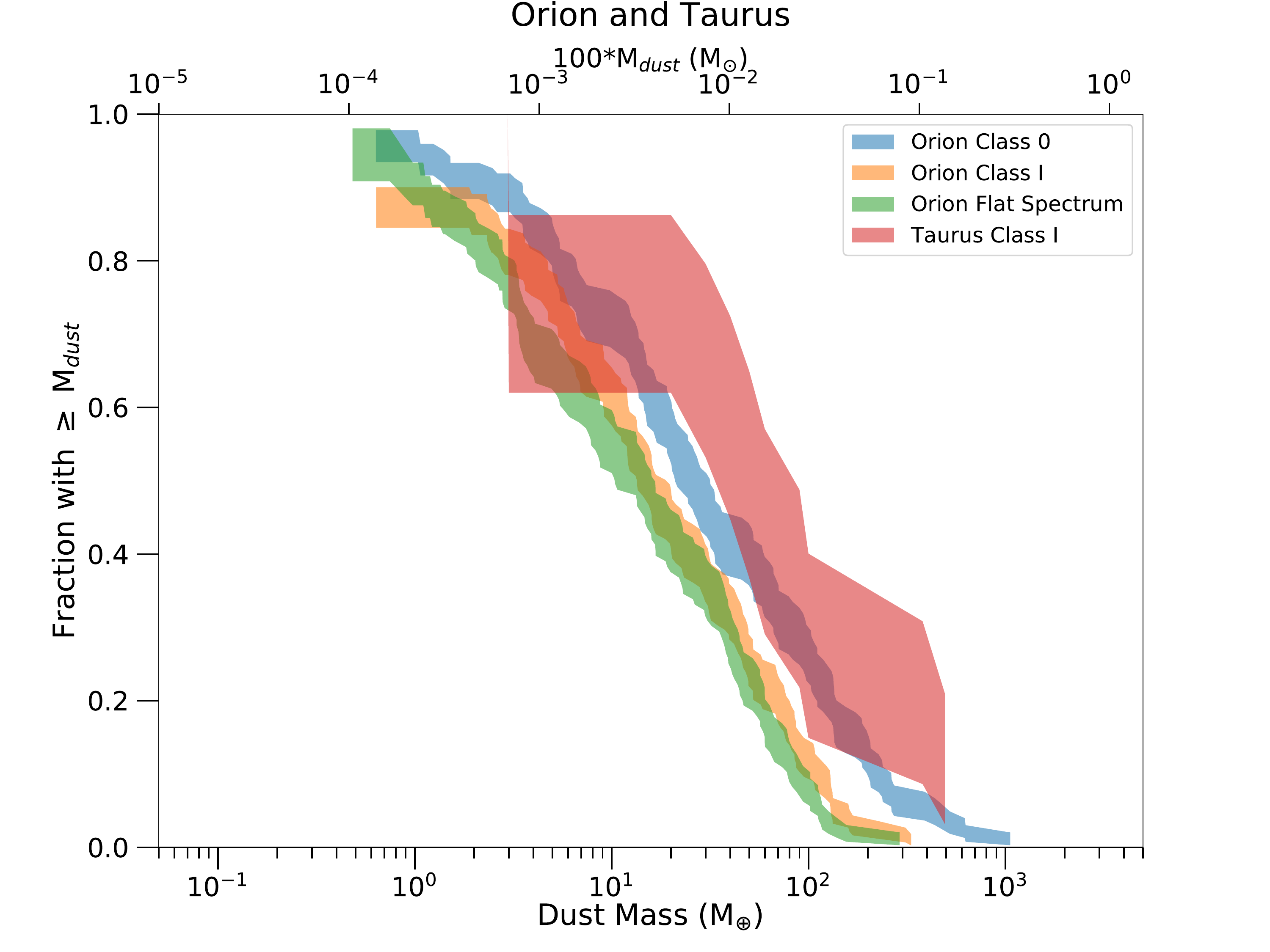}
\includegraphics[scale=0.31]{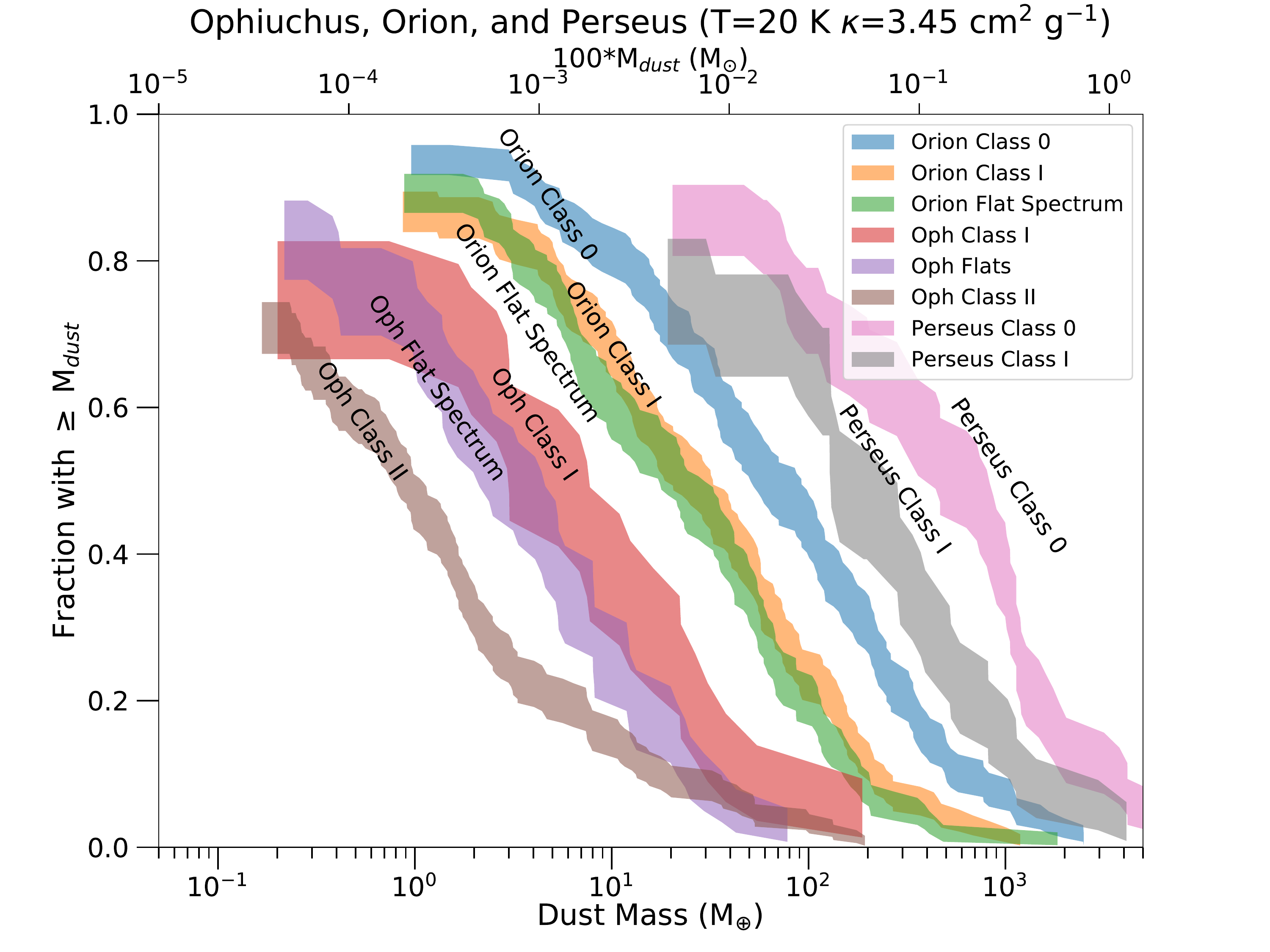}
\includegraphics[scale=0.31]{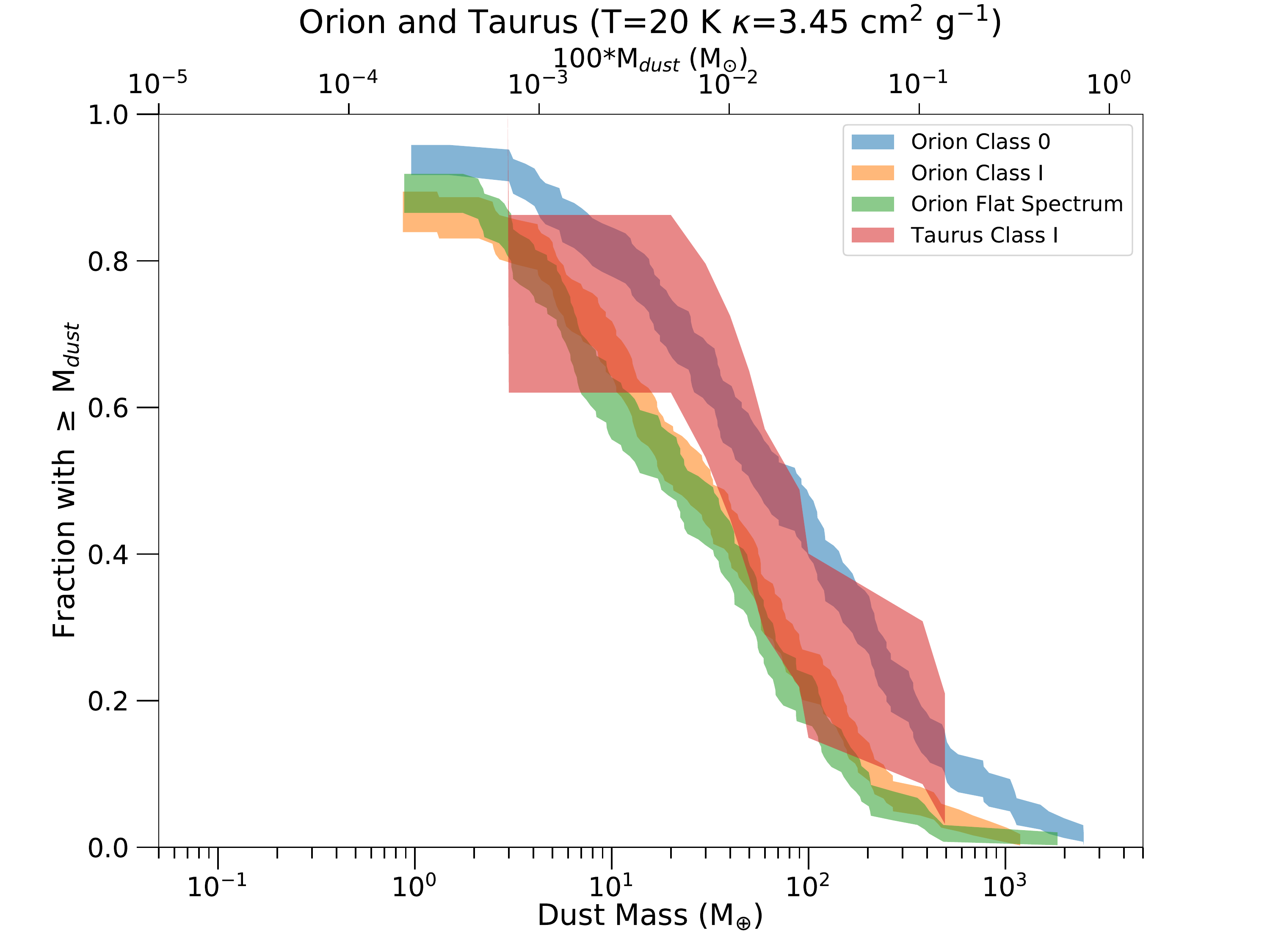}
\caption{
Cumulative distributions of dust disk masses within the Orion sample relative to other
protostellar disk surveys; Perseus and Ophiuchus in the left panel and Taurus in the right panel. 
The top panels use the dust opacity and temperature
scaling with luminosity defined in Section 2.4, while the bottom panels use the same dust opacity
law as the Ophiuchus/Class II disk samples and a temperature of 20~K. 
This distribution is for all Orion protostars regardless
of multiplicity because the other samples do not exclude multiples. Perseus appears
to have higher masses than Orion, but this is likely due to an underestimate of the dust opacity at
9~mm, leading to an overestimate of the masses. The Class I and Flat Spectrum
sources from Ophiuchus are significantly lower in mass with respect to Orion.
 The Class I protostars from Taurus (right panel) are in 
reasonable agreement with Orion, despite the different methods 
and small sample. 
}
\label{masses-ori-protos}
\end{figure}

\begin{figure}
\includegraphics[scale=0.31]{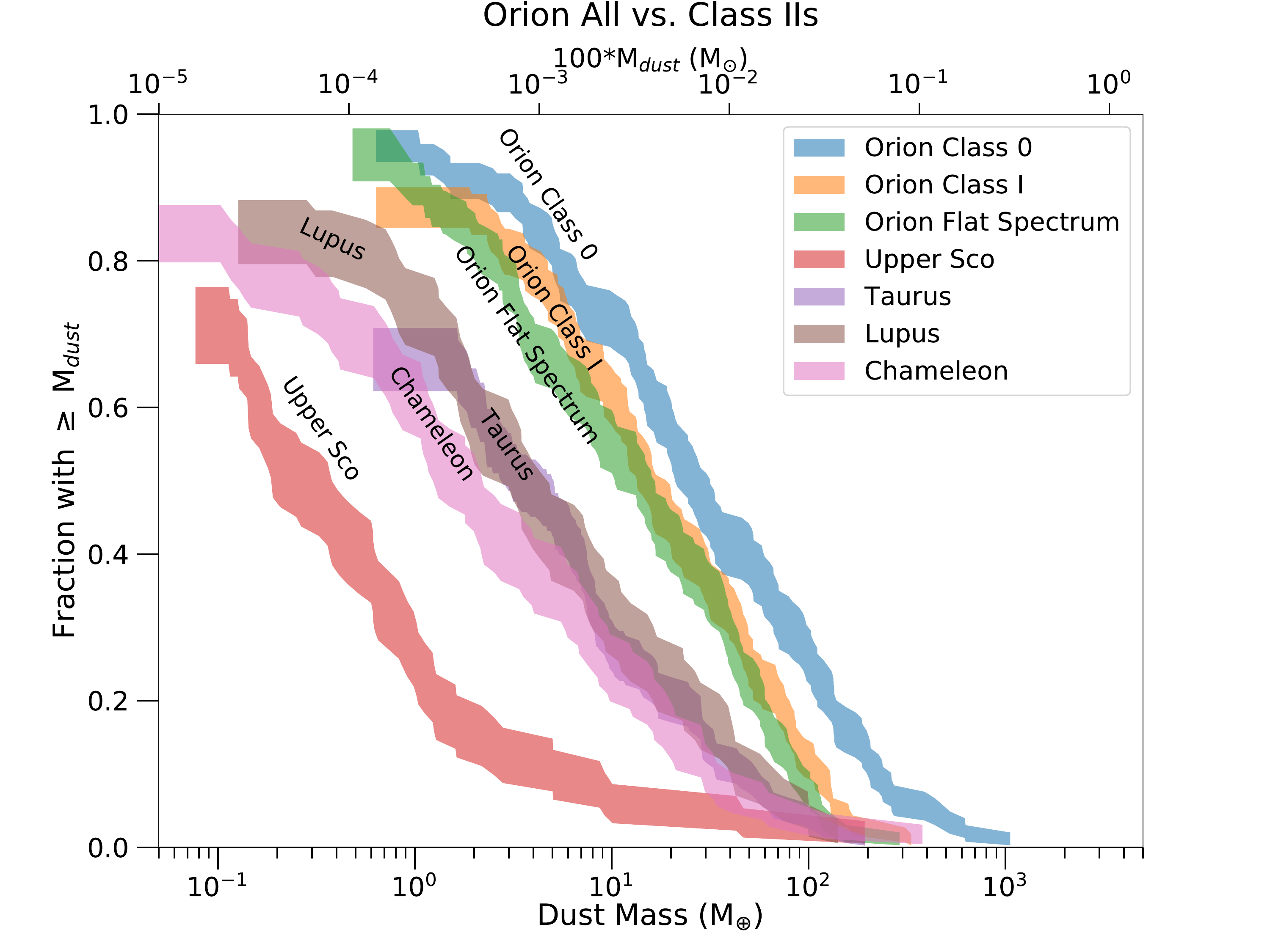}
\includegraphics[scale=0.31]{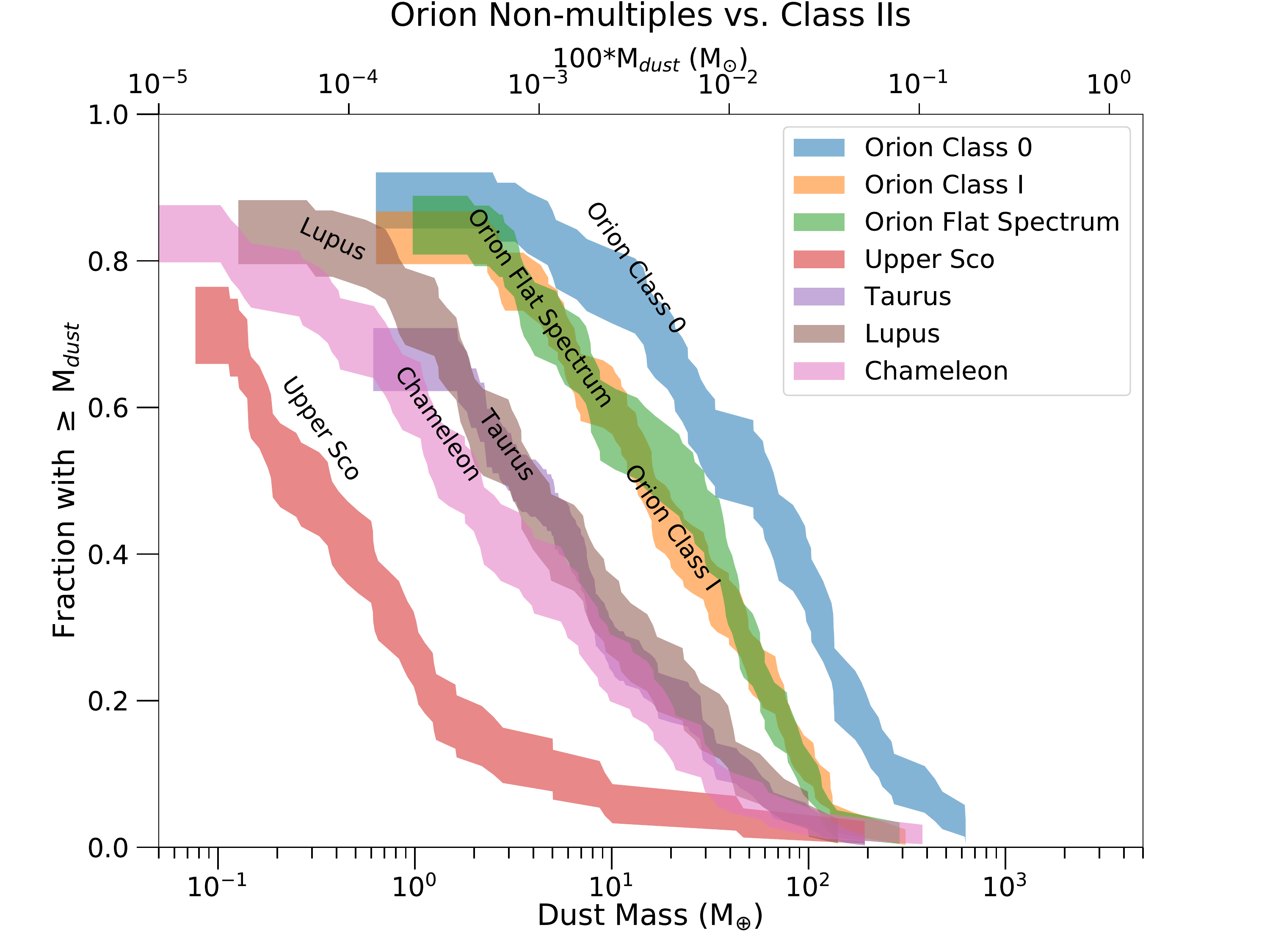}
\includegraphics[scale=0.31]{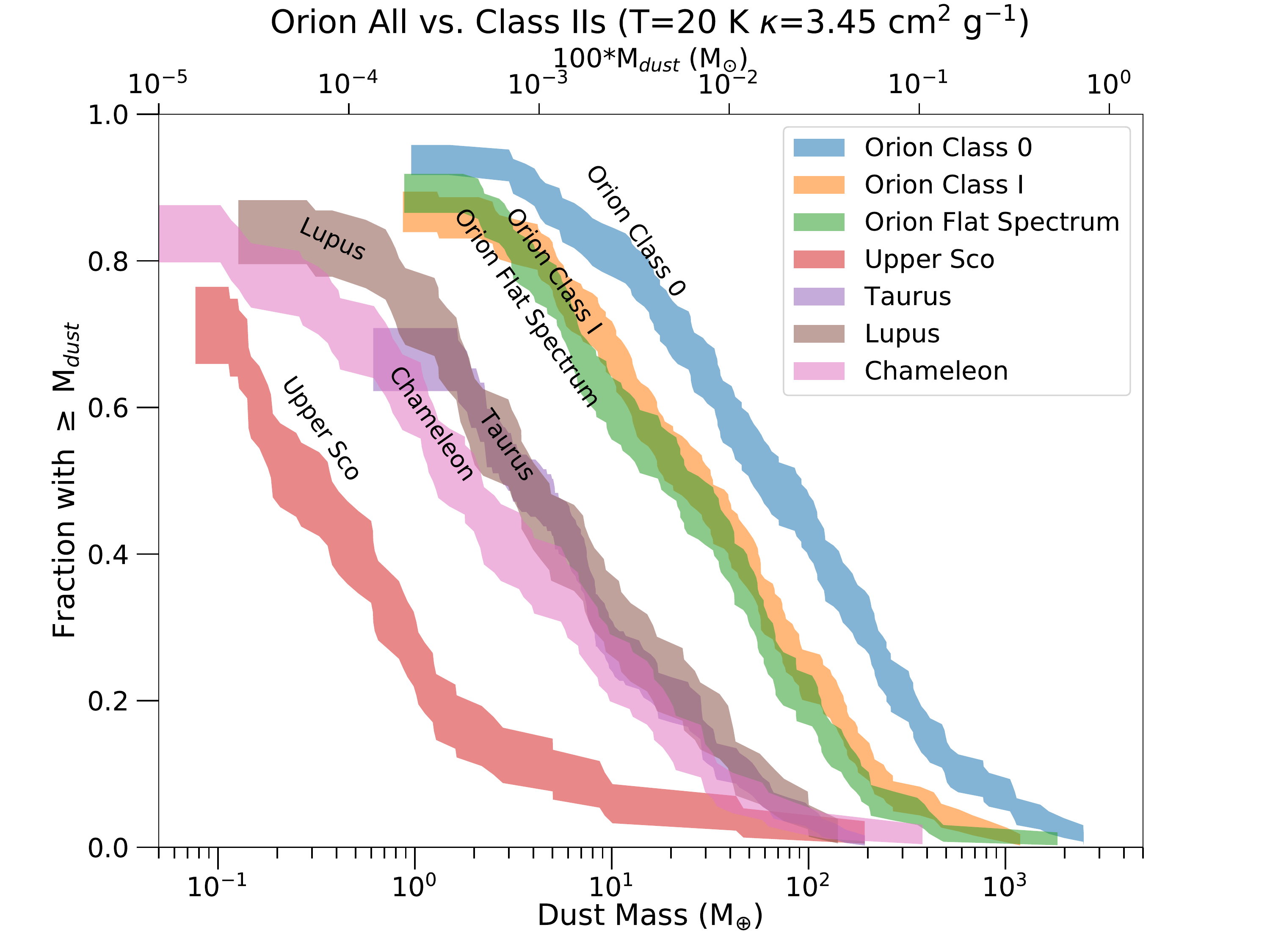}
\includegraphics[scale=0.31]{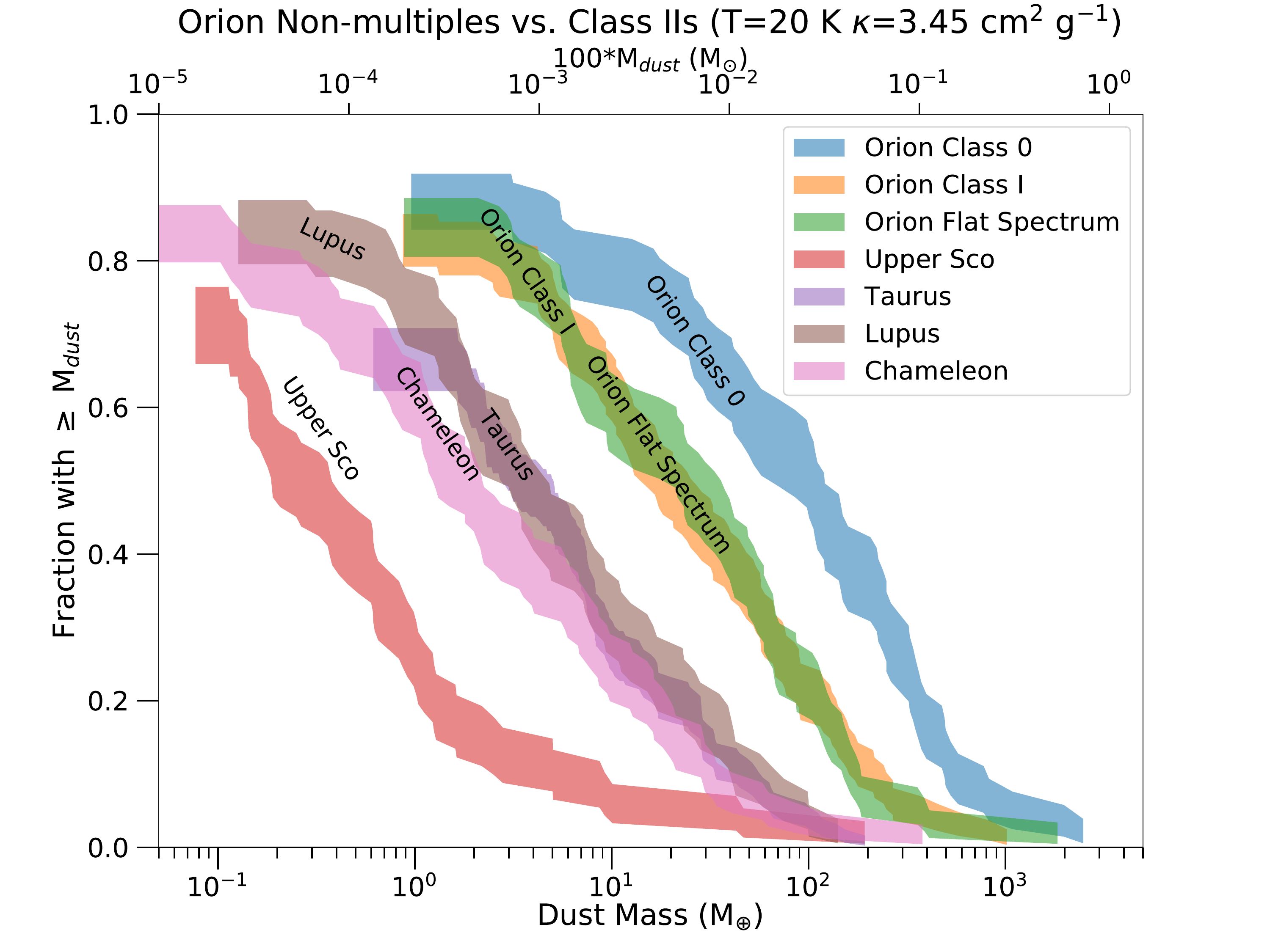}
\caption{
Cumulative distributions of dust disk masses within the Orion sample relative to
Class II disk surveys. The top panels use the dust opacity and temperature
scaling with luminosity defined in Section 2.4, while the bottom panels use the same dust opacity
law as the Class II disk samples and a temperature of 20~K. The left panels show
the full sample, while the right panels only show the non-multiple
sample. The protostellar sources in Orion all have significantly higher dust disk masses than 
the more-evolved Class II disks from multiple star-forming regions. This suggests
that the protostellar disks may be where planet formation begins, given the significantly
larger reservoir of dusty material.
}
\label{masses-ori-ppds}
\end{figure}

\begin{figure}
\includegraphics[scale=0.31]{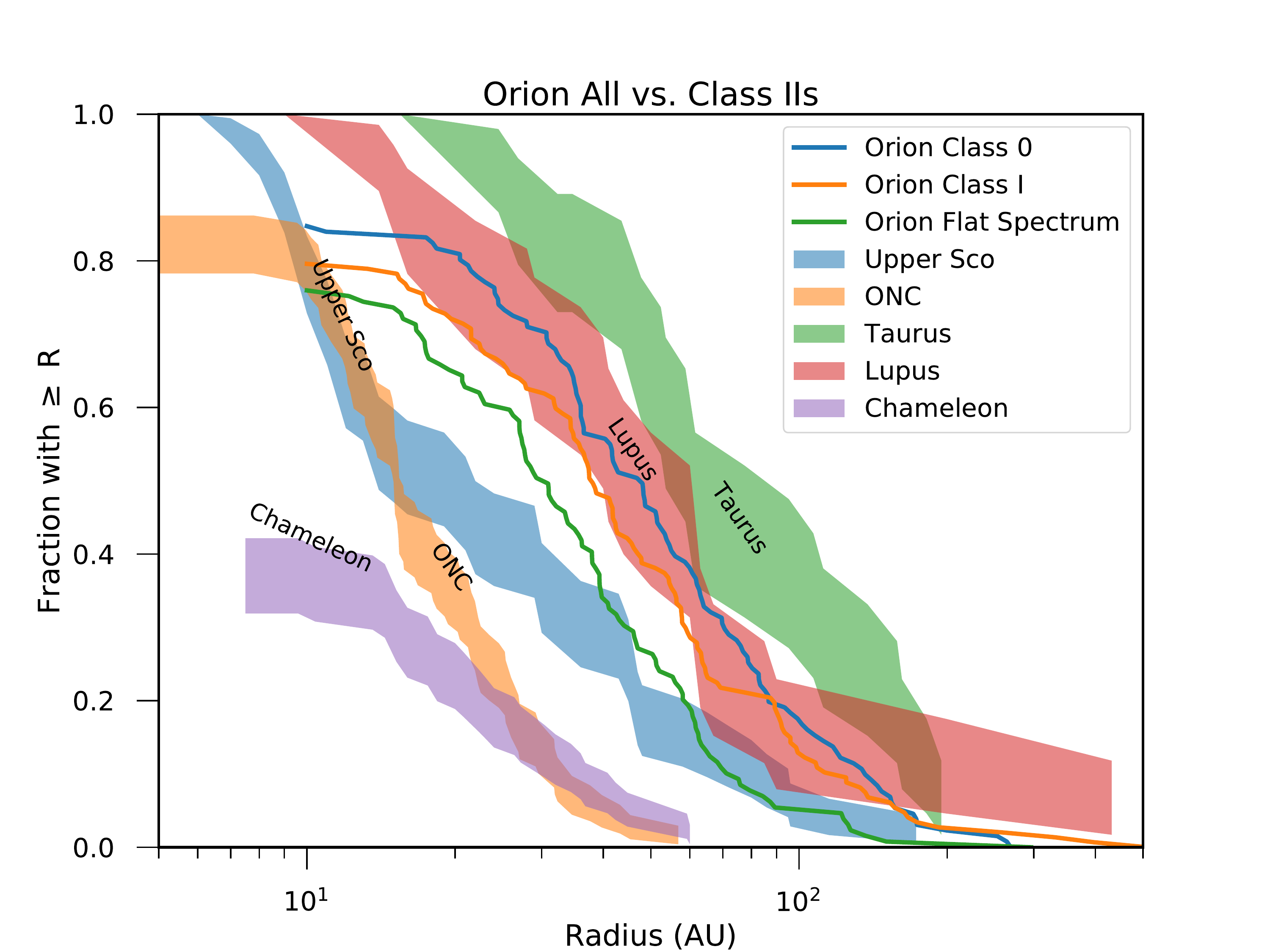}
\includegraphics[scale=0.31]{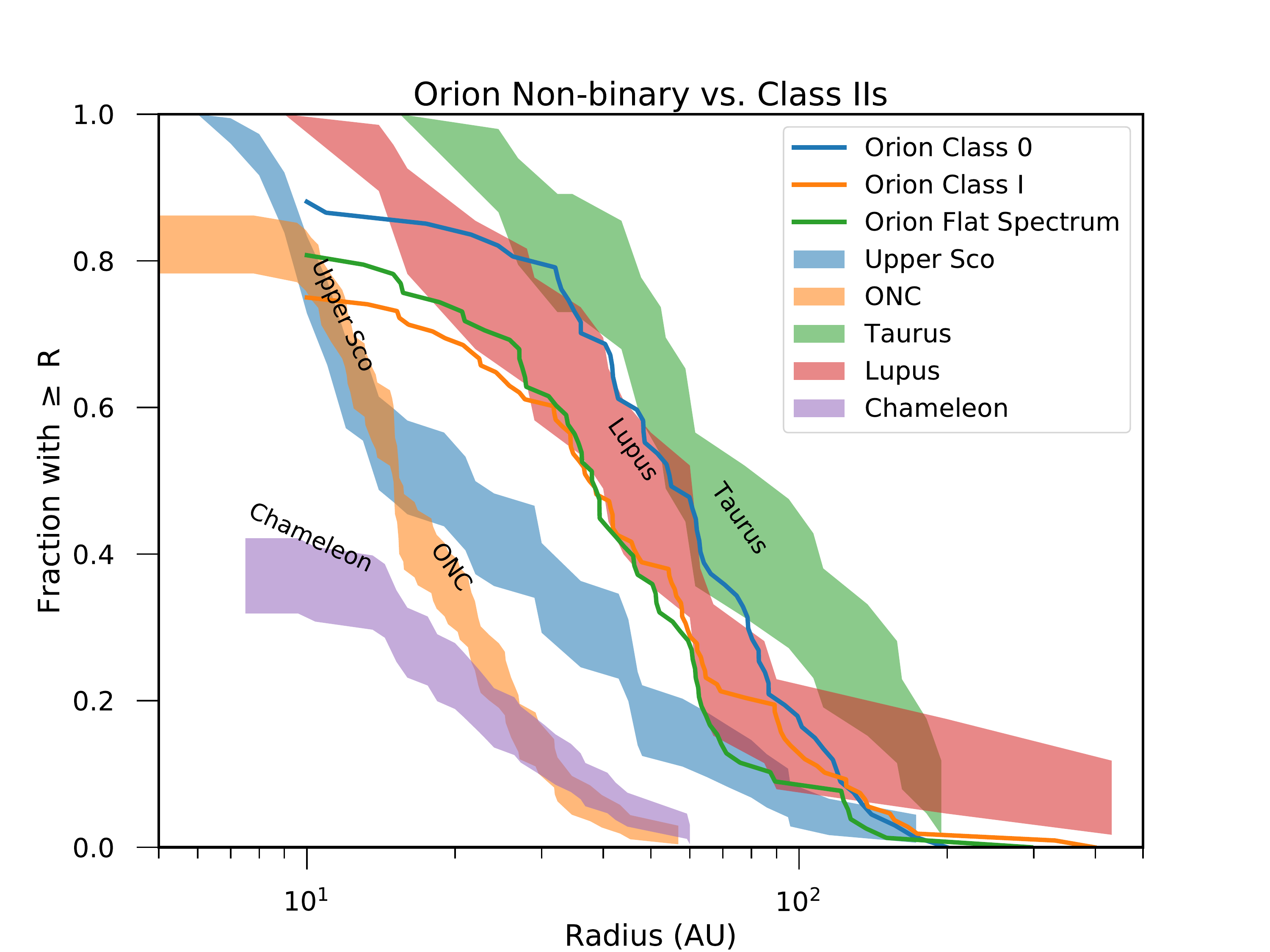}
\caption{
Cumulative distributions of disk radii within the Orion sample (lines) compared to 
Class II disk samples. The left panel shows the full sample,
 while the right panel only shows the non-multiple sample. The Orion protostellar disk
radii are larger than those of Class II disks in most regions, comparable to Lupus, but smaller than Taurus.
This, in addition to the radius comparison
within Orion, suggests that there may not be a monotonoic growth of disks 
from the protostellar to Class II disk phase.
}
\label{radii-ori-ppds}
\end{figure}

\clearpage
\appendix
\section{Distance Estimates}
Most protostars do not have a measurement 
of their parallax from Gaia DR2 due to their embedded nature, frequently resulting in
$>$10 magnitudes of extinction at visual wavelengths and/or significant confusion due to scattered light.
To estimate their distances, we rely on the nearby, more-evolved young stars 
that do have reliable parallax measurements. We use the input catalogs of 
\citet{mcbride2019}
for Orion A, \citet{kounkel2018} for Orion B, and \citet{megeath2012} for L1622, selecting sources that have 
Gaia detection with $\sigma_{\pi}$~$<$~0.2~mas. We multiplied the sample 10-fold 
through sampling the normal distribution of the parallaxes. We then used a fully connected neural network 
constructed in PyTorch with 1 hidden layer and 300 neurons to perform a 2-D extrapolation and
predict the most likely parallax for the position of each protostar. To convert to distance 
from parallax, we use the conversion of d=1000/($\pi$+0.03) where $\pi$ is the parallax in milliarcseconds, 
correcting for the systematic offset from \citet{lindegren2018}. We assume a flat uncertainty in distance of 10 pc, 
based on the FWHM distribution of distances in a given population in the input sample.

\begin{figure}
\begin{center}
\includegraphics[scale=0.50]{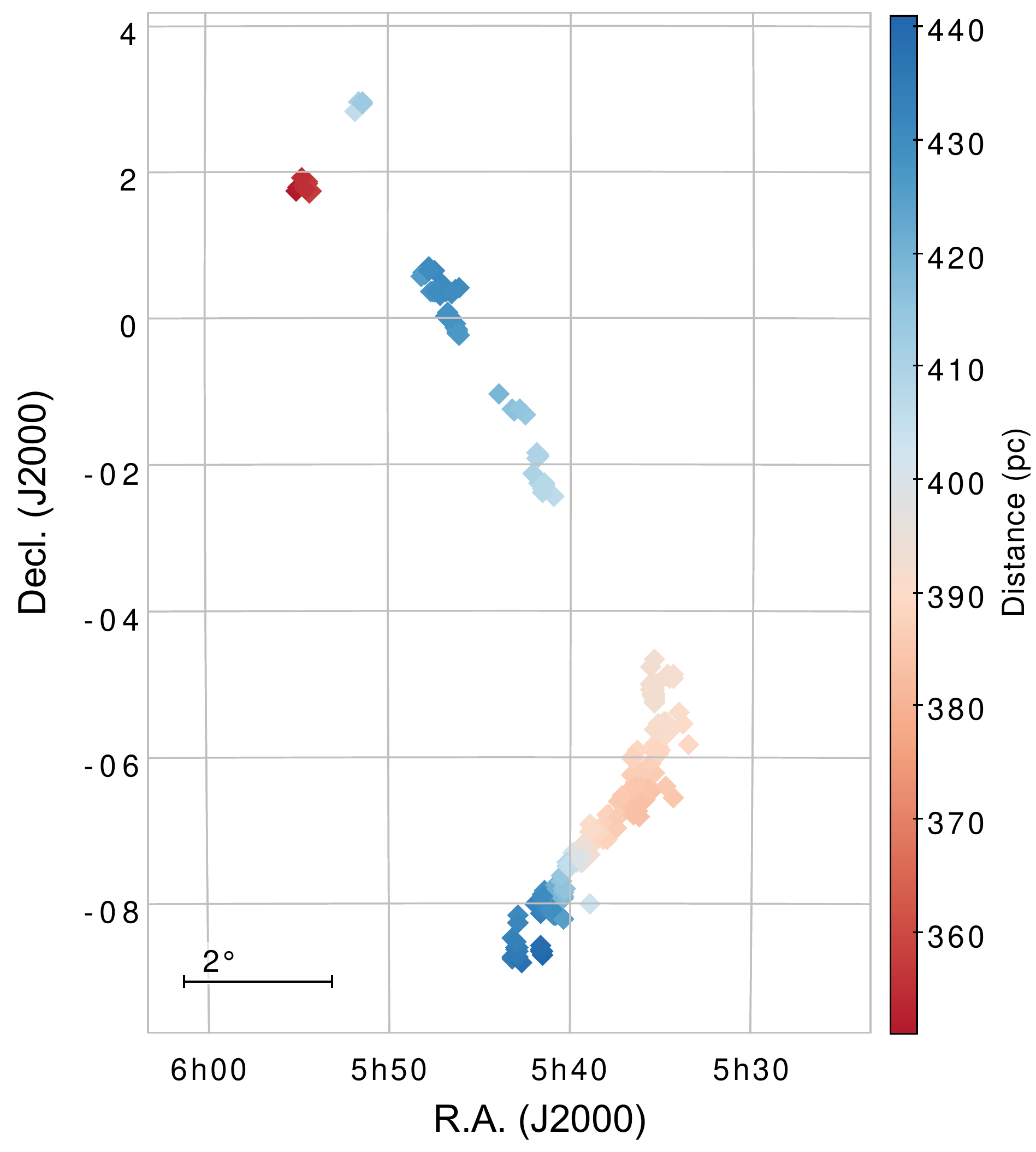}
\end{center}
\caption{
Plot of the estimated distance toward each protostar in the sample with the color scale
denoting the estimated distance. There is clearly a gradient of increasing distance north and
south of the Orion Nebula Cluster (Declination $\sim$5.5), and L1622 clearly stands out with a much closer
estimated distance than the rest of the sample.
}
\label{distances}
\end{figure}

We show the distribution of estimated distances throughout the clouds in Figure \ref{distances}.
The north-south distance gradient is obvious, but we also see that L1622 appears to be significantly 
closer than the rest of the locations within the Orion clouds. This is because L1622 
is not part of Orion~B, but rather one of the few remaining gaseous parts of Orion D. The distinction
from Orion B can also be seen in the radial velocities toward these regions where they are discontinuous
between L1622 and NGC 2068 in Orion B \citep{kounkel2018}, but agree with Orion D.
However, our sample contains only 9 protostars from this region (HOPS 1 through 7, HOPS-354, and HOPS-367), so its
existence as a separate entity from Orion B will not significantly affect our results.

\section{Average Disk Dust Temperature}

To determine the most appropriate dust temperature 
for estimating the dust disk masses from the 0.87~mm and 9~mm continuum emission, we 
used a grid of radiative transfer models to determine the average dust temperatures.
The model grid sampled the parameter space given in Table 11. The use of this model
grid enabled us to empirically derive an average dust temperature scaling based on the luminosity
of a protostellar system and the radius of the disk. We were also able to investigate the
variation of average dust temperature depending on dust disk mass. We used the Monte Carlo dust 
radiative transfer code RADMC-3D \citep{dullemond2012} 
to compute the temperature throughout the disk embedded within an envelope. The 
dust opacities in our model were parameterized by the maximum grain size (see Table 11) 
following the method outlined by \citet{woitke2016}.

Our models 
include a central protostar with an effective temperature of $T_{eff} = 4000$ K 
and a luminosity, $L_*$, that is varied in our grid. It also includes a flared 
disk with a power-law density distribution,
\begin{equation}
\Sigma(R) = \Sigma_0 \, \left( \frac{R}{1 \, \text{au}} \right)^{-\gamma},
\end{equation}
\begin{equation}
\rho(R, z) = \frac{\Sigma(R)}{\sqrt{2\pi} \, h(R)} \, \exp\left[-\frac{1}{2}\left(\frac{z}{h(R)}\right)^2\right],
\end{equation}
\begin{equation}
h(R) = h_0 \, \left( \frac{R}{1 \, \text{au}} \right)^{\beta},
\end{equation}
with $R$ and $z$ defined for cylindrical coordinates. The disk is truncated at an 
inner radius of 0.1 au and an outer radius of $R_{disk}$, which is allowed to vary in our grid. The 
total dust disk mass, $M_{disk}$, and the surface density power law exponent, $\gamma$, are also 
allowed to vary, while the scale height at 1 au, $h_0$ is fixed at 0.1 au, and the scale 
height power-law exponent, $\beta$, is fixed at 1.15. 
The overall surface density profile normalized such that it has the disk mass desired.

We also also include an envelope in our model with the density distribution for a 
rotating collapsing cloud of material from \citet{ulrich1976},
\begin{equation}
\rho(r, \mu) = \frac{\dot{M}}{4\pi}\left(G M_* r^3\right)^{-\frac{1}{2}} \left(1+\frac{\mu}{\mu_0} \right)^{-\frac{1}{2}} \left(\frac{\mu}{\mu_0}+2\mu_0^2\frac{R_c}{r}\right)^{-1},
\end{equation}
where $\mu = \cos \theta$, and $r$ and $\theta$ are defined in the typical 
sense for spherical coordinates. We truncate the envelope at an inner radius 
of 0.1 au and an outer radius of 1500 au. Moreover, the centrifugal radius, 
$R_c$, where the envelope begins to flatten, is defined to be the same as the 
disk radius. Our envelope includes an outflow cavity with a half-opening angle 
of 45$^{\circ}$ and a reduction of the density within the cavity of 50\%.
Note that the envelope density depends on the mass infall rate, while the total 
mass depends on that and the outer radius. If the density profile of the 0.001~\msun\ envelope
is extended out to 5000~au, the envelope dust mass would instead be 0.037~\msun\ 
(total mass of 3.7~\msun\ assuming a gas-to-dust ratio of 100:1). Thus, the inner envelope
density is fairly typical for Class 0 protostars in Orion \citep{furlan2016}.

We calculated the average dust temperatures directly from the density grid for each RADMC-3D 
model using the equation
\begin{equation}
T_{dust,average} = \frac{\Sigma \, T_i \, m_i \, w_i}{\Sigma \, m_i \, w_i},
\end{equation}
where T$_i$ is the temperature in each grid cell, m$_i$ is the mass in each grid cell, and $\tau_i$ is the 
optical depth to each grid cell at 0.87~mm, computed from the z direction. 
The weighting function, $w_i$ is defined as
\begin{equation}
w_i = \frac{\pi}{2} \arctan[-4\pi \, \log_{10}(\tau_i) - 2\pi] + 1.
\end{equation}
This weighting function asymptotes to constant values at both high and low values of $\tau_i$
such that regions of extremely high optical depth and extremely low optical depth are not
given disproportionate weight in the average dust temperature.

\begin{figure}
\includegraphics[scale=0.20]{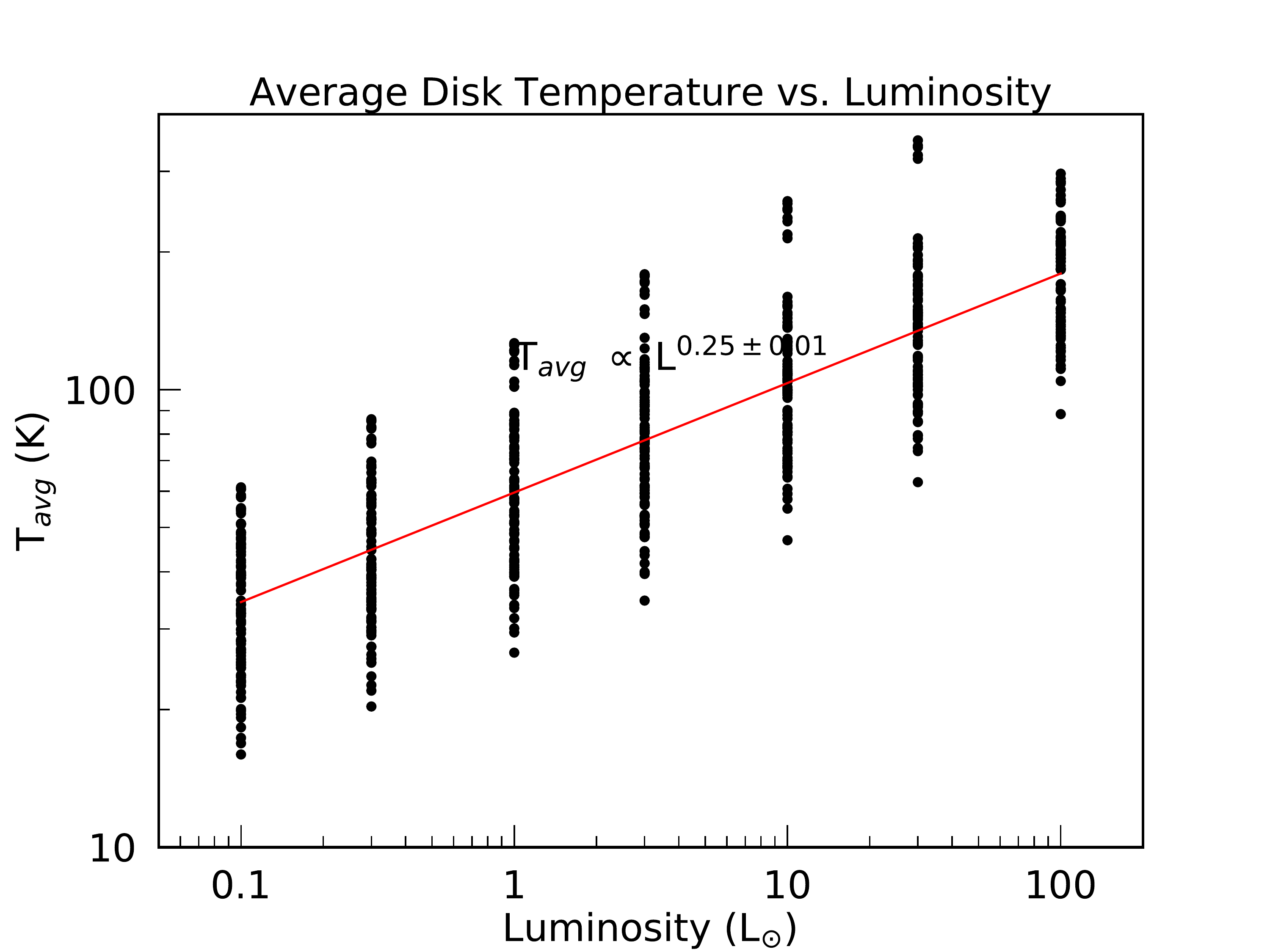}
\includegraphics[scale=0.20]{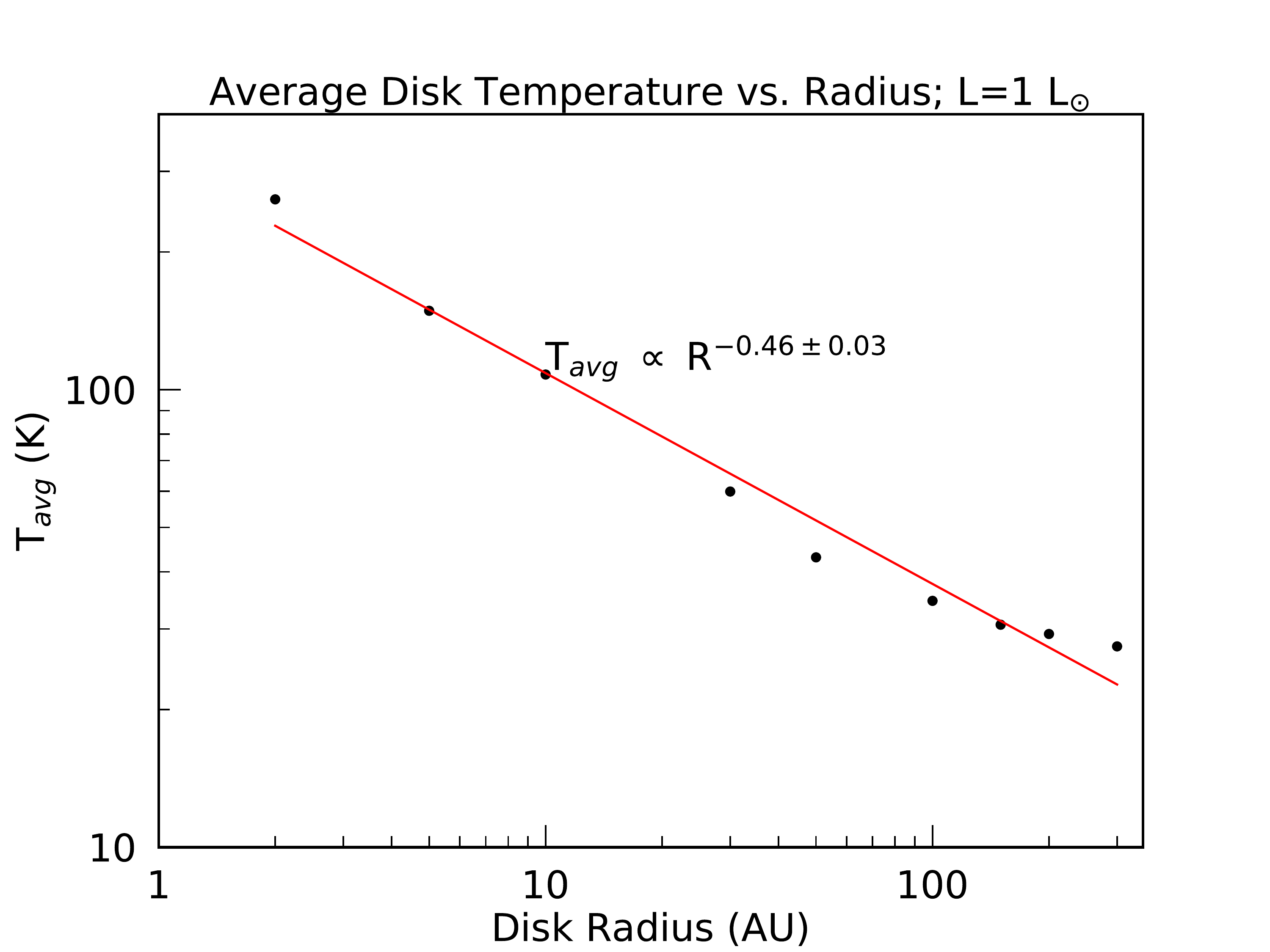}
\includegraphics[scale=0.20]{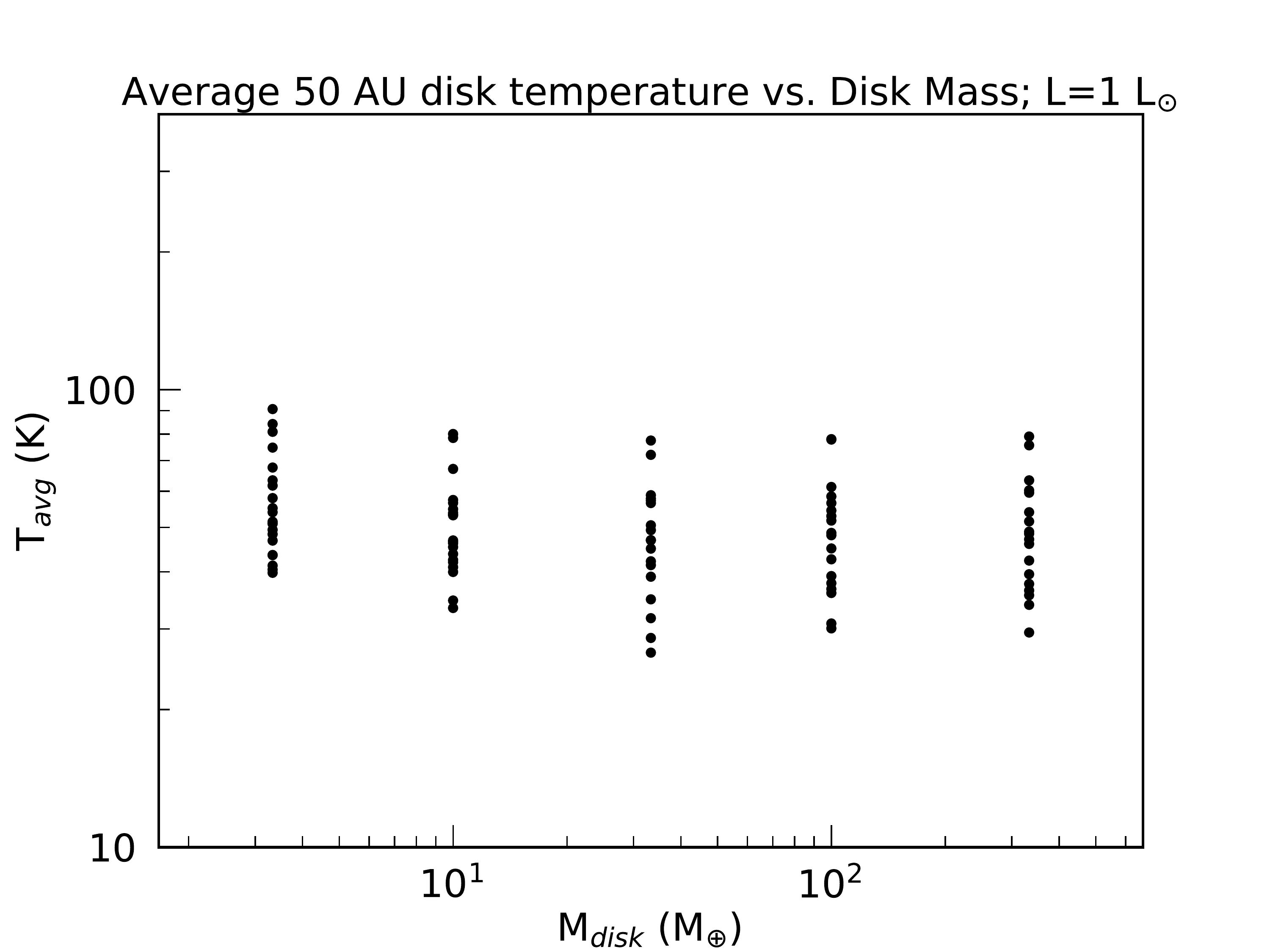}
\caption{
We show the average model disk temperatures as a function of luminosity (left), disk radius (middle), and
dust disk mass (right). These relationships are used to determine our assumption of average disk temperature and
how it should scale with disk radius and protostellar luminosity. In the left panel, we are including
all dust disk masses to show that the dependence of the average disk temperature on luminosity does not
strongly depend on dust disk mass. The middle panel only shows a single dust disk mass for clarity, but each point
is the average dust temperature for a disk with a given radius. Lastly, the right panel shows the average dust temperature
for disks with different masses, at a single luminosity of 1~\lsun. The spread at each dust disk mass is present
because we plot all the disk radii for each mass.
}
\label{model-plots}
\end{figure}

Using these models, we found that a 1~\lsun\ protostar has an 
average dust temperature of 43~K for a 50~au disk. We also confirmed that the
average disk dust temperature scales $\propto$~L$^{0.25}$ considering the full 
set of disk radii (left panel, Figure \ref{model-plots}). We further show in
the middle panel of Figure \ref{model-plots} that
the average dust temperature scales $\propto$R$^{-0.46}$, near the functional dependence 
expected from theory ($\propto$R$^{-0.5}$). It is also apparent that the average disk temperature is not
quite linear in log-log space, and the temperature profile is flattening slightly for the largest radii. 
This is due to the backheating from the surrounding envelope. We did
need to assume a dust disk mass of 33~\mearth\ to derive these relationships.
However, we show that the average temperature for a disk with a radius of 50~au does 
not strongly depend on the mass of the disk in the right panel of Figure \ref{model-plots}.

\clearpage
\begin{deluxetable}{lc}
\tablecaption{Model Grid Parameters}
\tablehead{\colhead{Parameter} & \colhead{Value}}
\startdata
$L_*$ [L$_{\odot}$] & 0.1, 0.3, 1, 3, 10, 30, 100, 300 \\
$M_{disk,dust}$ [M$_{\odot}$] & $1\times10^{-7}$, $3\times10^{-7}$, $1\times10^{-6}$, $3\times10^{-6}$,$1\times10^{-5}$, $3\times10^{-5}$, $1\times10^{-4}$, $3\times10^{-4}$, 0.001,  \\
$R_{disk}$ [au] & 2, 5, 10, 30, 50, 100, 150, 200, 300 \\
$\gamma$ & 1, 1.5 \\
$M_{env,dust}$ [M$_{\odot}$] & 0.0001, 0.001 \\
$a_{max}$ [$\mu$m] & 1, $10^3$, $10^4$ \\
$i$ [$^{\circ}$] & 0, 30, 60, 90 \\
$H.A.$ [hours] & -2.5, 0, 2.5 \\
\enddata
\tablecomments{The parameters are described as follows: $L_*$ is the total luminosity of the protostar
from the star and accretion, $M_{disk,dust}$ is the dust disk mass, $R_{disk}$ is the dust disk radius, $\gamma$ is the power-law of the surface density profile defined in Equation B1,$M_{env,dust}$ is the 
envelope dust mass, $a_{max}$ is the radius of the maximum dust grain size,
$i$ is the inclination of the system, and $H.A.$ is the hour angle at which the observations were
simulated.
}
\end{deluxetable}

\clearpage
\section{Images of all Protostars}

We provide the images for the entire sample of Orion protostars in Figures 16 through 51
as image cutouts around each protostar. The cutout images are generally 1\farcs25~$\times$~1\farcs25
except for the few sources that exceed this size (i.e., HOPS-136, HOPS-65). Figures 52 through 66 show
larger images that encompass wide multiple systems. The images for each figure are available from the Harvard
Dataverse (https://dataverse.harvard.edu) \citep{alma_images,vla_images}.

\section{Comparing Perseus Dust Disk Mass Methodologies to Orion}
\begin{figure}
\includegraphics[scale=0.65]{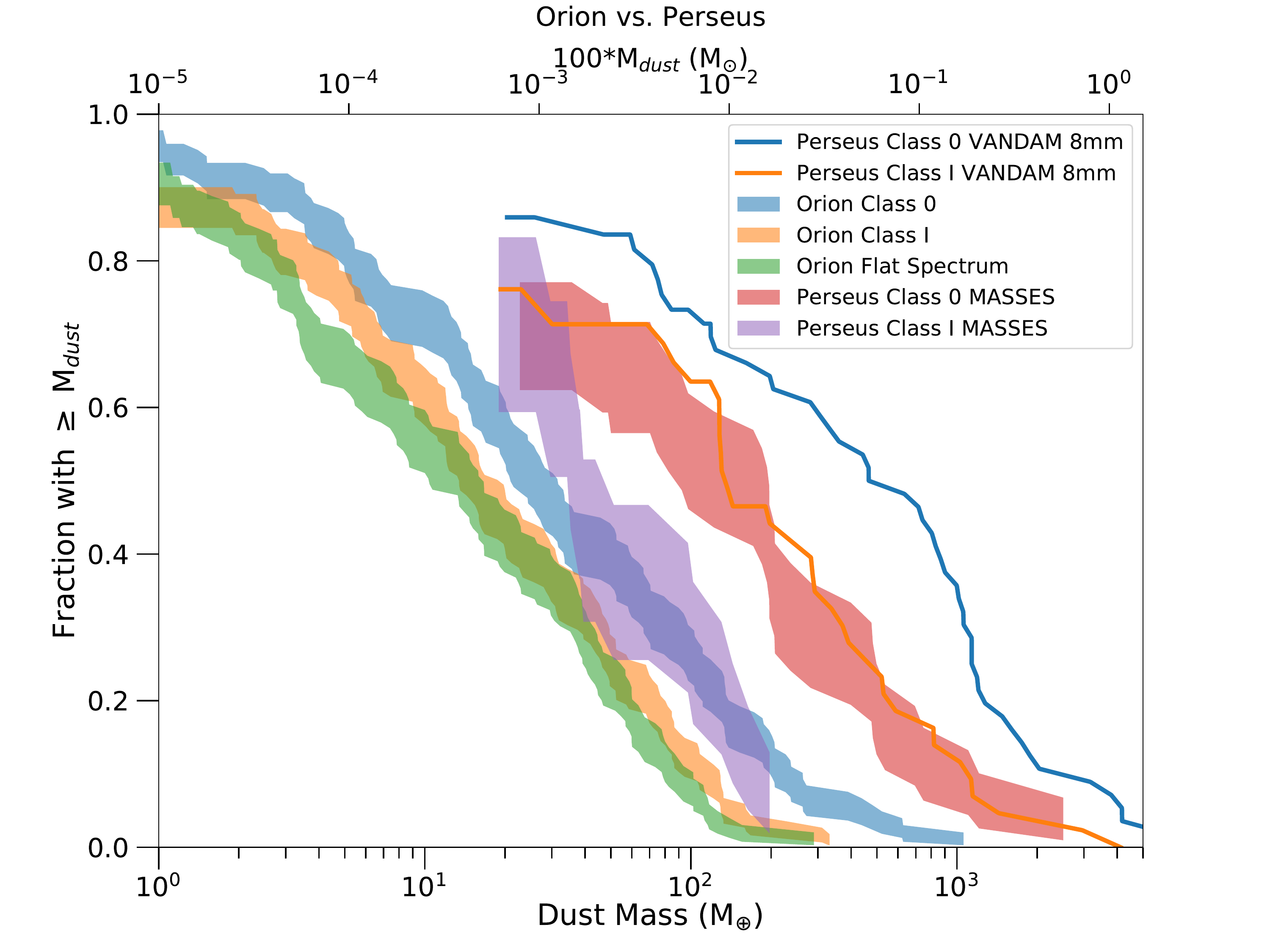}

\caption{
Cumulative distributions of dust disk masses within the full Orion sample relative
to the protostellar dust disk mass measurements in Perseus from both VANDAM 
\citep{tychoniec2018} and MASSES \citep{andersen2019}. The mass distributions
from \citet{tychoniec2018} are drawn as lines in this plot for clarity. This comparison
enables us to explore differences that might be due to the
wavelengths observed. The values from \citet[MASSES; ][]{andersen2019} indicate systematically
lower masses than those in \citet[VANDAM; ][]{tychoniec2018}, but are still not consistent
with the mass distributions of the Orion sample.
}
\label{masses-ori-perseus}
\end{figure}

Here we analyze the differences in the dust disk mass distributions
derived by \citet{andersen2019} and \citet{tychoniec2018} for the
Perseus protostars. \citet{andersen2019} used
an independent method to calculate dust disk mass from lower resolution data, finding a strong
correlation with the dust disk masses from \citet{seguracox2018} and \citet{tychoniec2018}.
\citet{andersen2019} utilized Subcompact array data from the SMA with $\sim$4\arcsec\ resolution,
assumed that the flux density at 50~k$\lambda$ is dominated by the disk, and corrected for the 
estimated contribution of the envelope at this angular scale \citep[see also][]{jorgensen2009}. The 
masses calculated by \citet{andersen2019} were consistent with those derived 
from the the 0\farcs2 resolution 9~mm data used by \citet{tychoniec2018}.
However, \citet{andersen2019} made different assumptions relative to this work and
\citet{tychoniec2018} for the calculation of dust temperatures; \citet{andersen2019} adopted average
dust temperatures of 30~K for the Class 0 protostars and 15~K for the Class I protostars, while
\citet{tychoniec2018} adopted 30~K for all protostars. We renormalized and scaled both the \citet{tychoniec2018} and
\citet{andersen2019} dust disk masses to be consistent with our luminosity-dependent dust temperature 
method. We also corrected for the updated distance to Perseus of $\sim$300~pc.
We plot the cumulative distributions for Class 0 and 
Class I protostars in Figure \ref{masses-ori-perseus}. Scaling the \citet{tychoniec2018} data to account
for the luminosity made little difference, but scaling the \citet{andersen2019} results shifted
the distributions to lower masses, but they are still not consistent with Orion.
It is unclear if the difference in the mass distribution between
Perseus and Orion results from a combination
of methodology, using unresolved observations \citep{andersen2019}, 
and/or wavelength \citep{tychoniec2018}, and the additional
uncertainty of the proper dust opacity at 8~mm.

\begin{figure}
\includegraphics[scale=0.31]{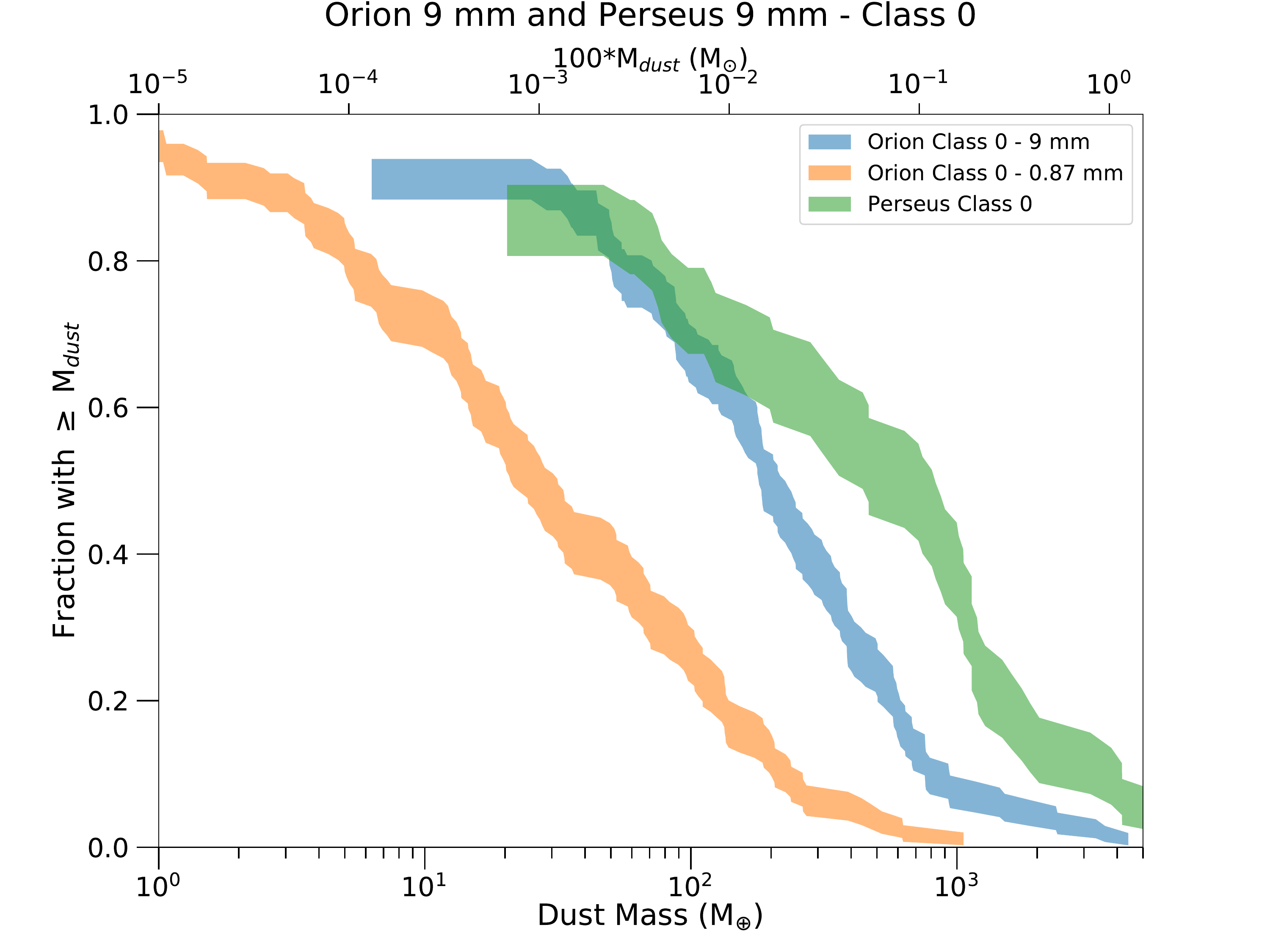}
\includegraphics[scale=0.31]{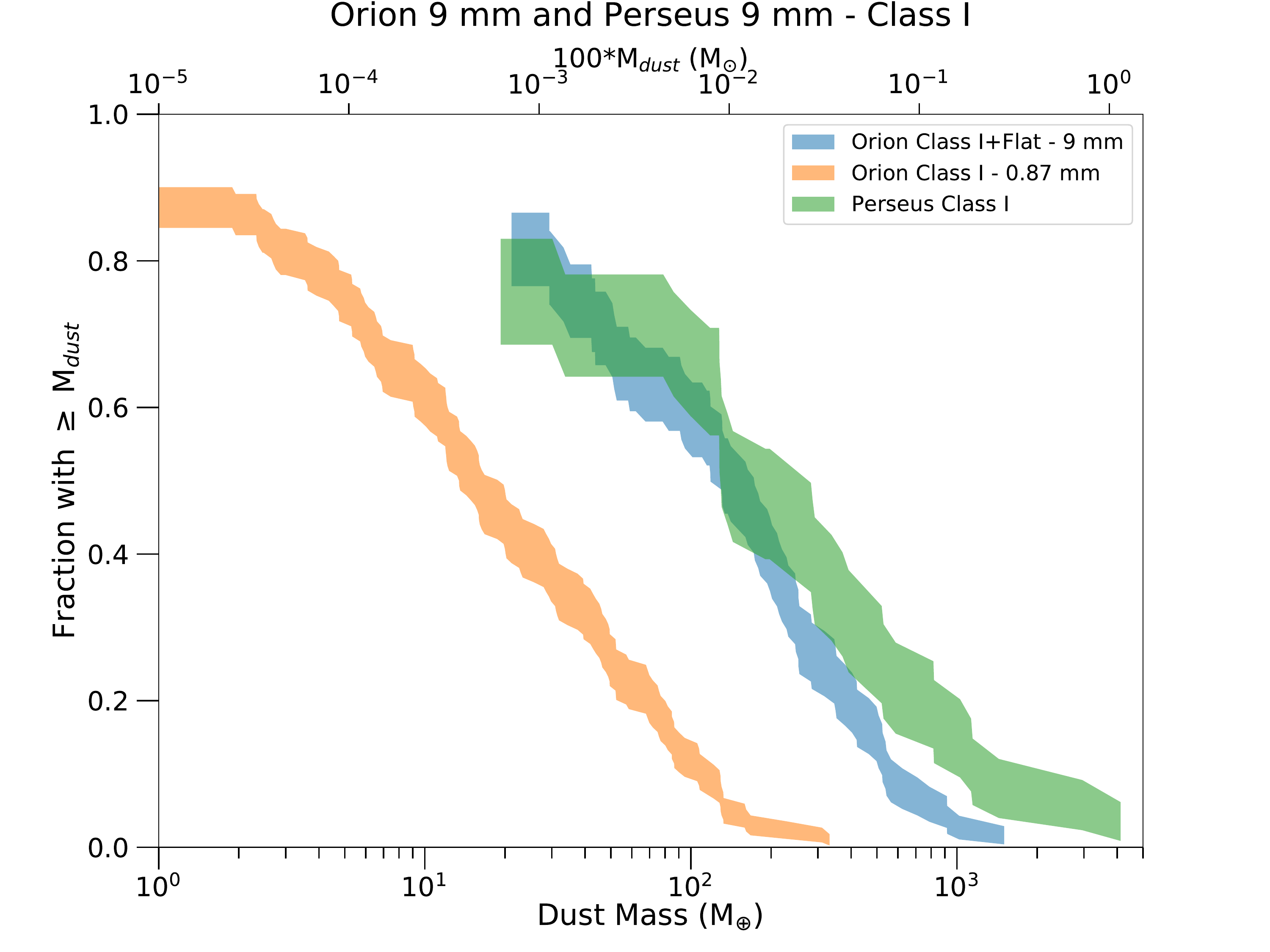}
\caption{
Cumulative distributions of dust disk masses within the Orion sample calculated from 
9~mm data
compared to the protostellar dust disk mass measurements in Perseus from VANDAM 
\citep{tychoniec2018}. The dust disk masses from the Orion 0.87~mm data
are also shown for comparison.
The Class 0 sources are shown in the left panel and the Class I (combined
with Flat Spectrum) in the right panel. Even without the free-free correction to the
Orion data, the Class 0 sources from Perseus are still calculated to have higher masses. The Class
I sources are more comparable, with some overlap in the distribution.
}
\label{masses-ori-perseus-9mm}
\end{figure}

To test if the dust opacity assumption at 9~mm is driving
the discrepancy between Perseus and Orion, we compared the Orion VLA dust disk masses (Table 8)
to the Perseus dust disk masses and show the result in Figure \ref{masses-ori-perseus-9mm}. 
The VLA 9~mm mass distributions for Orion are in much closer agreement with the Perseus
distributions than the 0.87~mm mass distribution,
but are still systematically shifted toward lower masses. The difference
is less extreme for the Class I sample, and the difference in wavelength from 8.1~mm to
9~mm could contribute to the disagreement. A log rank test indicates that the probability 
of the Class 0 and Class I samples for Perseus and Orion to be drawn from the same sample
is 1.6~$\times$~10$^{-6}$ and 0.01, respectively. However, these masses for the Orion disks are
upper limits because they do not have a free-free contribution removed, so the discrepancy could
be larger.

\begin{figure}
\includegraphics[scale=0.31]{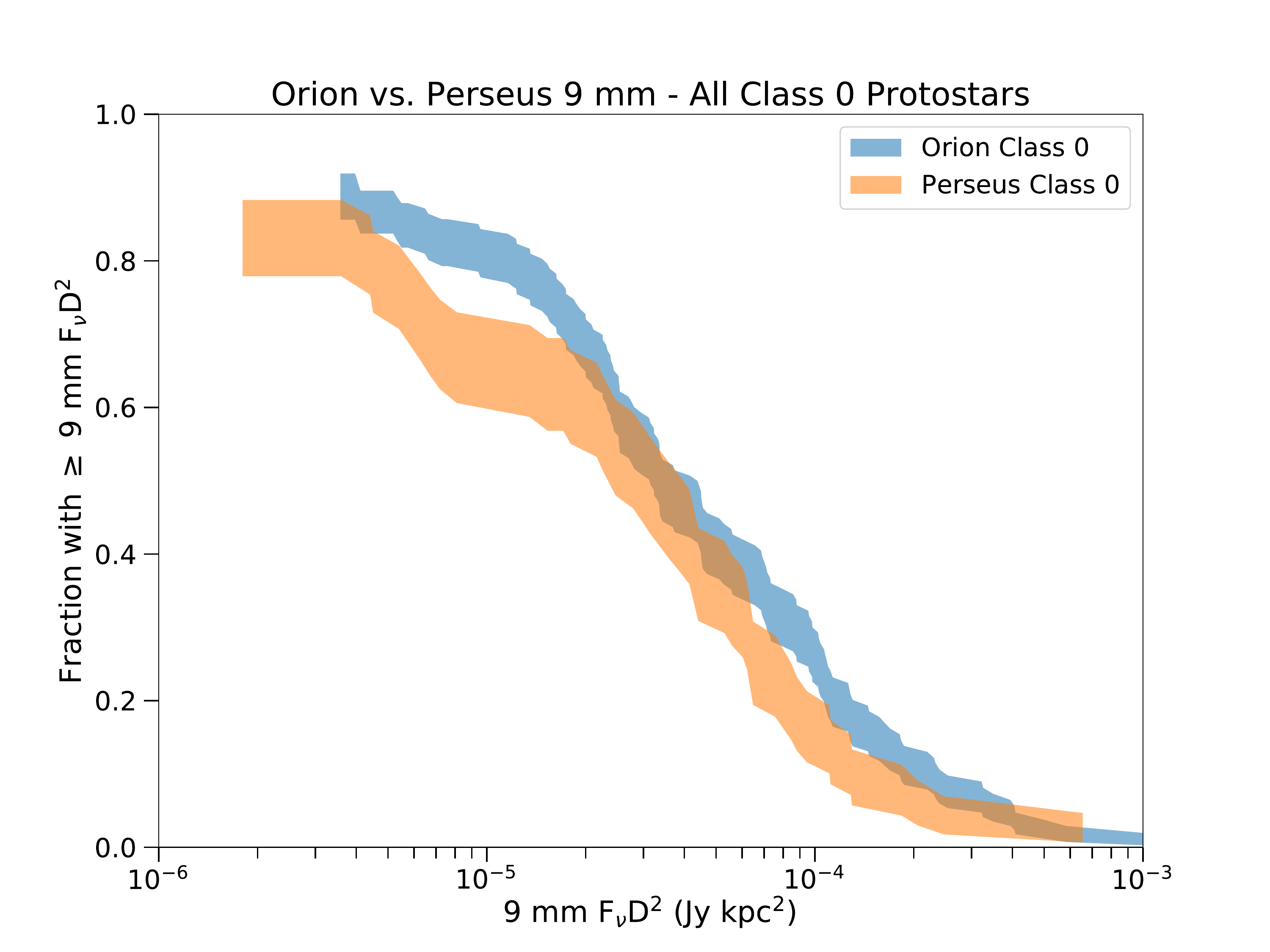}
\includegraphics[scale=0.31]{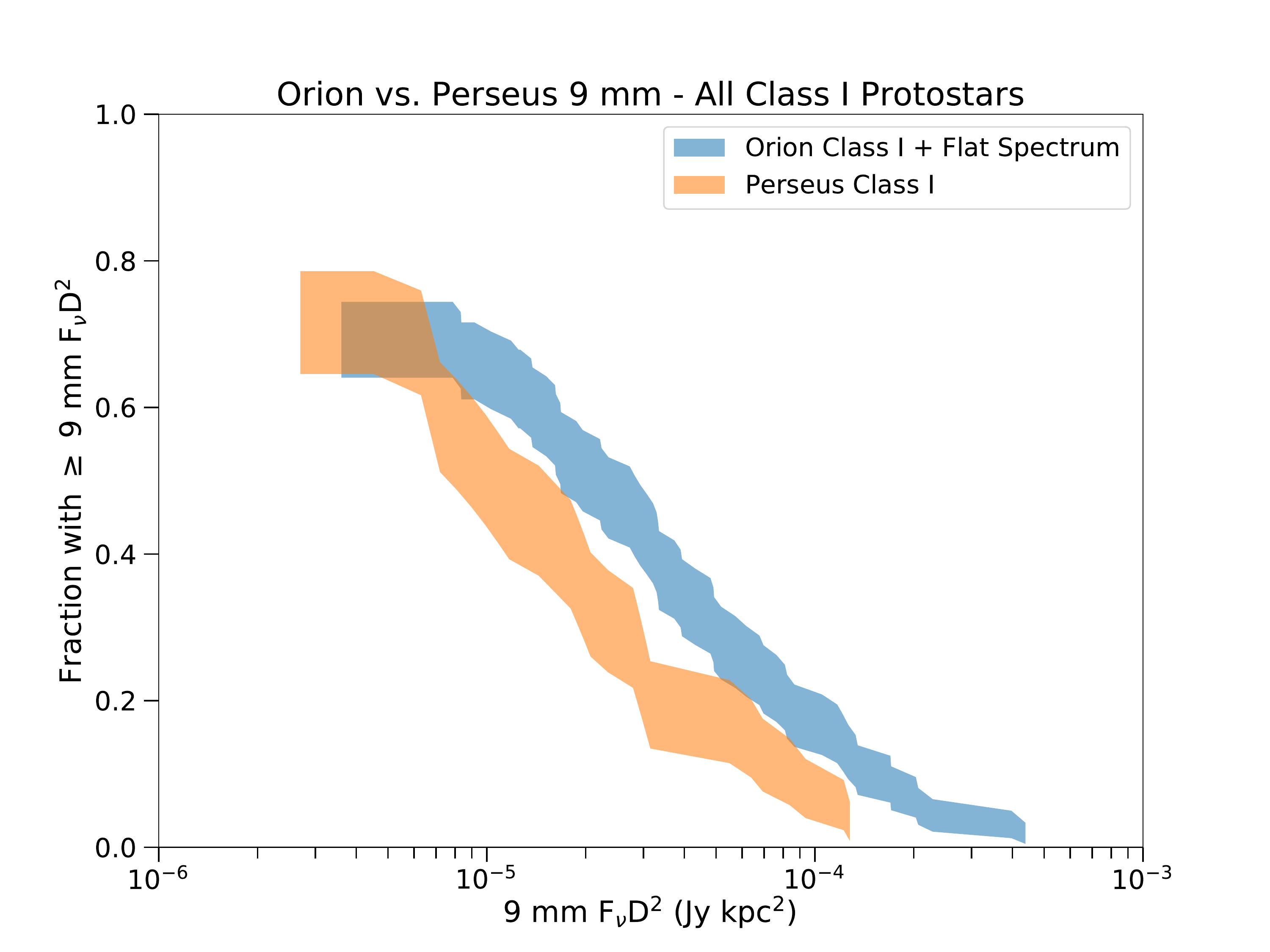}
\caption{
Cumulative distributions of 9~mm flux densities from the Orion and Perseus samples. 
The left panel shows Class 0 sources and the right panel shows the Class I sources.
Given that the Perseus Class I sample includes both Class I and Flat Spectrum sources,
we show both the distribution of 9~mm flux densities for the combined Class I and
Flat Spectrum samples. The Orion Class 0 flux densities at 9~mm are statistically indistinguishable from
the Perseus Class 0 flux densities at 9~mm, while the Class I flux densities at 9~mm from
Orion compared to Perseus are marginally consistent, having a probability of 0.01 
for being drawn from the same parent distribution.
}
\label{fluxes-ori-perseus-9mm}
\end{figure}

We also compared the Orion 9~mm flux densities to the Perseus 9~mm flux densities (both normalized by distance squared) in Figure
\ref{fluxes-ori-perseus-9mm} to compare the samples without the conversion to dust mass.
A log-rank test shows that
the Class 0 flux densities at 9~mm for Perseus and Orion are consistent with
being drawn from the same parent distribution with a probability of 0.12. The Perseus Class I 
and the Orion Class I + Flat Spectrum samples appear marginally inconsistent with being drawn from
the same parent distribution with a probability of 0.01. It is also apparent that the distributions
of masses for Class 0 and I protostars are shifted to higher masses for Perseus with respect to Orion, while the distributions
of 9~mm flux densities are lower. This can be best explained by the higher luminosities of the 
Orion protostars resulting in lower masses due to the higher average dust temperatures. The flux densities
compared here for both Perseus and Orion do not have correction for free-free emission, and since free-free
emission is correlated with bolometric luminosity \citep[e.g.,][]{tychoniec2018}, the higher luminosity protostars in
Orion are likely to have higher overall 9~mm flux densities due to increased free-free emission.

Regardless of the results from the statistical comparison of the samples,
we can see in Figure \ref{masses-ori-perseus-9mm} that the VLA 9~mm mass
distributions are shifted toward much higher masses than the ALMA 0.87~mm mass distributions. 
Thus, this is evidence that the dust opacity law at 9~mm
is significantly different from the adopted opacities of \citet{ossenkopf1994} and extrapolated
from 1.3~mm to 9~mm assuming a dust opacity spectral index of 1. The dust mass opacity at 9~mm
would need to be as much as $\sim$7$\times$ larger to bring the 0.87~mm and 9~mm distributions into closer
agreement. 

From this analysis, it is clear that the Perseus dust disk masses at 8.1~mm
may be significantly overestimated, and that further study of the Orion and Perseus
populations at comparable wavelengths and spatial resolution is needed to determine if the
mass distribution of Perseus is truly different from that of Orion. Moreover, additional investigation
into the dust opacities at centimeter wavelengths is also needed to help reconcile the differences between
observations of dust emission at very different wavelengths.

\clearpage

\input{tab1}

\input{tab2}

\input{tab3}

\clearpage
\input{tab4}

\clearpage
\input{tab5}

\input{tab6}
\input{tab7}
\input{tab8}
\input{tab9}
\input{tab10}

\end{document}

%% file: tab1.tex
\begin{deluxetable}{lllclll}
\tabletypesize{\scriptsize}
\tablewidth{0pt}
\tablecaption{ALMA Observations}

\tablehead{\colhead{Fields} & \colhead{Date} & \colhead{Duration} & \colhead{Maximum Baseline} & \colhead{Antennas} & \colhead{PWV}\tablenotemark{a} & \colhead{Calibrators}\\
                            &                & \colhead{(min)}     & \colhead{(m)}             &                    & \colhead{(mm)} &\colhead{Bandpass, Flux, Complex Gain}
}

\startdata
Orion B   & 2016 Sep 03 & 76 & 2483 & 36 & 0.39 & J0522-3627, J0750+1231, J0552+0313\\
Orion B   & 2016 Sep 04 & 84 & 2483 & 41 & 0.79 & J0510+1800, J0510+1800, J0552+0313\\
Orion A-1 & 2016 Sep 04 & 98 & 2483 & 41 & 0.73 &  J0510+1800, J0510+1800, J0541-0541\\
Orion A-1 & 2016 Sep 05 & 98 & 2483 & 41 & 0.53 &  J0510+1800, J0510+1800, J0541-0541\\
Orion A-2 & 2016 Sep 06 & 82 & 2483 & 39 & 0.42 &  J0510+1800, J0510+1800, J0541-0541\\
Orion A-2 & 2016 Sep 06 & 86 & 2483 & 34 & 0.43 &  J0510+1800, J0522-3627, J0541-0541\\
Orion A-1 & 2017 Jul 19 & 97 & 3697 & 41 & 0.47 & J0510+1800, J0423-0120, J0541-0541\\
Orion A-2 & 2017 Jul 19 & 84 & 3697 & 42 & 0.42 &  J0510+1800, J0423-0120, J0541-0541\\
Orion B   & 2017 Jul 19 & 75 & 3697 & 41 & 0.58 & J0510+1800, J0423-0120, J0552+0313\\
\enddata
\tablecomments{The Orion B fields include all HOPS protostars located at Declinations
greater than -4.5. The Orion A fields include HOPS protostars below Dec. = -4.5\degr;
fields denoted A-1 include protostars with numbers between 10 and 175, and fields denoted 
A-2 include HOPS protostars with numbers greater than 175.}
\tablenotetext{a}{Precipitable water vapor.}
\end{deluxetable}

%% file: tab2.tex
\startlongtable
\begin{deluxetable}{llllllll}
\tabletypesize{\scriptsize}
\tablewidth{0pt}
\tablecaption{ALMA Fields}

\tablehead{\colhead{Field} & \colhead{RA} & \colhead{Dec.} & \colhead{L$_{bol}$} & \colhead{T$_{bol}$} & \colhead{Distance} & \colhead{Class} & \colhead{Sources} \\
                           & \colhead{(J2000)} & \colhead{(J2000)}   & \colhead{(L$_{\sun}$)}  & \colhead{(K)}  & \colhead{(pc)}          &                   
}
\startdata
\input{table-alma-fields.tex}
\enddata
\tablecomments{The fields HOPS-395, HOPS-406, and HOPS-408 were observed twice due to a different target
name being assigned to the same position. The duplicate fields have a `-d' appended to their field name.
The field names that begin with S13 are taken from \citet{stutz2013}, which identified them as potential 
protostellar sources. However, no sources were detected within the S13 fields.}
\end{deluxetable}

%% file: table-alma-fields.tex
HH270VLA1      & 05:51:34.626 & +02:56:45.02 & 7.0 & 32.0 &  430.1 & 0    &  2\\
HH270mms1      & 05:51:22.631 & +02:56:06.73 & 8.3 & 72.0 &  405.7 & I    &  2\\
HH270mms2      & 05:51:22.559 & +02:55:42.78 & 4.7 & 249.0 &  413.3 & Flat &  1\\
HH111mms       & 05:51:46.238 & +02:48:29.65 & 23.0 & 78.0 &  410.6 & I    &  2\\
HOPS-367       & 05:54:36.259 & +01:53:54.02 & 0.0 & 249.4 &  354.9 & I    &  1\\
HOPS-7         & 05:54:20.045 & +01:50:42.75 & 0.5 & 58.0 &  357.1 & 0    &  1\\
HOPS-6         & 05:54:18.411 & +01:49:03.43 & 0.1 & 112.5 &  357.0 & I    &  0\\
HOPS-5         & 05:54:32.163 & +01:48:07.16 & 0.4 & 187.1 &  354.6 & I    &  1\\

%% file: tab3.tex
\startlongtable
\begin{deluxetable}{lcll}
\tabletypesize{\scriptsize}
\tablewidth{0pt}
\tablecaption{VLA Observations}

\tablehead{\colhead{Field(s)} & \colhead{Number of Fields} & \colhead{Date} & \colhead{Calibrators}\\
                              &                             & \colhead{(UT)} & \colhead{Bandpass, Flux, Complex Gain}
}

\startdata
\input{table-vla-obslog.tex}
\enddata
\tablecomments{
The standard observation length was 2.5 hours, and we achieved $\sim$1 hour on-source. For the observations
with Number of sources listed as 2, the observation length was 4 hours, and we were able to still reach
$\sim$1 hr on source for each source in the SB. Observing efficiency is improved due to sharing the
flux and bandpass calibration in addition to the fixed 10 minute overhead at the start of the observation.
}
\end{deluxetable}

%% file: table-vla-obslog.tex
HOPS-310     &   1 & 2016 October 20    & 3C84, 3C48, J0541-0541\\
HOPS-400     &   1 & 2016 October 21    & 3C84, 3C48, J0552+0313\\
HH111mms     &   1 & 2016 October 22    & 3C84, 3C48, J0552+0313\\
HOPS-401     &   1 & 2016 October 23    & 3C84, 3C48, J0552+0313\\
HOPS-399/372 &   1 & 2016 October 23    & 3C84, 3C48, J0552+0313\\
HOPS-370     &   1 & 2016 October 26    & 3C84, 3C48, J0541-0541\\
HOPS-203     &   1 & 2016 October 27    & 3C84, 3C48, J0541-0541\\
HOPS-50      &   1 & 2016 October 27    & 3C84, 3C48, J0541-0541\\

%% file: tab4.tex
\startlongtable
\begin{deluxetable}{llllllll}
\tabletypesize{\scriptsize}
\tablewidth{0pt}
\tablecaption{VLA Fields}

\tablehead{\colhead{Field} & \colhead{RA} & \colhead{Dec.} & \colhead{L$_{bol}$} & \colhead{T$_{bol}$} & \colhead{Distance} & \colhead{Class} & \colhead{Sources} \\
                           & \colhead{(J2000)} & \colhead{(J2000)}   & \colhead{(L$_{\sun}$)}  & \colhead{(K)}  & \colhead{(pc)}          &                   
}
\startdata
\input{table-vla-fields.tex}
\enddata
\end{deluxetable}

%% file: table-vla-fields.tex
HH270VLA1    & 05:51:34.731 & +02:56:46.80 & 7.0 & 32.0 &  430.1 & 0    &  3\\
HH270mms     & 05:51:22.624 & +02:55:53.40 & 8.3 & 72.0 &  405.7 & I    &  1\\
HH111mms     & 05:51:46.253 & +02:48:29.63 & 23.0 & 78.0 &  410.6 & I    &  2\\
HOPS-7       & 05:54:20.045 & +01:50:42.75 & 0.5 & 58.0 &  357.1 & 0    &  1\\
HOPS-354     & 05:54:24.252 & +01:44:19.39 & 6.6 & 34.8 &  355.4 & 0    &  1\\
HOPS-1       & 05:54:12.336 & +01:42:35.49 & 1.5 & 72.6 &  356.9 & I    &  1\\
HOPS-406     & 05:47:43.361 & +00:38:22.45 & 0.5 & 24.6 &  430.3 & 0    &  3\\
HOPS-343     & 05:47:59.030 & +00:35:32.85 & 3.9 & 82.1 &  427.6 & I    &  1\\

%% file: tab5.tex
\begin{deluxetable}{lcccc}
\tabletypesize{\scriptsize}
\tablewidth{0pt}
\tablecaption{Targeted Field Detection Summary}

\tablehead{\colhead{Sample} & \colhead{ALMA 0.87~mm}& \colhead{ALMA 0.87~mm} & \colhead{VLA 9~mm}& \colhead{VLA 9~mm} \\
                           &\colhead {Targeted}   & \colhead{Detected}     & \colhead{Targeted} & \colhead{Detected}
}
\startdata
Class 0       &  94 & 86 (125) & 88 & 76 (122)\\
Class I       & 128 & 109 (130) & 10 & 10 (43)\\
Flat Spectrum & 103 & 91  (118) & \nodata      & \nodata (26)  \\
OMC1N         & \nodata & \nodata (\nodata) & 4 & 4 (16)\\
Other         &   3 & 0  & \nodata & \nodata (25)\\
\hline
Total         & 328 & 286 (379) & 102  & 90 (232)\\
\enddata
\tablecomments{The numbers listed refer whether or not a source was detected in
apparent association with the HOPS protostar
targeted. The numbers in parentheses refer to the number of continuum
sources detected toward protostars in a given category. For OMC1N, we only targeted this 
region with 4 VLA fields and detected sources in all 4 fields. See Section 3.1 for
further details on the detection statistics.}
\end{deluxetable}

%% file: tab6.tex
\begin{longrotatetable}
\begin{deluxetable}{lllllllllllll}
\tabletypesize{\tiny}
\tablewidth{0pt}
\tablecaption{ALMA Source Properties}

\tablehead{\colhead{Source} & \colhead{RA}      & \colhead{Dec.}      & \colhead{ALMA Field}     & \colhead{$\Delta\phi_{A}$}  &  \colhead{VLA Field} & \colhead{$\Delta\phi_{V}$} &\colhead{Class}      & \colhead{F$_{\nu}$} & \colhead{Peak F$_{\nu}$} & \colhead{RMS} & \colhead{Decon. Size} & \colhead{Decon. PA}\\
                            & \colhead{(J2000)} & \colhead{(J2000)}   &                          & \colhead{(\arcsec)}            &                      & \colhead{(\arcsec)}          &                     & \colhead{(mJy)}      & \colhead{(mJy~bm$^{-1}$)} & \colhead{(mJy)}      & \colhead{(\arcsec)}    & \colhead{(\degr)}         
}
\startdata
\input{table-alma-data.tex}
\enddata
\end{deluxetable}
\end{longrotatetable}

%% file: table-alma-data.tex
HH270VLA1-A & 05:51:34.587 & +02:56:46.01 & HH270VLA1 & 1.1 & HH270VLA1 & 2.3 & 0    & 18.90$\pm$0.68 & 14.79 & 0.32 & 0.08$\times$0.03 & 143.9\\
HH270VLA1-B & 05:51:34.600 & +02:56:45.88 & HH270VLA1 & 0.9 & HH270VLA1 & 2.2 & 0    & 17.71$\pm$0.73 & 12.90 & 0.32 & 0.10$\times$0.04 & 141.2\\
HH270mms1-B & 05:51:22.632 & +02:56:06.80 & HH270mms1 & 0.1 & HH270mms  & 13.4 & I    & 9.57$\pm$2.56 & 7.06 & 1.06 & 0.08$\times$0.07 & 2.7\\
HH270mms1-A & 05:51:22.717 & +02:56:04.98 & HH270mms1 & 2.2 & HH270mms  & 11.7 & I    & 258.94$\pm$5.25 & 68.77 & 1.11 & 0.27$\times$0.14 & 87.4\\
HH270mms2  & 05:51:22.572 & +02:55:43.16 & HH270mms2 & 0.4 & HH270mms  & 10.3 & Flat & 75.44$\pm$2.17 & 14.20 & 0.38 & 0.25$\times$0.20 & 47.3\\
HH111MMS-B & 05:51:46.066 & +02:48:30.80 & HH111mms  & 2.8 & HH111mms  & 3.0 & I    & 25.44$\pm$4.56 & 21.91 & 2.43 & 0.07$\times$0.04 & 146.2\\
HH111MMS-A & 05:51:46.253 & +02:48:29.65 & HH111mms  & 0.2 & HH111mms  & 0.0 & I    & 694.09$\pm$15.68 & 101.69 & 2.28 & 0.55$\times$0.17 & 7.6\\
HOPS-367   & 05:54:36.287 & +01:53:54.06 & HOPS-367  & 0.4 & None      & \nodata & I    & 1.01$\pm$0.21 & 0.94 & 0.22 & \nodata & \nodata\\

%% file: tab7.tex
\begin{longrotatetable}
\begin{deluxetable}{lllllllllllll}
\tabletypesize{\tiny}
\tablewidth{0pt}
\tablecaption{VLA Source Properties}

\tablehead{\colhead{Source} & \colhead{RA}      & \colhead{Dec.}      & \colhead{ALMA Field}     & \colhead{$\Delta\phi_{A}$}  &  \colhead{VLA Field} & \colhead{$\Delta\phi_{V}$}  & \colhead{Class}     & \colhead{F$_{\nu}$} & \colhead{Peak F$_{\nu}$} & \colhead{RMS} & \colhead{Decon. Size} & \colhead{Decon. PA}\\
                            & \colhead{(J2000)} & \colhead{(J2000)}   &                          & \colhead{(\arcsec)}            &                      & \colhead{(\arcsec)}           &                     & \colhead{(mJy)}      &\colhead{(mJy~bm$^{-1}$)} &\colhead{($\mu$Jy)} & \colhead{(\arcsec)}    & \colhead{(\degr)}     
}
\startdata
\input{table-vla-data.tex}
\enddata
\end{deluxetable}
\end{longrotatetable}

%% file: table-vla-data.tex
HH270VLA1-A & 05:51:34.587 & +02:56:46.03 & HH270VLA1 & 1.1 & HH270VLA1 & 2.3 & 0    & 0.243$\pm$0.019 & 0.166  & 7.5 & 0.08$\times$0.03 & 143.0\\
HH270VLA1-C & 05:51:34.595 & +02:56:45.89 & HH270VLA1 & 1.0 & HH270VLA1 & 2.2 & 0    & 0.530$\pm$0.029 & 0.212  & 7.5 & 0.11$\times$0.09 & 50.3\\
HH270VLA1-B & 05:51:34.601 & +02:56:45.89 & HH270VLA1 & 0.9 & HH270VLA1 & 2.2 & 0    & 0.088$\pm$0.006 & 0.160  & 7.5 & \nodata & \nodata\\
HH270mms1-B & 05:51:22.632 & +02:56:06.80 & HH270mms1 & 0.1 & HH270mms  & 13.4 & I    & $\leq$0.02 & $\leq$0.02  & 8.9 & \nodata & \nodata\\
HH270mms1-A & 05:51:22.717 & +02:56:04.98 & HH270mms1 & 2.2 & HH270mms  & 11.7 & I    & 0.907$\pm$0.025 & 0.463  & 8.2 & 0.09$\times$0.06 & 98.6\\
HH270mms2  & 05:51:22.572 & +02:55:43.16 & HH270mms2 & 0.4 & HH270mms  & 10.3 & Flat & $\leq$0.02 & $\leq$0.02  & 8.3 & \nodata & \nodata\\
HH111MMS-B & 05:51:46.066 & +02:48:30.81 & HH111mms  & 2.8 & HH111mms  & 3.0 & I    & 0.172$\pm$0.016 & 0.142  & 7.9 & 0.04$\times$0.02 & 82.0\\
HH111MMS-A & 05:51:46.253 & +02:48:29.64 & HH111mms  & 0.2 & HH111mms  & 0.0 & I    & 2.183$\pm$0.032 & 0.870  & 7.9 & 0.13$\times$0.09 & 7.5\\

%% file: tab8.tex
\begin{longrotatetable}
\begin{deluxetable}{lllllllllllll}
\tabletypesize{\tiny}
\tablewidth{0pt}
\tablecaption{ALMA and VLA Derived Parameters}

\tablehead{\colhead{Source} & \colhead{L$_{bol}$}  & \colhead{T$_{bol}$} & \colhead{Class} & \colhead{R$_{2\sigma,ALMA}$} & \colhead{R$_{2\sigma,VLA}$} & \colhead{M$_{D,ALMA}$} & \colhead{M$_{D,VLA}$} & \colhead{Q$_{ALMA}$} & \colhead{Q$_{VLA}$} & \colhead{Sp. Index} & \colhead{Sp. Index}\\
                            & \colhead{(L$_{\sun}$)} & \colhead{(K)}     &                 & \colhead{(au)}            & \colhead{(au)}           & \colhead{(M$_{\earth}$)}  & \colhead{(M$_{\earth}$)}  & & &  \colhead{(0.87 - 9~mm)} &  \colhead{(8.1 - 10~mm)}   
}
\startdata
\input{table-alma-vla-data-derived.tex}
\enddata
\end{deluxetable}
\end{longrotatetable}

%% file: table-alma-vla-data-derived.tex
HH270VLA1-A & 7.3 & 32.0 & 0    & 30.7$\pm$10.0 & 28.5$\pm$10.0 & 13.4$\pm$0.5 & 242.4$\pm$19.0 & 27.696 & 1.585 & 1.9$\pm$0.07 & 1.0$\pm$0.63\\
HH270VLA1-C & 7.3 & 32.0 & 0    & \nodata & 39.1$\pm$10.0 & \nodata & 528.6$\pm$29.0 & \nodata & 0.688 & \nodata & -0.4$\pm$0.48\\
HH270VLA1-B & 7.3 & 32.0 & 0    & 36.5$\pm$10.0 & \nodata      & 12.6$\pm$0.5 & 87.8$\pm$6.0 & 28.387 & \nodata & 2.3$\pm$0.07 & 3.1$\pm$0.52\\
HH270mms1-B & 7.7 & 72.0 & I    & 27.9$\pm$18.9 & \nodata & 6.0$\pm$1.6 & $\leq$21.8 & 62.810 & \nodata & \nodata & \nodata \\
HH270mms1-A & 7.7 & 72.0 & I    & 92.3$\pm$10.0 & 32.4$\pm$10.0 & 161.3$\pm$3.4 & 793.9$\pm$23.1 & 1.729 & 0.467 & 2.4$\pm$0.06 & 1.9$\pm$0.23\\
HH270mms2  & 4.6 & 249.0 & Flat & 87.4$\pm$10.0 & \nodata & 56.7$\pm$1.6 & $\leq$24.9 & 5.804 & \nodata & \nodata & \nodata \\
HH111MMS-B & 22.0 & 78.0 & I    & 23.0$\pm$14.6 & 14.3$\pm$10.0 & 12.2$\pm$2.3 & 118.5$\pm$11.1 & 24.084 & 2.761 & 2.1$\pm$0.11 & 1.9$\pm$0.79\\
HH111MMS-A & 22.0 & 78.0 & I    & 192.2$\pm$10.0 & 44.3$\pm$10.0 & 332.0$\pm$7.5 & 1504.3$\pm$22.1 & 0.508 & 0.180 & 2.5$\pm$0.06 & 1.3$\pm$0.12\\

%% file: tab9.tex
\begin{deluxetable}{lccccccc}
\tabletypesize{\scriptsize}
\tablewidth{0pt}
\tablecaption{Disk Masses and Radii Distribution Properties}

\tablehead{\colhead{Sample} & \colhead{Mean $M_{dust}$} & \colhead{$\sigma$(log$_{10}$$\left(\frac{M_{dust}}{M_{\earth}}\right)$)} & \colhead{Median $M_{dust}$} & \colhead{Mean R$_{Disk}$} & \colhead{$\sigma$(log$_{10}$$\left(\frac{R_{disk}}{1~AU}\right)$)} & \colhead{Median $R_{disk}$} & Ref.\\
                           & \colhead{($M_{\earth}$)}   &    & \colhead{(M$_{\earth}$)} & \colhead{(AU)} & & \colhead{(AU)}                  
}
\startdata
All Sources (including multiples)\\
\hline   
Class 0                   & 25.9$^{+7.7}_{-4.0}$ & 0.80$^{+0.003}_{-0.010}$ & 25.7$^{+102.9}_{-6.7}$  & 44.9$^{+5.8}_{-3.4}$ & 0.38$^{+0.002}_{-0.004}$ & 48.1$^{+79.6}_{-24.5}$ & 1\\ 
Class I                   & 14.9$^{+3.8}_{-2.2}$ & 0.76$^{+0.008}_{-0.019}$ & 15.6$^{+49.5}_{-4.9}$   & 37.0$^{+4.9}_{-3.0}$ & 0.42$^{+0.001}_{-0.005}$ & 38.1$^{+64.0}_{-17.5}$ & 1\\ 
Flat Spectrum             &  11.6$^{+3.5}_{-1.9}$ & 0.81$^{+0.011}_{-0.016}$ & 13.8$^{+43.8}_{-3.3}$  & 28.5$^{+3.7}_{-2.3}$ & 0.38$^{+0.001}_{-0.007}$  & 30.9$^{+51.3}_{-13.0}$ & 1\\ 
Orion Class 0  (T = 20K, $\kappa$=3.45) & 111.6$^{+36.5}_{-19.0}$ & 0.87$^{+0.010}_{-0.017}$ & 116.7$^{+420.8}_{-30.3}$   & & & & 1\\
Orion Class I  (T = 20K, $\kappa$=3.45)    & 42.1$^{+13.2}_{-7.4}$ & 0.91$^{+0.017}_{-0.029}$ & 48.5$^{+164.4}_{-10.8}$  & & & & 1\\
Orion Flat Spectrum (T = 20K, $\kappa$=3.45) & 34.3$^{+11.1}_{-5.9}$ & 0.87$^{+0.007}_{-0.017}$ & 40.8$^{+124.9}_{-9.8}$ & & & & 1\\
Northern ISF              &  14.5$^{+6.1}_{-2.6}$ & 0.77$^{+0.001}_{-0.013}$ & 14.7$^{+58.4}_{-5.0}$  & 36.3$^{+7.5}_{-3.6}$ & 0.41$^{+0.001}_{-0.009}$ & 38.4$^{+63.1}_{-18.0}$ & 1\\
\enddata
\tablecomments{
The mean disk masses and standard deviation of the log-normal distribution is
derived from fitting the cumulative distribution function derived from
the survival analysis with a Gaussian cumulative distribution function.
The median disk dust masses include non-detections, while the median radii do not
include non-detections. The sub- and superscripts on the median values correspond
to the first and third quartiles of the distributions and are absolute values, not relative
to the median.
Multiple sources refer to any system with a companion detected
with the VLA or ALMA within 10000~AU.
References: This work (1), \citet{williams2019} (2), \citet{sheehan2017b} (3),
\citet{tychoniec2018} (4), \citet{pascucci2016} (5), \citet{ansdell2016} (6), \citet{tripathi2017} (7),
\citet{eisner2018} (8), and \citet{barenfeld2016} (9).
}
\end{deluxetable}

%% file: tab10.tex
\begin{deluxetable}{lllll}
\tabletypesize{\scriptsize}
\tablewidth{0pt}
\tablecaption{Disk Mass and Radius Sample comparison}

\tablehead{\colhead{Sample} & \colhead{Disk Mass} &\colhead{Number} & \colhead{Disk Radius} &\colhead{Number}\\
&\colhead{Probability} & \colhead{Srcs} & \colhead{Probability} & \colhead{Srcs}}
\startdata
All Sources (including multiples) \\
\hline   
Class 0 vs. Class I                      & 0.0014   (0.017)    & 133, 150 (125, 113)  & 0.63 (0.59)   & 131, 147 (104, 112)\\
Class 0 vs. Flat Spectrum                & 0.000014 (0.00045)  & 133, 132 (125, 111)  & 0.0002 (0.00054) & 131, 129 (104, 88)\\
Class I vs. Flat Spectrum                & 0.061    (0.045)    & 150, 132 (113, 111)  & 0.003 (0.0089) & 147, 129 (112, 88)\\
Northern ISF vs. L1641 and Southern ISF  & 0.71                & 76, 235              & 0.78   & 72, 203\\
Northern ISF vs. Orion B                 & 0.062               & 76, 113              & 0.54   & 72, 105\\
Orion B vs. L1641 and Southern ISF       & 0.021               & 113, 235             & 0.24  & 105, 203\\
\hline
Non-multiple Sources & \\
\enddata

\tablecomments{The values in parentheses represent the probability
values from the Anderson-Darling test that was used as an additional
check on the robustness of the log-rank test for the cases where 
distributions may have crossed, limiting the effectiveness of the
log-rank test. The only case where the Anderson-Darling test and
the log-rank test differ significantly is Class 0 vs. Flat Spectrum
for non-multiple sources.}
\end{deluxetable}